\documentclass[twocolumn,aps,superscriptaddress,nofootinbib]{revtex4}

\usepackage{amsmath,amssymb}
\usepackage{graphicx}
\usepackage{dcolumn}
\usepackage{bm}
\usepackage[usenames]{color}
\usepackage[breaklinks=true,colorlinks=true,linkcolor=blue,urlcolor=blue,citecolor=blue]{hyperref}
\usepackage{ulem}
\usepackage{amsmath}
\usepackage{ulem}

\begin{document}

\title{ {Driven p}olymer translocation into a channel: Iso-flux tension propagation
theory and Langevin dynamics simulations}

\author{Jalal Sarabadani}
\email{jalal@ipm.ir}
\affiliation{School of Nano Science, Institute for Research in Fundamental
Sciences (IPM), 19395-5531, Tehran, Iran.}
\author{Ralf Metzler}
\affiliation{Institute for Physics \& Astronomy, University of Potsdam,
14476 Potsdam, Germany}
\author{Tapio Ala-Nissila}
\affiliation{Department of Applied Physics and QTF Center of Excellence,
Aalto University, P.O. Box 11000, FI-00076 Aalto, Espoo, Finland.}
\affiliation{Interdisciplinary Centre for Mathematical Modelling and
Department of Mathematical Sciences, 
Loughborough University, Loughborough, Leicestershire LE11 3TU, UK.}

\begin{abstract}
Iso-flux tension propagation (IFTP) theory and Langevin dynamics
(LD) simulations are employed to study the dynamics of channel-driven
polymer translocation in which a polymer translocates into a narrow channel and the
monomers in the channel experience a driving force $f_{\rm c}$. In the high
driving force limit, regardless of the channel width, IFTP theory predicts
$\tau\propto f_{\textrm{c}}^{\beta}$ for the translocation
time, where $\beta=-1$ is the force scaling exponent. Moreover, LD data show
that for a very narrow channel fitting only a single file of monomers,
the entropic force due to the subchain inside the channel does not play
a significant role in the translocation dynamics, and the force exponent $\beta=
-1$ regardless of the force magnitude. As the channel width increases
the number of possible spatial configurations of the subchain inside the
channel becomes significant, and the resulting entropic force causes
the force exponent to drop below unity.
\end{abstract}

\maketitle

\section{Introduction} \label{intro}

Starting with the seminal experimental works of Bezrukov {\it et al.} \cite{Bezrukov}
and Kasianowicz {\it et al.} \cite{KasiPNAS1996} as well as the theoretical study of
Sung and Park \cite{SungPRL1996}, the understanding
of the physical mechanisms behind the process of polymer translocation through
a nanopore has attracted considerable interest both from experimental
\cite{mellerPRL2001,BrantonPRL2003,Storm2003,Keyser_2006,RadenovicNanoLett2007,
Keyser_2009,RadenovicNanoscale2012,RadenovicNutreNanoTech2013,Bulushev_2015} 
and theoretical \cite{Muthukumar_book,Tapio_review,jalalJPCM2018,MilchevJPCM2011,
SungPRL1996,MuthukumarJCP1999,Chuang2001,SlaterRatchet,Kantor_PRE2004,tobias,roya,
GrosbergPRL2006,Luo2006,SakauePRE2007,aksimentievNanolett2008,slaterPRE2009,kaifu,
kaifu1,kaifu2,kaifu3,SakauePRE2010,
slaterPRE2010,rowghanian2011,golestanianPRL2011,RalfJCP2011,SakauePRE2012,
ikonen2012a,ikonen2012b,slaterJCP2012,hamidJCP2013,jalalJCP2014,
MichelettiMacLett2015,Aniket_2015,jalalJCP2015,menais_2017,JalalEPL2017,
MichelettiPNAS2017,jalalSciRep2017,menais_2018,jalalPolymers2018,
jalalPolymers2019,Aniket_2020,jalalJPCM2020_1,jalalJPCM2020_2,jalalPRR2021} 
groups. The process of { {polymer translocation}} has a wide
variety of applications in many different areas, ranging from gene transfer
between bacteria \cite{bact}, RNA transport through nuclear membrane pores
\cite{nuclear}, over DNA sequencing and drug delivery 
\cite{Branton_Naturebio,Deamer_Naturebio,drug_DNA}
to the filtration of polyelectrolytes by entropic ratchets \cite{SlaterRatchet}.

The translocation of a polymer can be either driven \cite{SakauePRE2007,
rowghanian2011,jalalJPCM2018} or unbiased 
\cite{Chuang2001,Luo2006,slaterPRE2009,slaterPRE2010}.
There exist several scenarios for the driven case, e.g., the external driving
force can be localized in the nanopore \cite{SakauePRE2007,rowghanian2011}, be
due to interaction of chaperones \cite{tobias,roya} or active rods with the
{\it trans} side subchain \cite{jalalPRR2021}, or can be applied on the
head monomer of the polymer (the end-pulled case \cite{JalalEPL2017}) by a
magnetic or optical tweezers \cite{Keyser_2006,Keyser_2009,Bulushev_2014,
Bulushev_2015,Bulushev_2016}, or even by an atomic force microscope (AFM)
\cite{RitortJPCMatt2006}. In the localized case in which the driving acts
on the monomer(s) inside the nanopore, the driving force may alternate
\cite{aksimentievNanolett2008,FaloPRE2015,LeeChemComm2012,MellerBiophysJ2003},
and the process be influenced by flickering of the nanopore \cite{jalalJCP2015,
golestanianPRL2011}. The dynamics of the translocation process for all of the
above scenarios has been theoretically described by the iso-flux tension
propagation (IFTP) theory \cite{SakauePRE2007,rowghanian2011,jalalJPCM2018,
jalalJCP2015,jalalPRR2021}. Particularly, in Ref. \cite{jalalJCP2015} the IFTP approach
to pore driven polymer translocation through a flickering nanopore under an
alternating external driving force was studied for three different regimes
of weak, moderate, and strong driving.

In the weakly driven or unbiased cases the entropic force, arising from
the fact that the polymer chain can assume different configurations in
space, plays an important role \cite{SungPRL1996} and has a significant
contribution to the translocation dynamics. The spatial configurations of
the chain can be controlled either by external fields and/or by the
presence of a confining geometry \cite{FleerBook,DeGennesBook}. 
An interesting realization occurs when the nanopore is
replaced by a long nanochannel, and the polymer is attracted into the
channel by an external force acting on all monomers inside the channel
(the "chain sucker") \cite{RalfJCP2011}. As over time more monomers
migrate into the channel, the net external driving force inside the channel naturally
increases. In this case, due to the external channel driving force and
the confinement by the channel the {\it trans}-side subchain possesses
less spatial configurations with respect to the free space case.
Consequently, the entropic force depends on both the channel driving
force $f_{\rm c}$ and the channel diameter $D$, as we will demonstrate below.

\begin{figure*}
\includegraphics[width=1.0\textwidth]{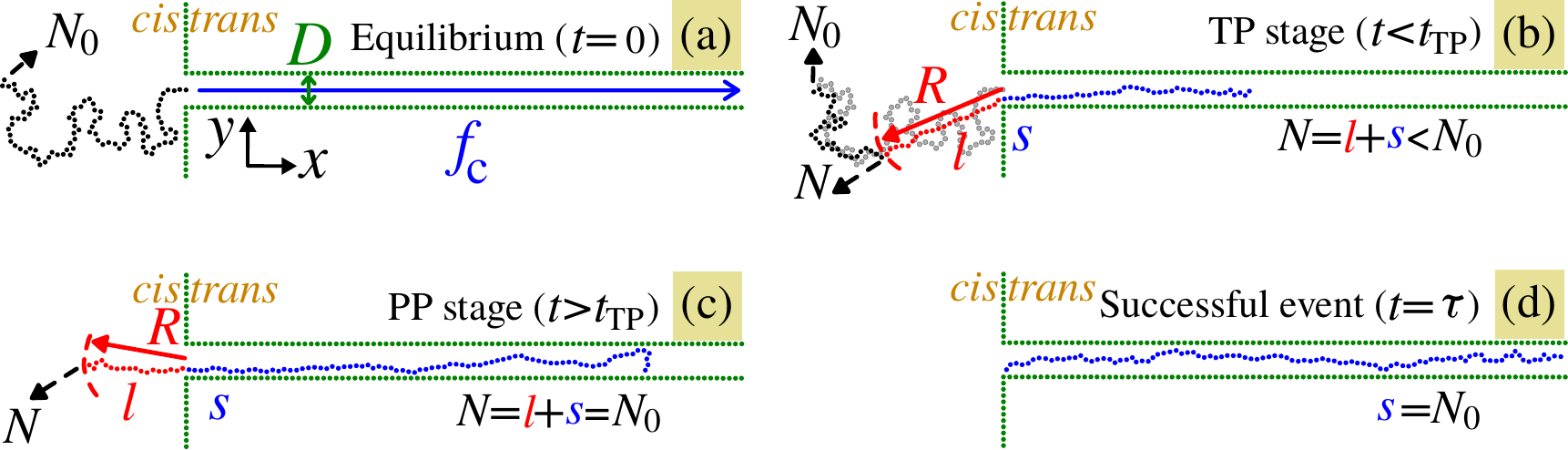}
\caption{(a) Chain configuration on the {\it cis} side after equilibration and
before the filling process. The chain contour length and
the channel width are $N_0 = 100$ and $D$, respectively. The external driving
force, $f_{\textrm{c}}$, acts on the monomer(s) inside the channel (defined to be the {\it
trans} side here) during the translocation process. (b) The system configuration
during the tension propagation (TP) stage when $t <  {t}_{\textrm{TP}}$.
$R$ is the tension front location, $s$ is the number of monomers on the {\it
trans} side (translocation coordinate), $l$ is the number of beads under
the propagating tension on the {\it cis} side, and $N$ is the number of all
beads that are already influenced by the tension force. During the TP stage $N=
l+s<N_0$, as the tension has not yet reached the chain end on the {\it cis}
side. The chain part in gray color is the polymer configuration at time zero,
shown for comparison of the polymer configuration at time $t$ with
the one at time zero. (c) This panel is the same as (b) but for the post
propagation (PP) stage when $t >  {t}_{\textrm{TP}}$. Here, the whole chain
including the last bead of the {\it cis}-side chain end has already
experienced the tension force and $N= l + s = N_0$. (d) Final configuration
of the chain at the end of the translocation process.}
\label{fig:schematic}
\end{figure*}

For polymer traversing a pore of finite length $L$, the total translocation time
$\tau$ comprises three separate parts and can be written as 
$\tau = \tau_1 + \tau_2 + \tau_3$. Here $\tau_1$, $\tau_2$, and $\tau_3$ 
correspond to pore filling, traversing through the pore, and pore emptying, respectively
\cite{Luo2007,Luo_SoftMatter_2012}. In the present work we focus on the case where the pore length $L$
is much longer than the polymer length $N_0$, {\it i.e.} $L \gg N_0$ and thus the pore filling time $\tau_1$ (see Fig. \ref{fig:schematic})
is identified as the translocation time. Accordingly, the channel is considered to be the {\it trans} compartment
in this process \cite{RalfJCP2011}. By using
extensive Langevin dynamics (LD) simulations and the IFTP theory
\cite{SakauePRE2007,rowghanian2011,jalalJCP2014, jalalJCP2015}
we show that in the strong channel driving force limit and in the
absence of the entropic force (or when the entropic force is much weaker than
the driving) the scaling form for the filling time is given by $\tau
\propto  f_{\textrm{c}}^{\beta}$, where $\tau$, $f_{\textrm{c}}$ and $\beta =
-1$ are the average translocation time, the force acting on each individual
monomer inside the channel, and the force scaling exponent, respectively.
This dependence is the same as in the short-pore driven translocation in the high force
limit \cite{jalalJPCM2018}. We also verify this result with LD simulations for a narrow channel that 
allows single-file translocation only. Further, we demonstrate that when the channel diameter
 $D$ increases, the increased chain fluctuations inside the channel can cause deviations
 from the narrow-pore limit and the magnitude of the exponent $\beta$ decreases.

With experimental setups similar to the scenario considered here the genomic 
information has been visualized by studying DNA conformations inside long 
confining micro or nanochannels \cite{Tegenfeldt_ABC2004,Tegenfeldt_PNAS2004,%
Tegenfeldt_PNAS2005,Tegenfeldt_book,Tegenfeldt_CSR2010}. 
We expect that comparison of such experiments with the observations reported 
here may lead to a further quantitative testing of the predictions of 
translocation theory.

The outline of the paper is as follows. We present details on the
simulation methods in Sec.~\ref{LD_SIM} and introduce the IFTP approach
in Sec.~\ref{IFTP}. Section~\ref{results} is devoted to the results,
comparing IFTP predictions with the behavior observed in the LD simulations.
Finally, summary and conclusions are in Sec.~\ref{conclusion}.

\section{Langevin dynamics simulations}
\label{LD_SIM}

 {For numerical accuracy and efficiency we consider here a 2D system which is
well justified for confined systems in particular. We note that the theory contains the
general Flory scaling exponent that has different values in 2D and 3D.}
The 2D system under consideration is composed of a linear,
flexible self-avoiding polymer modeled by a bead-spring model
\cite{grest} and a rigid membrane including a long channel as depicted
in Fig.~\ref{fig:schematic}. The interaction between any two bonded monomers
is the sum of the Weeks-Chandler-Anderson (WCA) and finitely extensible
nonlinear elastic (FENE) potentials. The WCA potential is the repulsive
\begin{equation}
U_{\textrm{WCA}}(r)=
\begin{cases}
U_{\textrm{LJ}}(r)-U_{\textrm{LJ}}(r_{\textrm{c}}), & \text{if $r \leq r_{\textrm{c}}$};\\
0, & \text{if $r > r_{\textrm{c}}$,}
\end{cases}
\label{WCAPot}
\end{equation}
where $r_{\textrm{c}}=2^{1/6}$ is the cutoff radius and $U_{\textrm{LJ}}(r)$
is the LJ potential
\begin{equation}
\label{LJPot}
U_{\textrm{LJ}}(r)=4\varepsilon\left[\left(\frac{\sigma}{r}\right)^{12}-
\left(\frac{\sigma}{r}\right)^{6}\right],
\end{equation}
with the depth $\varepsilon$ of the potential well, $\sigma$ is the radius of
each monomer, and the monomer-monomer distance is $r$. Nearest neighbor
monomers interact via the FENE potential
\begin{equation}
\label{FENEPot}
U_{\textrm{FENE}}(r)=-\frac{1}{2}kR^2_0\ln\left(1-\frac{r^2}{R_0^2}\right),
\end{equation}
where $R_0$ and $k$ are the maximum distance allowed between the neighboring
monomers and the effective spring constant, respectively. All non-bonded interactions
are given by the WCA potential. The contour length of the polymer in
LJ units is $N_0=100$ and the channel length $L=200$ is larger than
$N_0$. 

The simulation box dimensions are $L_x=400$ and $L_y=300$ in the $x$ and $y$
directions, respectively, with periodic boundaries in the $y$ direction. 
The membrane walls inside and outside of the pore are made of spatially fixed (frozen) 
particles located at distance $\sigma$ from each other (small green circles in 
Fig.~\ref{fig:schematic}). 
{The interactions between monomers and frozen particles (membrane and nanochannel particles) are governed by the repulsive WCA potential.}
In Fig.~\ref{fig:schematic}(a) $D$ denotes the channel 
width that varies as $D=2$, $3$, $4.5$, and $6$ in the LD simulations. The external 
driving force $f_{\textrm{c}}$ (depicted in Fig.~\ref{fig:schematic}(a) in blue color), 
acts horizontally on each individual monomer inside the channel and assumes the 
values $f_{\textrm{c}}=0.2$, $0.3$, $0.5$, and $1.2$ (in LJ units). The resulting 
dynamical equation for the position vector $\mathbf{r}_i$ of monomer $i$ then reads
\begin{equation}
\label{LD}
M\ddot{\mathbf{r}}_i=-\eta \dot {\mathbf{r}}_i-(\nabla U_i)+\boldsymbol{\xi}_i(t),
\end{equation}
where $M$ is the monomer mass, $\eta$ the solvent friction coefficient, and
$U_i$ is the sum of all interactions of the $i$th monomer. Here, $\boldsymbol
{\xi}_i$ represents thermal white noise  {vector} with zero mean $\langle\boldsymbol{
\xi}_i(t)\rangle=  {{\bm 0}}$ and correlation $\langle\boldsymbol{\xi}_i(t) {\cdot} \boldsymbol{
\xi}_j(t')\rangle= {4}\eta k_{\textrm{B}}T\delta_{ij}\delta(t-t')$, where $k_{\textrm{B}}$ and $T$ are the Boltzmann
constant and temperature, respectively, $\delta_{ij}$ is the Kronecker, and
$\delta(t-t')$ the Dirac delta function. Moreover, $\sigma$, $M$, and
$\varepsilon$ are used as the units of length, mass, and energy, respectively.
The diameter of the monomers is $\sigma=1$, and this also determines the size
of each channel or membrane particle. The mass of each monomer and other 
particles in the system is given by $M=1$. We choose $\varepsilon=1$, $\eta=0.7$, 
and $k_{\textrm{B}}T=1.2$. The time unit for the simulations is defined as 
$\tau_0=\sqrt{M\sigma^2/\varepsilon}$. The maximum allowed distance between 
two bonded monomers is $R_0=1.5$, and $k=30$ is the spring constant. 

All LD 
simulations were performed by using the LAMMPS \cite{lammps} package. 
In our model the size of each bead is about $1.5$ nm in real units
which corresponds approximately to the Kuhn length of a single-stranded DNA, 
the mass of each bead is about $936$ amu, and the strength of interaction 
is $3.39 \times 10^{-21}$ J at $T=295$ K (room temperature). Therefore, the force 
and time scales in LJ units are approximately $2.3$ pN and $32.1$ ps, respectively.

Within IFTP theory below we use dimensionless quantities, denoted by the
tilde, as $\tilde{Y}\equiv Y/Y_u$. Here the denominator indicates the
units of time, length, monomer flux, velocity, force, and friction as $t_u
\equiv\eta\sigma^2/(k_{\rm B}T)$, $s_u\equiv\sigma$, $\phi_u\equiv k_{\rm
B}T/(\eta\sigma^2)$, $v_u\equiv\sigma/t_u=k_{\rm B}T/(\eta\sigma)$, $f_u
\equiv k_{\rm B}T/\sigma$, and $\Gamma_u\equiv\eta$, respectively. The
parameters without the tilde symbol are expressed in LJ units.

\begin{figure*}
    \begin{minipage}{0.27\textwidth}
    \begin{center}
        \includegraphics[width=1.0\textwidth]{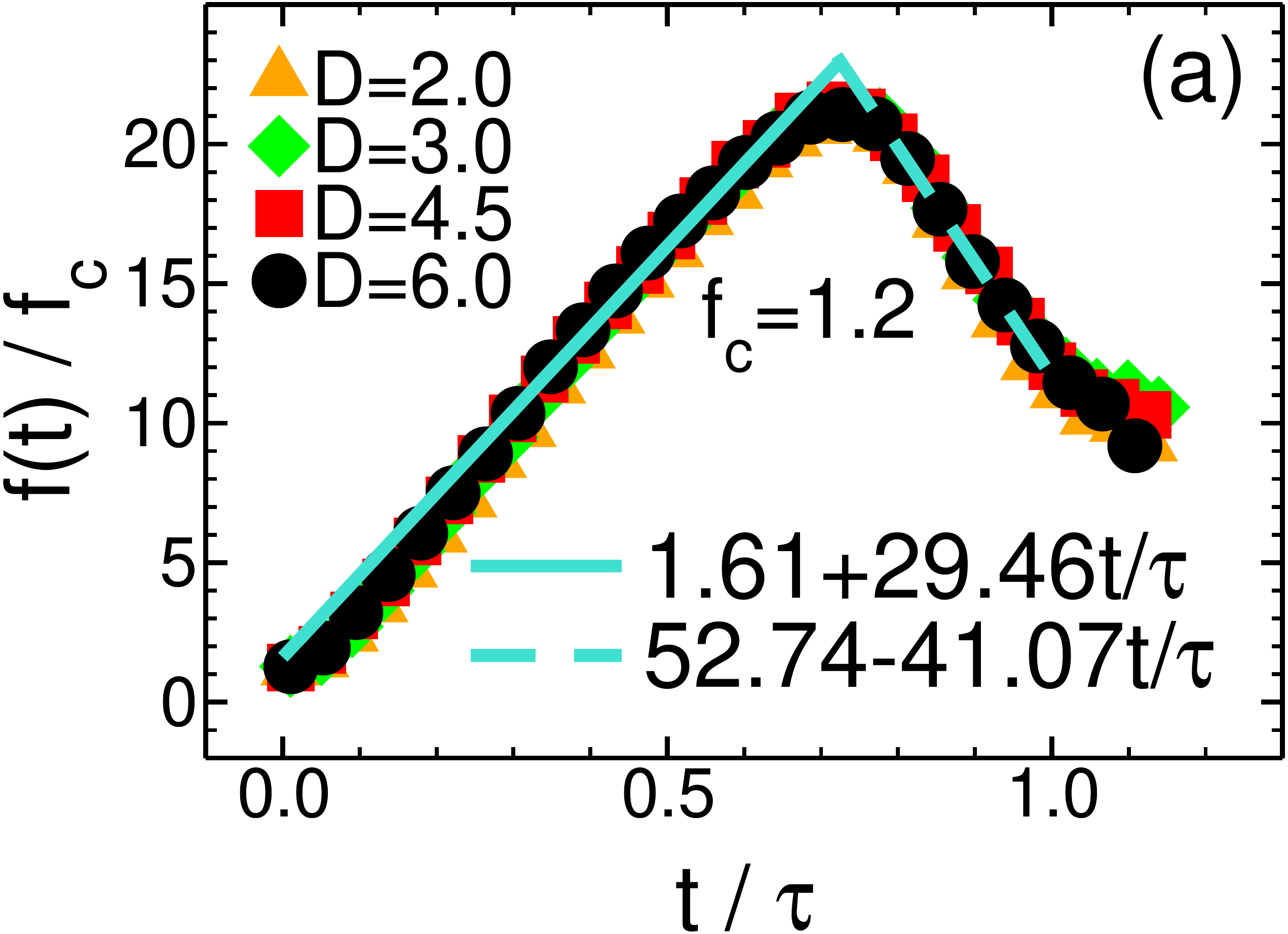}
    \end{center}\end{minipage} \hskip-0.07cm
        \begin{minipage}{0.23\textwidth}
    \begin{center}
        \includegraphics[width=1.0\textwidth]{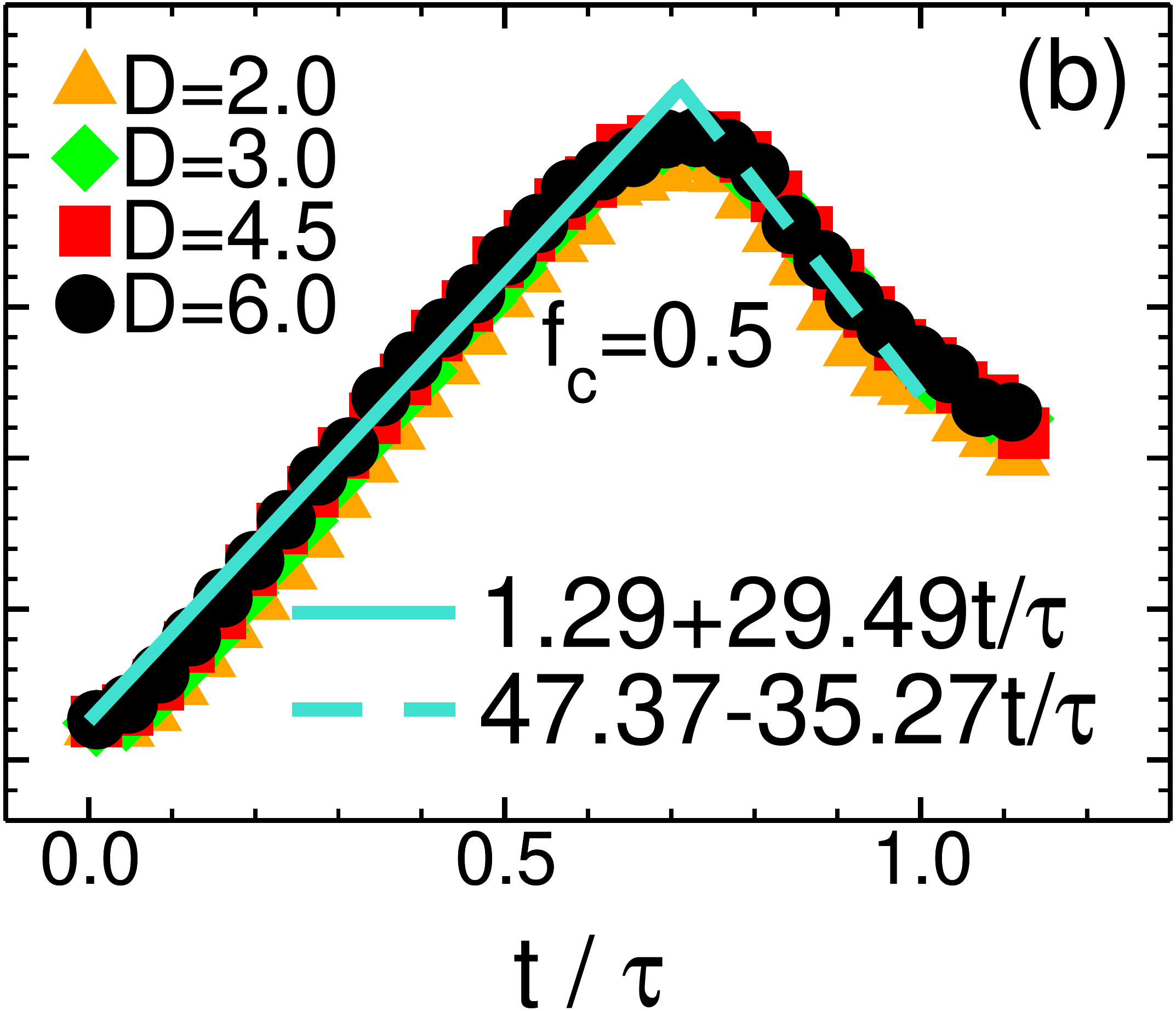}
    \end{center}\end{minipage} \hskip-0.07cm
        \begin{minipage}{0.23\textwidth}
    \begin{center}
        \includegraphics[width=1.0\textwidth]{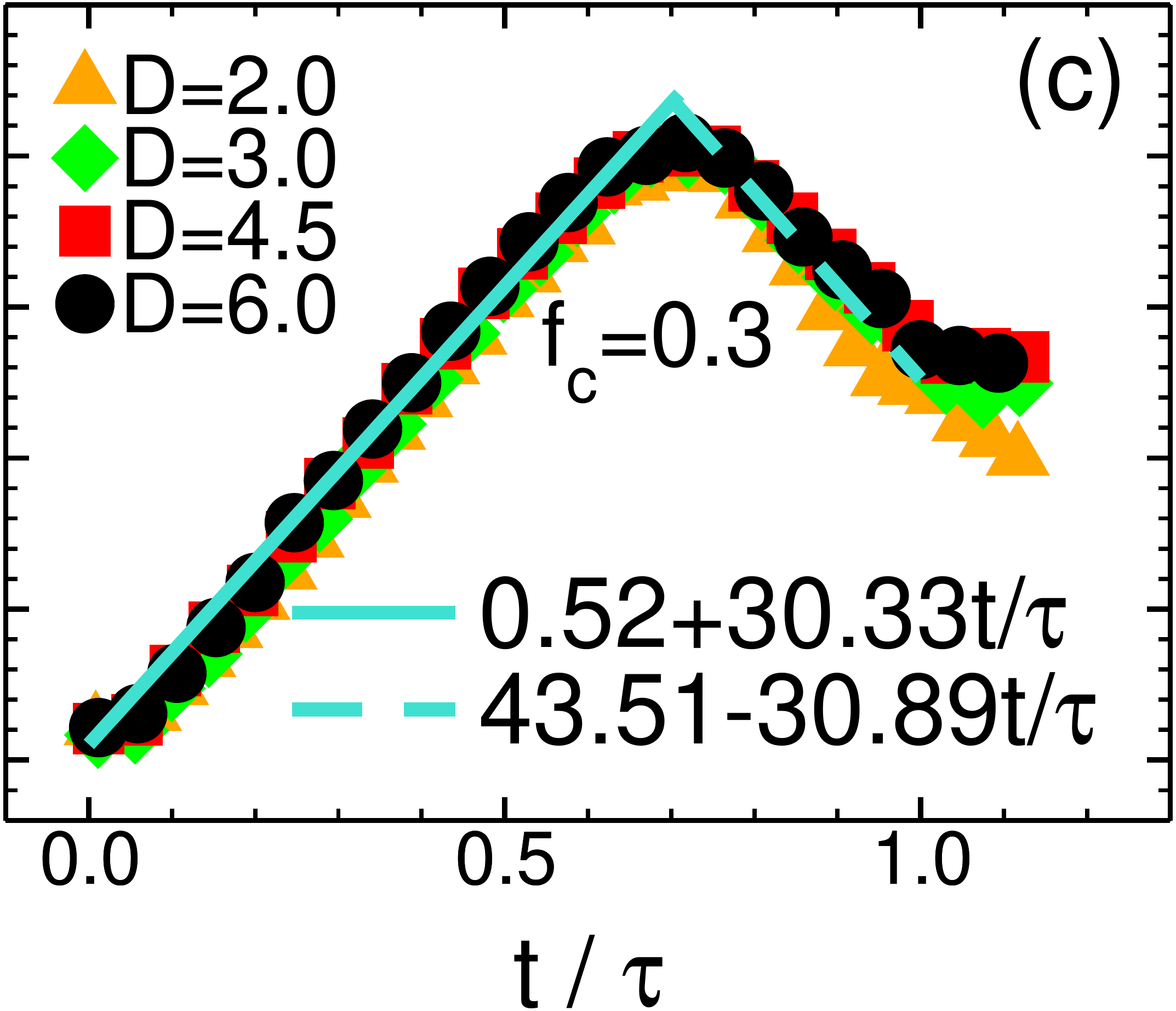}
    \end{center}\end{minipage} \hskip-0.07cm
    	\begin{minipage}{0.23\textwidth}
    \begin{center}
        \includegraphics[width=1.0\textwidth]{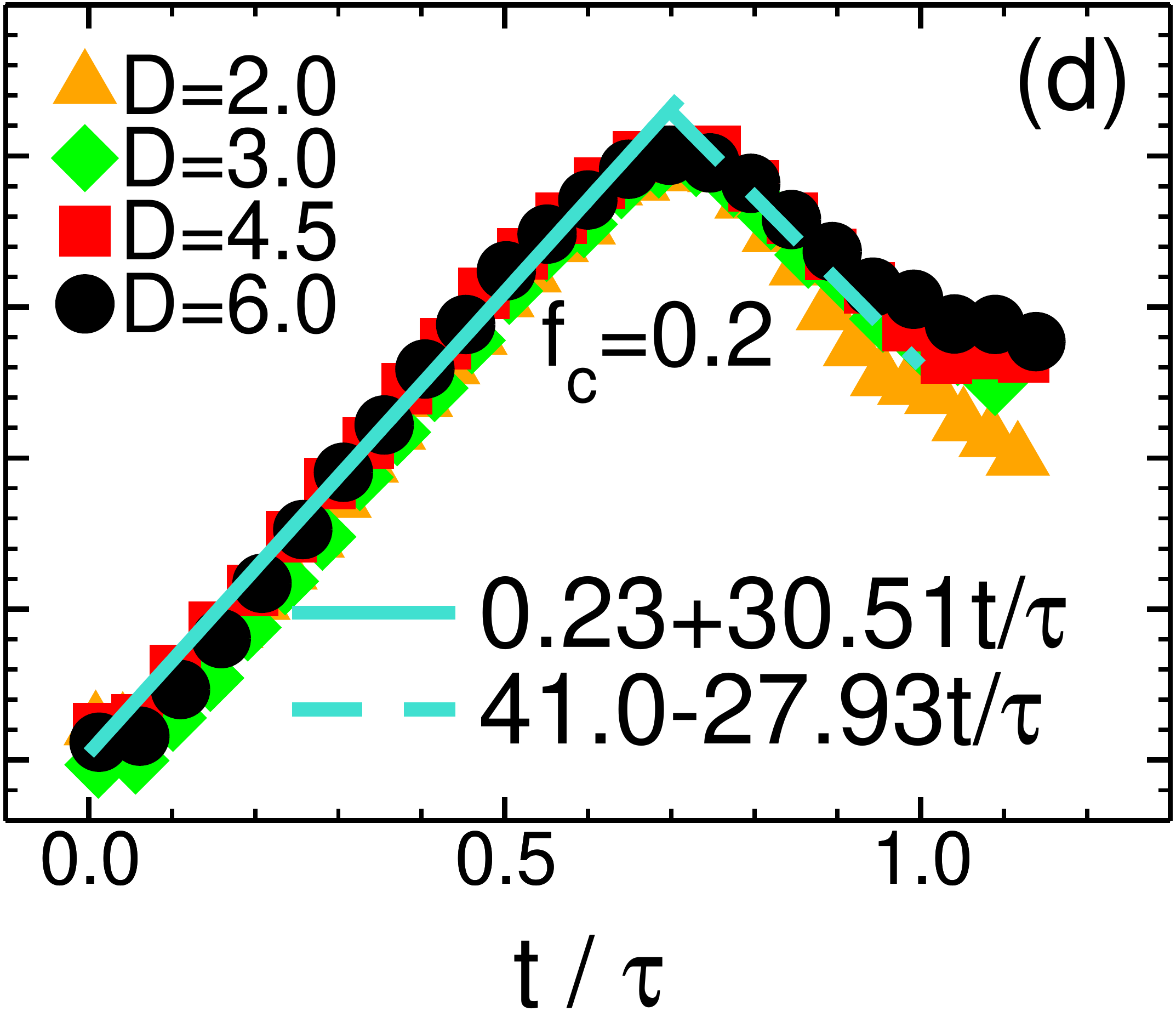}
    \end{center}\end{minipage}
\caption{(a) Normalized effective force $f(t)/f_{\textrm{c}}$ as function of
normalized time $t/\tau$, for fixed channel driving force $f_{\textrm{c}}=
1.2$, evaluated for the channel widths $D=2.0$ (orange filled triangles),
3.0 (green filled diamonds), 4.5 (red filled squares), and 6.0 (black filled
circles). Turquoise solid and dashed lines are fitting curves to the
normalized forces in the TP and PP stages, respectively (see text). Panels
(b), (c), and (d) are the same as panel (a), but for $f_{\textrm{c}}=0.5$,
0.3 and 0.2, respectively.}
\label{fig:force}
\end{figure*}

Before the actual translocation process, the head monomer of the polymer
chain is fixed in the center of the entrance of the channel (pore) and
the polymer is equilibrated. Then the fixed head monomer is released,
and at the same time the external channel driving force $f_{\textrm{c}}$
is switched on. To obtain sufficient statistics, averaging is performed
over $1000$ uncorrelated trajectories.
Panel (a) in Fig.~\ref{fig:schematic} shows a typical equilibrium state of the
polymer chain (from LD simulations) just before starting the translocation
process. Panels (b) and (c) correspond to the tension propagation (TP) and
post propagation (PP) stages, respectively. Finally, panel (d) presents
the configuration of the chain inside the channel just at the termination
of translocation process which equals the filling time $\tau_1 = \tau$ here. In  the TP stage presented in panel (b)
$t < t_{\textrm{TP}}$, as the chain translocates into the channel by the
external channel driving force $f_{\textrm{c}}$ acting on each individual
monomer inside the nanopore, tension propagates along the backbone of
the subchain on the {\it cis} side. The snapshot, at time $t$, in panel (b)
shows that $s$ monomers (in blue color) have translocated into the channel,
while $l$ monomers (in red color) on the {\it cis} side have been influenced
by the tension force and are moving towards the channel, i.e., $N=s+l$ is
the number of all monomers that have been influenced by the tension force
up to time $t$. The remaining $N_0-N$ monomers are at equilibrium and have
zero net velocity (monomers in black color). In the TP stage, as the tension
force has not reached the chain end, we thus have $N=s+l< N_0$.

The picture above shows that the subchain on the {\it cis}
side can be divided into two parts, a mobile part (in red color) and an
equilibrium part (in black color). The location of the tension front, that
is the border between the mobile and equilibrium parts, is specified by
$R$. The gray monomers represent the configuration of the polymer at time
zero, for comparison of the polymer configurations at times $t$ and zero.
The snapshot of the system (from LD simulation) is shown in panel (c) for $t>
t_{\textrm{TP}}$. At time $t_{\textrm{TP}}$, the tension propagation time,
the tension reaches the chain end on the {\it cis} side, and by that time
all monomers of the chain have already been influenced by the tension force,
i.e. $N=l+s=N_0$. Finally, panel (d) illustrates the configuration of the
chain just at the termination of the translocation process, when the entire
polymer chain has been translocated inside the {\it trans}-side nanochannel.

\section{Iso-flux tension propagation theory}
\label{IFTP}

The quantitatively accurate IFTP theory can be employed to describe the
dynamics of the translocation process for the channel-driven case.  To this
end we generalize the IFTP theory by considering the effective
force acting on the monomer(s) located just at the entrance of the channel. The
dynamics of the translocation process can be unraveled by solving the equation of
motion for the time evolution of the translocation coordinate $\tilde{s}$
within the iso-flux (IF) approximation for the monomer flux $\tilde{\phi}=
\textrm{d}\tilde{s}/\textrm{d}\tilde{t}$. In the IF approximation the monomer
flux in the mobile part of chain ({\it cis} side, red monomers in panels (b)
and (c) in Fig.~\ref{fig:schematic}) is spatially constant but evolves in
time. 

To obtain the equation of motion for $\tilde{s}$, the tension force
$\tilde{\mathbb{F}}(\tilde{x},\tilde{t})$ at the distance $\tilde{x}$ from
the channel entrance on the {\it cis} side is needed. Integrating the local
force-balance equation $\textrm{d}\tilde{\mathbb{F}}(\tilde{x}',\tilde{t})
=-\tilde{\phi}(\tilde{t})\textrm{d}\tilde{x}' $ over the distance from the
channel entrance at $\tilde{x}'=0$ to the distance $\tilde{x}' = \tilde{x}$
from the channel entrance, the tension force at distance $\tilde{x}$ on the
{\it cis} side is written as $\tilde{\mathbb{F}} (\tilde{x} , \tilde{t})
=\tilde{\mathbb{F}}_0-\tilde{x}\tilde{\phi}(\tilde{t})$, where
$\tilde{\mathbb{F}}_0=\tilde{f}(\tilde{t})-\tilde{\eta}_{\textrm{c}}
\tilde{\phi} (\tilde{t}) $ is the tension force just at the entrance of
channel on the {\it cis} side in terms of the channel friction coefficient
$\tilde{\eta}_{\textrm{c}}$ and the effective force $\tilde{f} (\tilde{t})$
acting on the monomer(s) just inside the entrance of the channel
(see the discussion below and the schematic in Fig.~\ref{fig:force}).
Using the fact that the tension force vanishes at the tension front,
i.e., $\tilde{\mathbb{F}} (\tilde{x} = \tilde{R} , \tilde{t}) = 0 $, the
tension force just at the entrance of channel is related to $\tilde{R}$
via $\tilde{\mathbb{F}}_0=\tilde{R}(\tilde{t})\tilde{\phi} (\tilde{t})$.
Combining this with the definition of the monomer flux, the equation of
motion for the translocation coordinates $\tilde{s}$ is obtained as
\begin{equation}
\left[\tilde{R}(\tilde{t})+\tilde{\eta}_{\textrm{c}}\right]\frac{
\textrm{d}\tilde{s}}{\textrm{d}\tilde{t}}=\tilde{f}(\tilde{t}).
\label{BD_force}
\end{equation}
The expression in the brackets is the effective friction and is the sum of
friction due to the {\it cis}-side mobile subchain $\tilde{R} (\tilde{t})$
and the channel friction $\tilde{\eta}_{\textrm{c}}$. To solve the above
equation, the time evolution of $\tilde{f} (\tilde{t})$ and $\tilde{R}
(\tilde{t})$ must be known. We first explain how $\tilde{f}
({t})$ can be obtained numerically and then the equation of motion for the time evolution
of $\tilde{R} (\tilde{t})$ can be extracted within the IF approximation.

We performed extensive LD simulations to obtain the effective
driving force $f(t)$, which is the force experienced by the monomer inside
the channel entrance (i.e., the $x$ component of its position is in the
region $-0.5<x<0.5$). The channel driving force $f_{\textrm{c}}$, the force
$f_{\textrm{LJ}}$ due to the other monomers inside the channel in the region
of $x> 0.5$, and the interaction between the channel walls particles
and the monomer(s) inside the entrance of the channel contribute to the
total effective driving force $f(t)$. The contribution of $f_{\textrm{LJ}}$
to the driving is the combination of the FENE-bond force and non-bonded
repulsive LJ interactions (WCA). 

Panel (a) in Fig.~\ref{fig:force} presents
the normalized effective driving force $f (t) / f_{\textrm{c}}$ as a
function of the normalized time $t/\tau$, where $\tau$ is the translocation
time, for constant channel driving force $ f_{\textrm{c}} = 1.2 $ and for
the channel diameters $D=2$, $3$, $4.5$, and $6$. For each set of
parameters the effective driving force is obtained by averaging over $1000$
uncorrelated trajectories. Panels (b), (c) and (d) are the same as panel (a)
but for the channel driving forces $f_{\textrm{c}}=0.5$, $0.3$,
and $0.2$, respectively. We can see from Fig.~\ref{fig:force} that
data for fixed $f_{\rm c}$ collapse on a master curve. The effective force
$f(t)$ first linearly grows and reaches its maximum and then decreases. The
maximum occurs at the tension propagation time $t_{\textrm{TP}}$ for all
curves. This has been checked by considering the bond length distributions
in Appendix~\ref{blength} (see Fig.~\ref{fig:bond}) and also by looking at
the average velocity distributions of monomers in Appendix~\ref{velocity}
(see Fig.~\ref{fig:velocity}) at different instants during the translocation
process. Both bonds and velocity distributions confirm that for all sets
of parameters the PP stage starts around time $t_{\textrm{TP}} \approx
0.7 \tau$. As Fig.~\ref{fig:force} shows, in the TP and PP stages the
driving force can be written as
\begin{eqnarray}
\tilde{f}_{\textrm{TP}}(\tilde{t})/\tilde{f}_{\textrm{c}}&=&a+b\tilde{t}/
\tilde{\tau},
\nonumber\\
\tilde{f}_{\textrm{PP}}(\tilde{t})/\tilde{f}_{\textrm{c}}&=&c-d~\tilde{t}/
\tilde{\tau} ,
\label{f_TP_PP}
\end{eqnarray}
respectively. For strong channel driving force $f_{\textrm{c}}=1.2$
in panel (a), the values of the constants are $a=1.61$, $b=29.46$,
$c=52.74$ and $d=41.07$, and for weaker channel driving forces the
values of $a, b, c$ and $d$ can be found in the corresponding panels. At
the tension propagation time $t_{\textrm{TP}}$, the values of the
effective forces in the TP and PP stages should be equal (continuity
condition), i.e. $\tilde{f}_{\textrm{TP}}(\tilde{t}=t_{\textrm{TP}})
=\tilde{f}_{\textrm{PP}}(\tilde{t}=t_{\textrm{TP}})$. Consequently,
$\tilde{t}_{\textrm{TP}}=Q\tilde{\tau}$, where $Q=(c-a)/(b+d)$, with the
approximate value of $0.7$.

\begin{figure*}[t]\begin{center}
    \begin{minipage}{0.264\textwidth}
    \begin{center}
        \includegraphics[width=0.85\textwidth]{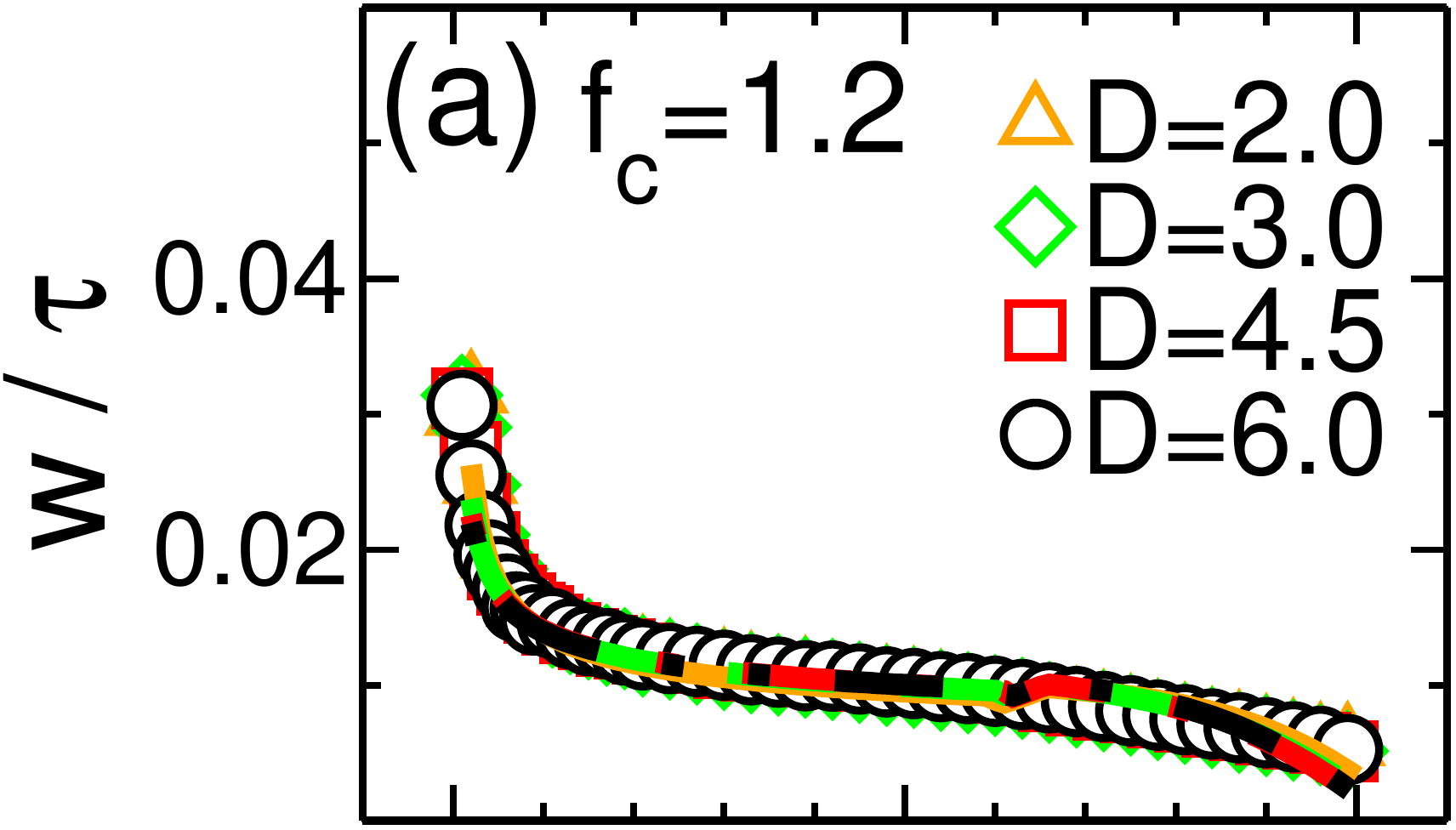}
    \end{center}\end{minipage} \hskip-0.6cm
        \begin{minipage}{0.2\textwidth}
    \begin{center}
        \includegraphics[width=0.85\textwidth]{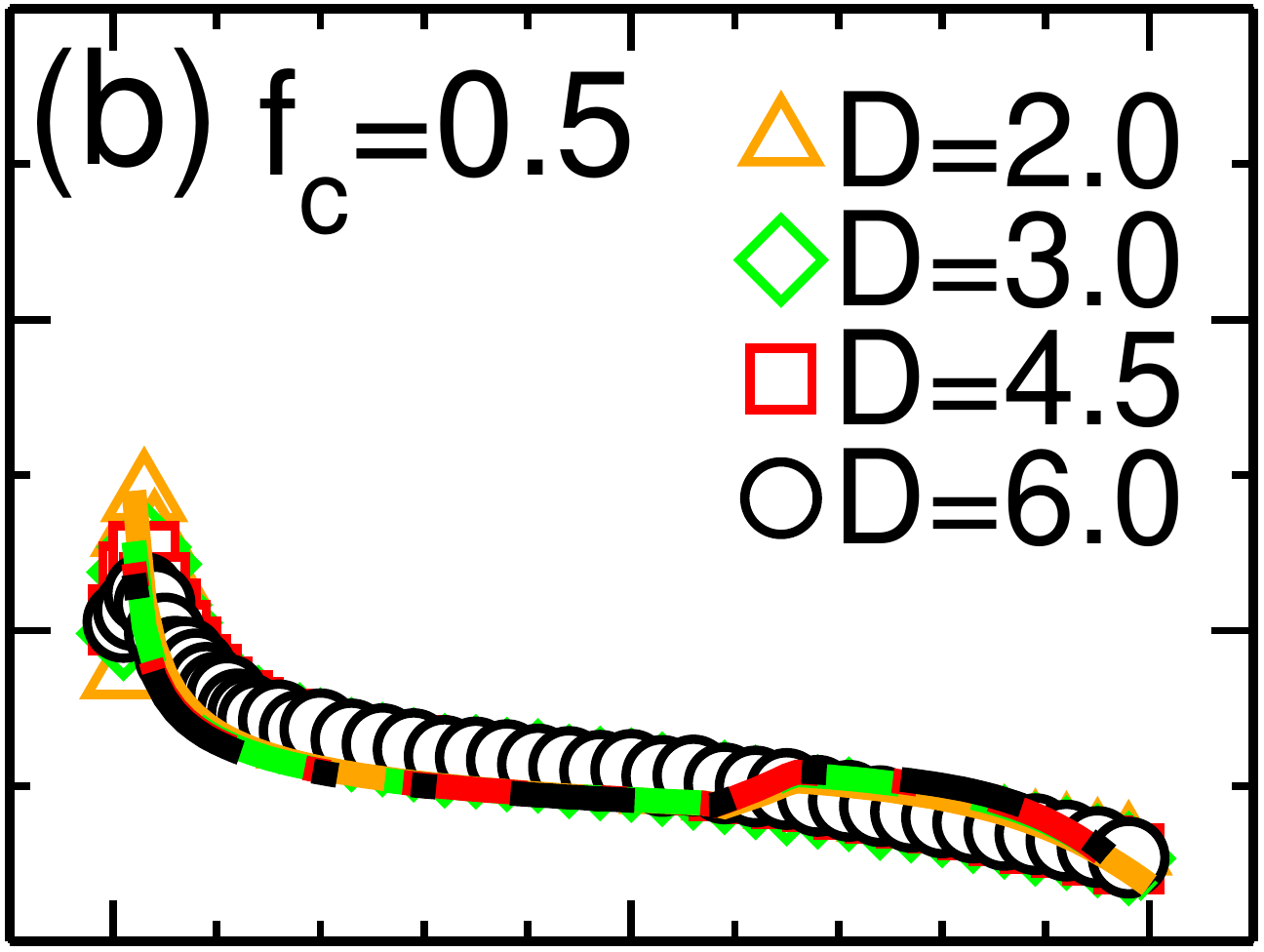}
    \end{center}\end{minipage} \hskip-0.52cm
        \begin{minipage}{0.2\textwidth}
    \begin{center}
        \includegraphics[width=0.85\textwidth]{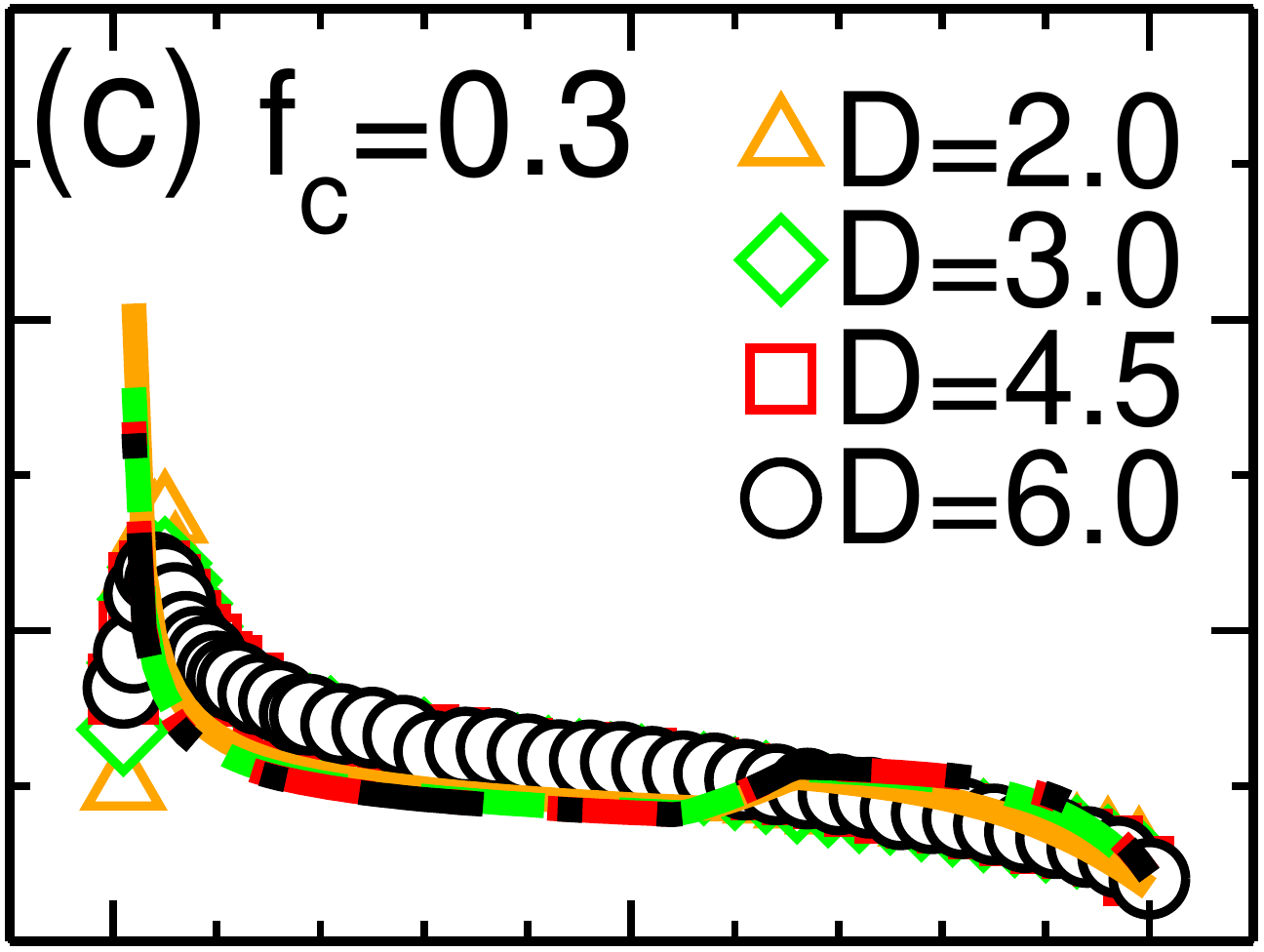}
    \end{center}\end{minipage} \hskip-0.52cm
    	\begin{minipage}{0.2\textwidth}
    \begin{center}
        \includegraphics[width=0.85\textwidth]{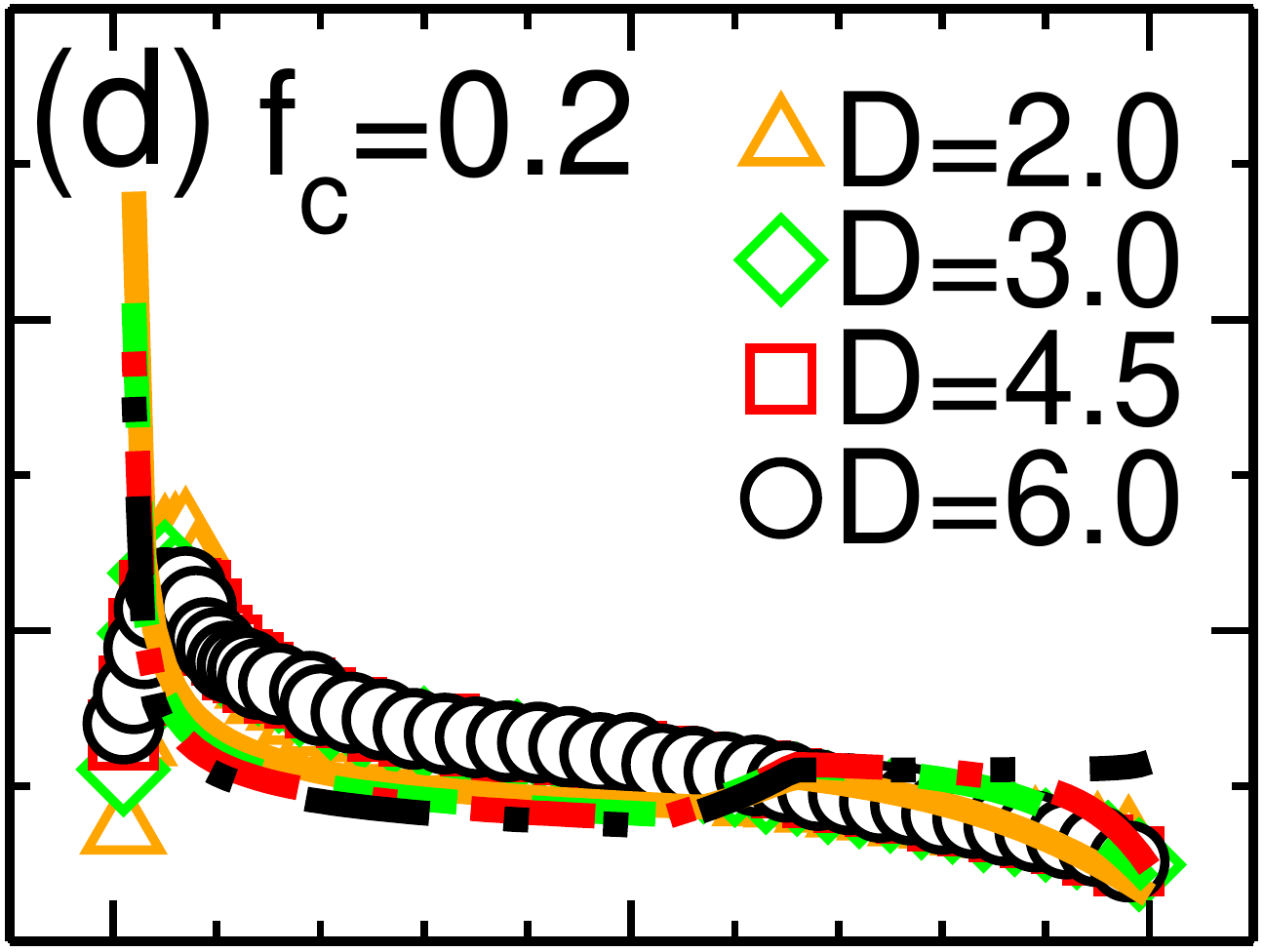}
    \end{center}\end{minipage}
    \begin{minipage}{0.264\textwidth}
    \vspace{+0.1cm}
    \begin{center}
        \includegraphics[width=0.85\textwidth]{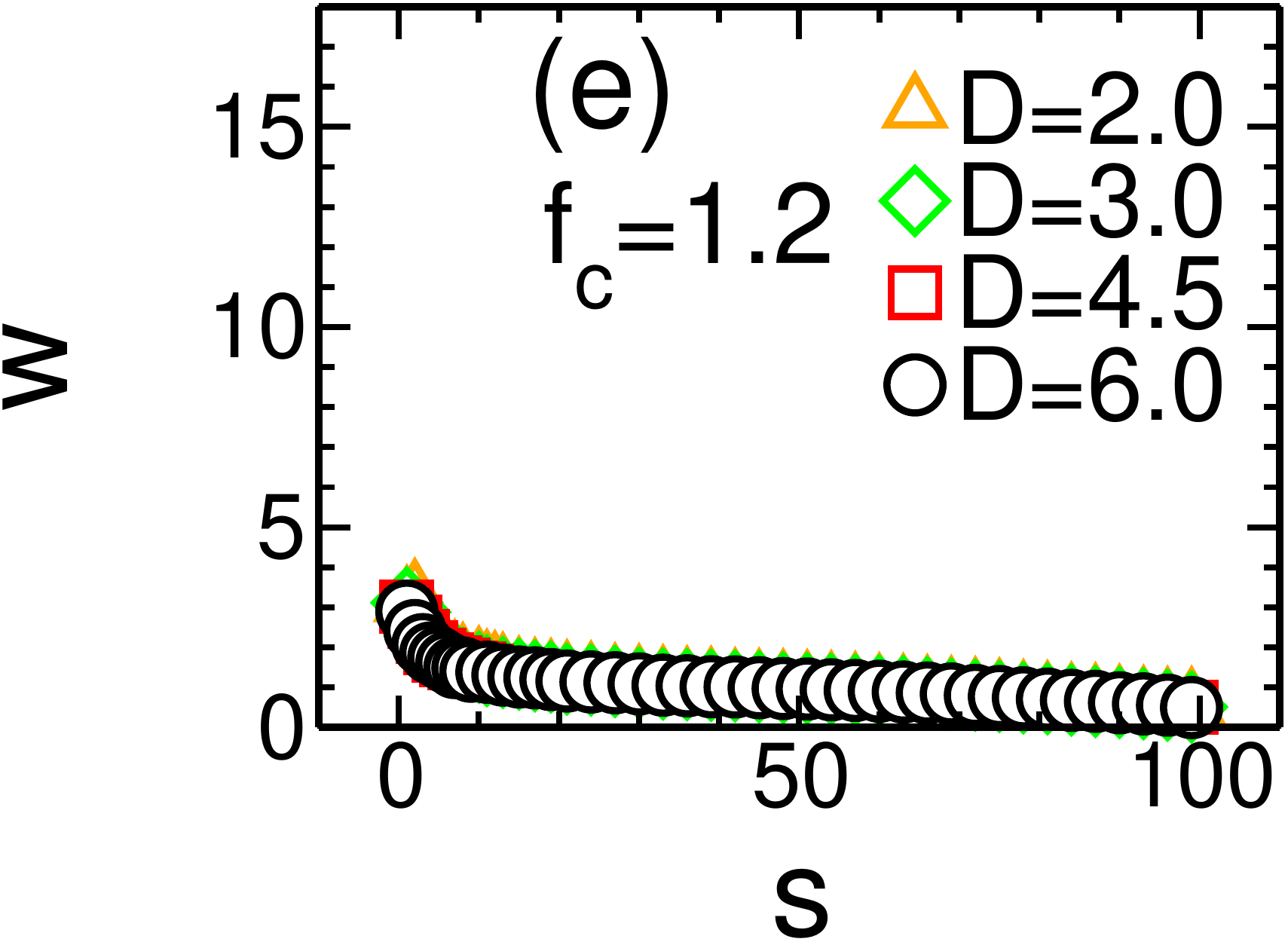}
    \end{center}\end{minipage} \hskip-0.6cm
        \begin{minipage}{0.2\textwidth}
    \vspace{+0.1cm}
    \begin{center}
        \includegraphics[width=0.85\textwidth]{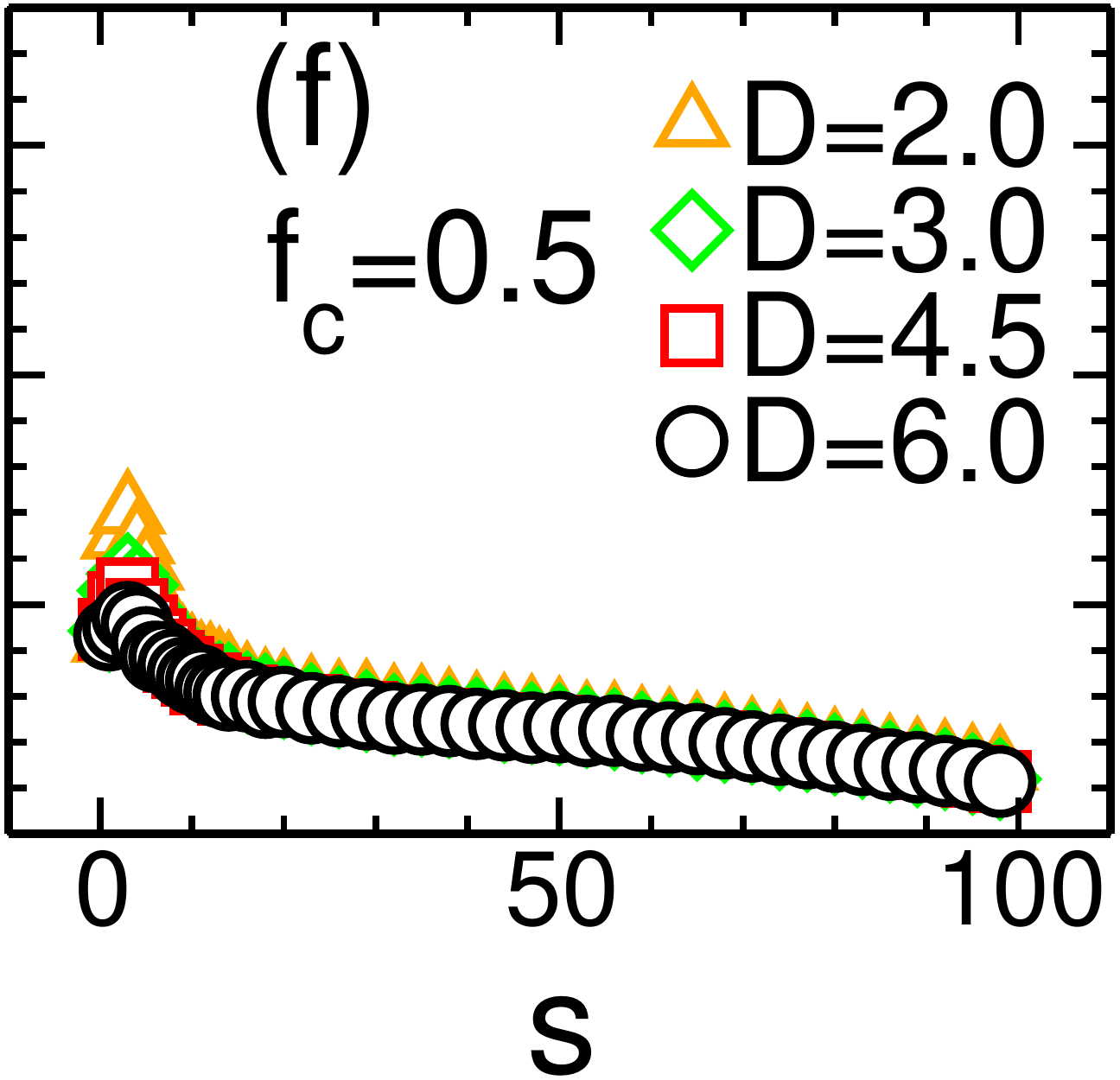}
    \end{center}\end{minipage} \hskip-0.52cm
        \begin{minipage}{0.2\textwidth}
	\vspace{+0.1cm}    
    \begin{center}
        \includegraphics[width=0.85\textwidth]{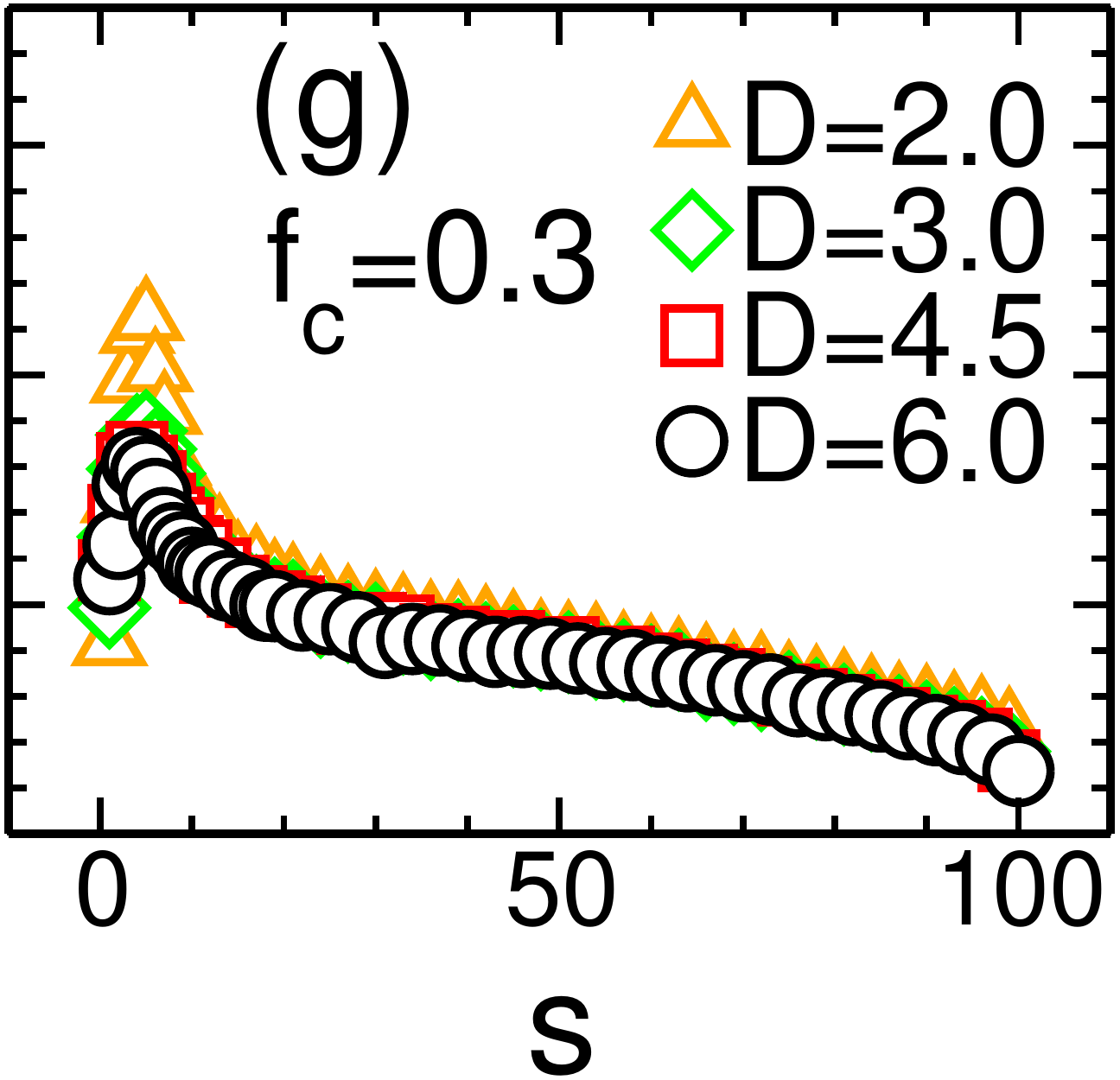}
    \end{center}\end{minipage} \hskip-0.52cm
    	\begin{minipage}{0.2\textwidth}
	\vspace{+0.1cm}    
    \begin{center}
        \includegraphics[width=0.85\textwidth]{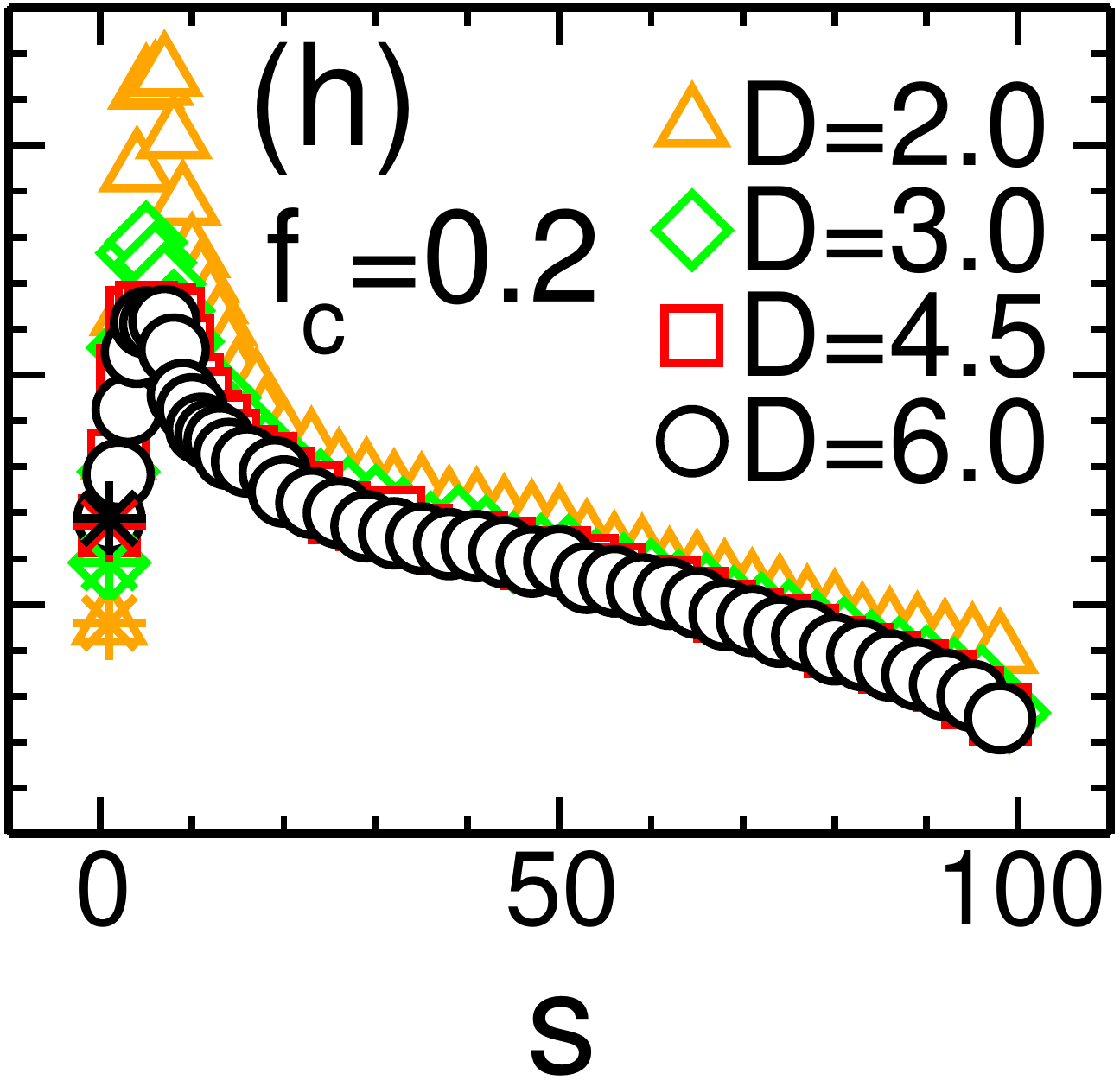}
    \end{center}\end{minipage}
\caption{(a) Normalized waiting time $w/\tau$ as function of
the translocation coordinate $s$, for fixed value $f_{\textrm{c}}=1.2$
of the channel driving force and the channel widths $D=2.0$ (orange
triangles), 3.0 (green diamonds), 4.5 (red squares), and 6.0 (black
circles). Solid orange, green dashed, red dashed-dotted, and black
dashed-dotted-dotted lines represent the IFTP results in the strong
stretching regime for channel widths $D=2.0$, 3.0, 4.5 and 6.0,
respectively (not shown in the legend). Panels (b), (c), and (d) are
the same as panel (a), but for $f_{\textrm{c}}=0.5$, 0.3, and 0.2,
respectively. Panels (e), (f), (g) and (h) are the same as (a), (b),
(c), and (d), respectively, but present the unnormalized waiting time
distribution $w$ as a function of $s$. 
{The values of the channel friction coefficients in the IFTP theory 
have been chosen as $\eta_{\textrm{c}} =4.3,3.0,2.5$ and 2.3 corresponding 
to different channel widths $D=2,3,4.5$ and 6, respectively \cite{ikonen2012b}.}
}
\label{fig:wtfc}
\end{center}
\end{figure*}

Based on the time evolution of the total effective driving force
$\tilde{f} (\tilde{t})$ one can solve the effective force-balance equation
(\ref{BD_force}), provided that the time evolution of the tension front
location $\tilde{R} (\tilde{t})$ is known. The equation of motion for
$\tilde{R} (\tilde{t})$ in the TP stage is obtained by comparing the average
spatial configurations of the chain at time zero (typical chain
configuration in gray color at time zero in Fig.~\ref{fig:schematic}(b))
and at time $\tilde{t}$ (typical chain configuration in blue, red and black
colors in Fig.~\ref{fig:schematic}(b)). The number of those monomers that at
equilibrium (at time zero) are located between the entrance of the channel
and the tension front (at time $\tilde{t}$) is, on average, the same as the
number of monomers influenced by the tension force by the time $\tilde{t}$,
i.e. $N= \tilde{s} + \tilde{l}$, the sum of the monomers in blue color
inside the channel and the mobile monomers in red color in the {\it cis}
side at time $\tilde{t}$. Then in equilibrium the size of the subchain that
occupies the region between the entrance of the channel and the tension
front is $\tilde{R}$. Therefore, according to Flory theory one can write the
end-to-end distance devoted to this portion of the polymer chain as $\tilde{R}
=A_{\nu} N^{\nu}$. Combining this end-to-end distance with the fact that
$N=\tilde{s}+\tilde{l}$, and after taking the time derivative of
both sides of $\tilde{R}=A_{\nu}(\tilde{s}+\tilde{l})^{\nu}$ in the
strong stretching (SS) regime where the mobile subchain on the {\it cis}
side is fully straightened ($\tilde{l} = \tilde{R}$), the equation of motion
for the tension front location in the TP stage is obtained as
\begin{equation}
\dot{\tilde{R}} (\tilde{t}) =  \frac{ \nu A_{\nu}^{1/\nu} \tilde{R} (\tilde{t}) ^{(\nu-1)/\nu} \tilde{\phi} (\tilde{t}) }{ 1- \nu A_{\nu}^{1/\nu} \tilde{R} (\tilde{t}) ^{(\nu-1)/\nu}  } ,
\label{R_TP}
\end{equation}
where the definition $\tilde{\phi}=\textrm{d}\tilde{s}/\textrm{d}\tilde{t}$
of the monomer flux has been used. Conversely, in the PP stage, in which the
tension has reached the chain end on the {\it cis} side, the closure relation
is $N=\tilde{s}+\tilde{l}=N_0$. Performing the time derivative on both sides
of this closure relation in the SS regime, in which $\tilde{l}=\tilde{R}$,
yields the equation of motion for the tension front,
\begin{equation}
\dot{\tilde{R}}(\tilde{t})=-\tilde{\phi}(\tilde{t}).
\label{R_PP}
\end{equation}
To have a full solution, in the TP stage Eqs.~(\ref{BD_force}) and
(\ref{R_TP}) need to be solved self-consistently, while the PP stage
is based on Eqs.~(\ref{BD_force}) and (\ref{R_PP}).

\section{Results}
\label{results}

This section is devoted to the presentation of the results from the IFTP
theory and LD simulations for the translocation dynamics both at the
monomer and global levels. To this end the waiting time distribution,
translocation time, monomer mean-square displacement (MSD), and the monomer
density are investigated.

\subsection{Waiting time distribution}
\label{WT}

One of the most important quantities that reveals the dynamics of the
translocation process at the monomer level is the distribution $w$ of
waiting times (WT), the time that each bead spends at the entrance of the
nanochannel (specifically when the $x$ component of the position vector
of the corresponding bead is in the region $-0.5\leqslant x\leqslant
+0.5$) during the translocation process. Figure~\ref{fig:wtfc}(a) shows
the normalized WT distribution $w/\tau$ as function of the translocation
coordinate $s$, for fixed value $f_{\textrm{c}}=1.2$ of the driving force
and the channel widths $D=2.0$ (orange triangles), 3.0 (green diamonds),
4.5 (red squares), and 6.0 (black circles). Solid orange, green dashed,
red dashed-dotted and black dashed-dotted-dotted lines represent the IFTP
results in the SS regime for channel widths $D=2.0$, 3.0, 4.5 and 6.0,
respectively (not shown in the legend). Panels (b), (c) and (d) are the same
as panel (a), but for $f_{\textrm{c}}$= 0.5, 0.3 and 0.2, respectively. 
{The values of the channel friction coefficients that have been 
used in the IFTP theory are $\eta_{\textrm{c}} =4.3,3.0,2.5$ and 2.3 corresponding 
to different channel widths $D=2,3,4.5$ and 6, respectively \cite{ikonen2012b}.}

We can conclude from the data that the results of the IFTP theory increasingly deviate from the
LD data as the value of the driving force decreases. This happens due to
the fact that here we only consider the SS regime
(the mobile subchain in the {\it cis} side is fully straightened) in IFTP
theory in the absence of the entropic force. This discrepancy can be resolved in
the weak and intermediate force regimes by taking the entropic force into account.
There the spatial shape of the
mobile subchain on the {\it cis} side assume the so-called trumpet (TR) and
stem-flower (SF) shapes \cite{rowghanian2011,jalalJCP2014}, respectively,
However, this is beyond the
scope of the current study. However, the global dynamics of the translocation
process is well explained by the IFTP theory in the high force limit (here,
$f_{\textrm{c}} = 1.2$) by the SS regime.

In panels (e), (f), (g)
and (h) of Fig. \ref{fig:wtfc}, which correspond to panels (a), (b), (c) and (d), respectively,
the unnormalized WT distribution $w$ is shown as function of $s$. As
seen in panel (e), regardless of the channel width, for the highest value
$f_{\textrm{c}}=1.2$ of the driving force WT monotonically decreases as $s$
increases. For smaller values of the driving force, $f_{\textrm{c}}\leqslant
0.5$ the WT curve is no longer monotonic, and a hump appears for small
$s$. This effect is more pronounced either by decreasing the value of the
driving force, from panel (f) to panel (h), or at constant driving force,
i.e. in each panel of (f), (g), and (h), by decreasing the channel width
from $D=6$ to $D=2$.

The reason for the monotonicity of the WT curve in panel (e) is that for strong
driving with $f_{\textrm{c}} = 1.2$, at the beginning of the translocation
process the tension propagates very fast and the subchain in the vicinity
of the channel entrance resists the driving force. This leads
to the large value of the WT at the beginning of the process.  At later
moments, when more monomers traverse the channel entrance the total net
force acting on the {\it trans}-side subchain increases, and the value
of the WT decreases. Therefore, for strong driving the WT monotonically
decreases. On the other hand for weaker driving force in panels (f) to
(h), at the very beginning of the translocation process the entropic force
due to spatial fluctuations of the {\it cis}-side chain configurations
that resists translocation, is more pronounced for the wide channels,
leading to the larger WT for the channel with $D=6$ and smaller values for
$D=2$ (star symbols in panel (h)). Then with time more monomers experience
the friction due to the mobile part of the {\it cis}-side subchain, while
the driving is not strong and $s$ does not increase fast enough.  
Consequently, the WT increases at the starting
moments of the translocation process. At the later time as enough monomers
have already traversed the channel entrance, the net effective driving force
becomes large enough to accelerate the translocation process and therefore the
WT decreases. At constant driving force, the non-monotonicity in the WT
curves is more pronounced with decreasing channel width. This is due to
the fact that decreasing channel width increases the effective channel friction
making the translocation dynamics slower and thus the
hump in the WT curve is more pronounced.

\subsection{Scaling of translocation time}
\label{ScalingTime}

\begin{figure}[t]\begin{center}
        \begin{minipage}{0.5\columnwidth}
    \begin{center}
        \includegraphics[width=0.98\columnwidth]{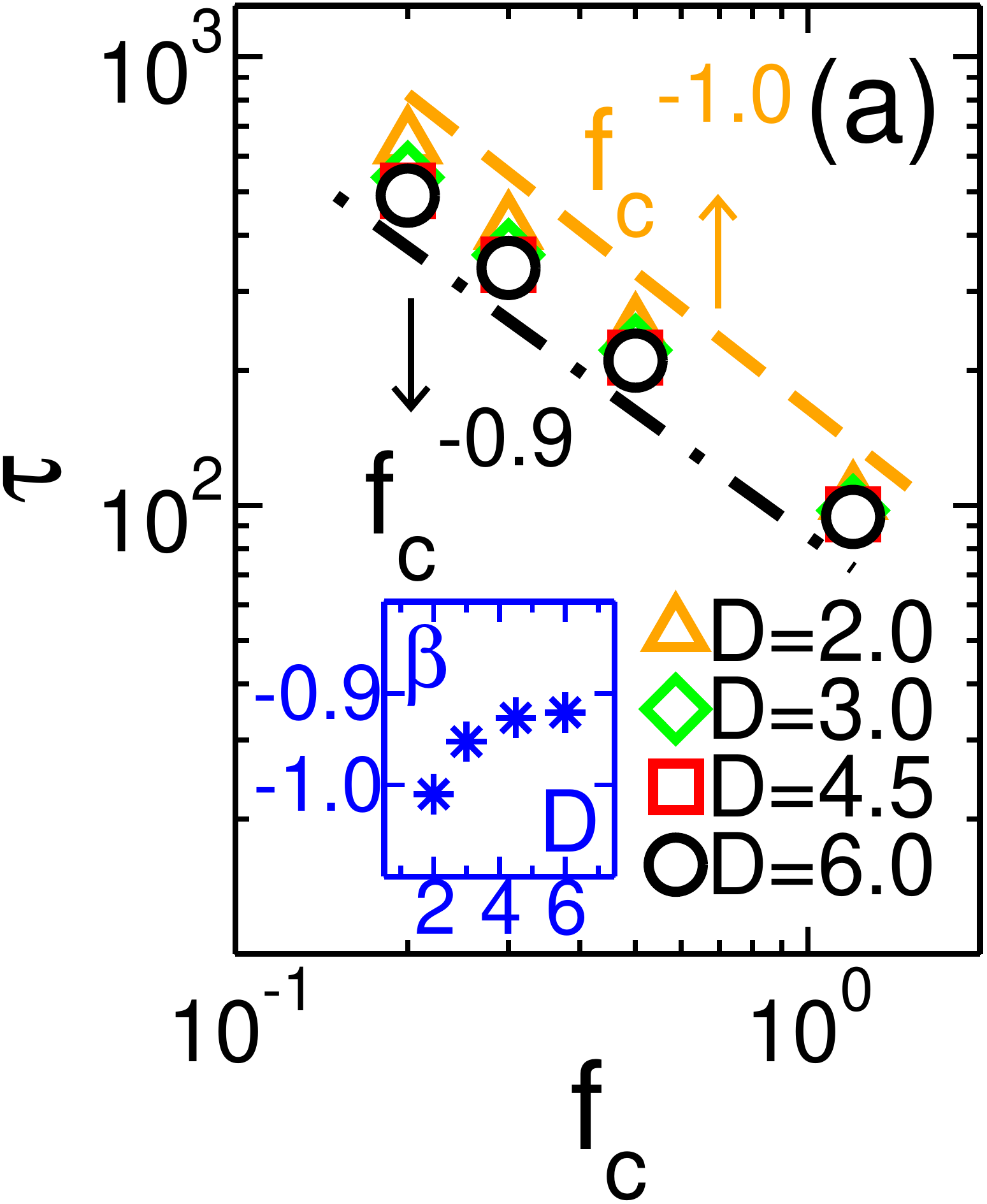}
    \end{center}\end{minipage}  \hspace{-0.2cm}
    \begin{minipage}{0.5\columnwidth}
    \begin{center}
        \includegraphics[width=0.98\columnwidth]{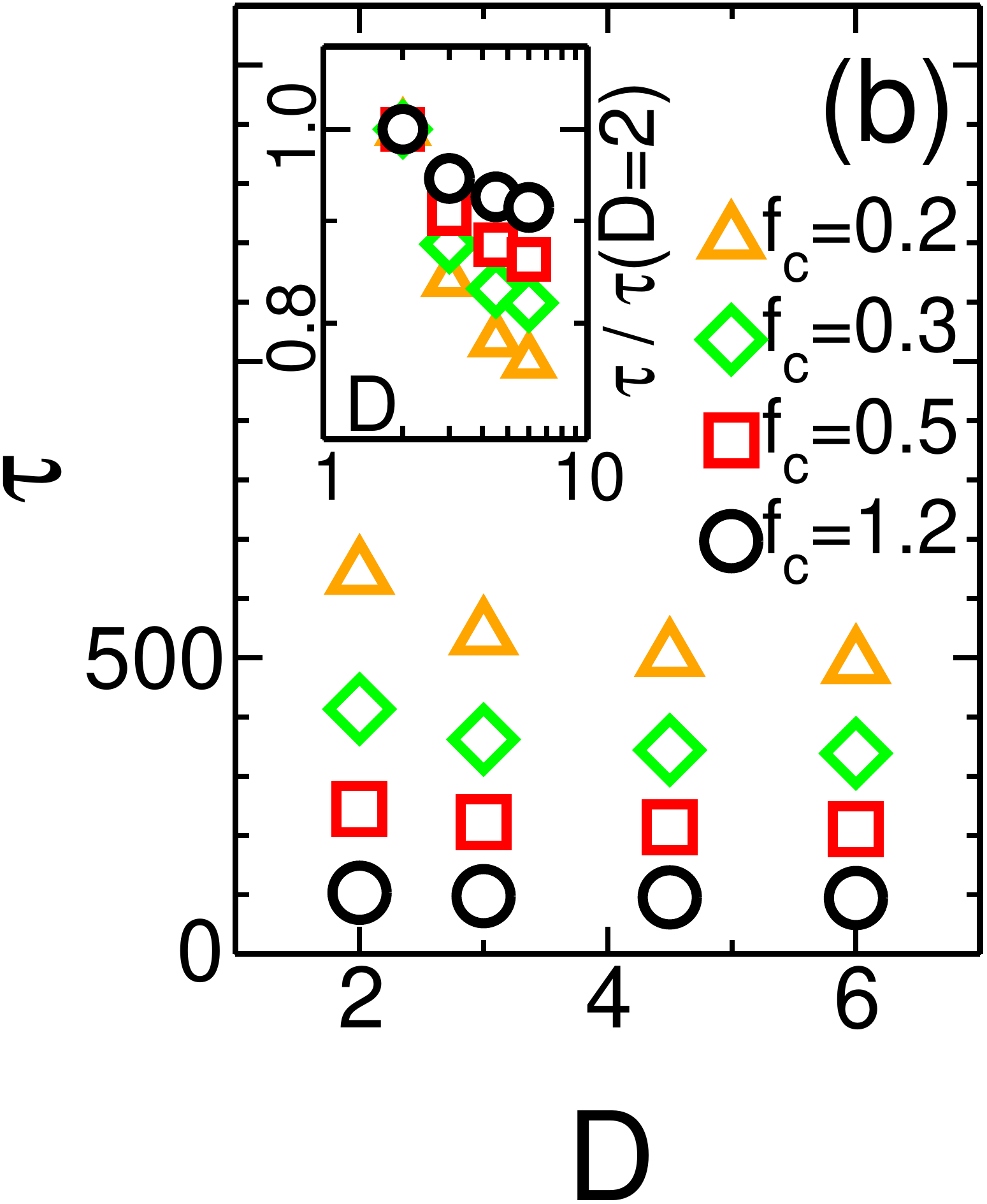}
    \end{center}\end{minipage} 
\caption{(a) Translocation time $\tau$ as function of the channel driving
force $f_{\textrm{c}}$, for channel widths $D=2.0$ (orange triangles),
3.0 (green diamonds), 4.5 (red squares), and 6.0 (black circles). The orange
dashed and black dashed-dotted lines are fits to the data with $D=2.0$ and
6.0, respectively, but shifted for better visibility. The inset shows the force
exponent $\beta$ of the scaling form $\tau\propto f_{\textrm{c}}^{\beta}$ as
function of the channel width $D$.  Panel (b) shows $\tau$ as a function of
$D$, for $f_{\textrm{c}}= 0.2$ (orange triangles), 0.3 (green diamonds),
0.5 (red squares) and 1.2 (black circles). The inset shows the normalized
translocation time $\tau/\tau(D=2)$ as function of $D$.}
\label{fig:tau}
\end{center}
\end{figure}

The fundamental quantity that reveals the global dynamics of the
translocation process is the translocation time $\tau$, which is the average
time for the whole chain to enter the channel and fully translocate to
the {\it trans}-side channel of the system. In Fig.~\ref{fig:tau} we show the translocation
times $\tau$ from LD simulations. In panel (a) $\tau$ is shown as
a function of the driving force $f_{\textrm{c}}$ for the channel diameters
$D=2$ (orange triangles), 3 (green diamonds), 4.5 (red squares) and
6 (black circles). The orange dashed and black dashed-dotted lines are
fits to the data with $D=2$ and 6, respectively, but shifted for better
visibility. The inset shows the force exponent $\beta$ from the scaling
relation $\tau\propto f_{\textrm{c}}^{\beta}$ as function of $D$. Th exponent $\beta$
approaches $-1$ as the channel width decreases to 2. In panel (b) $\tau$ is
shown as function of $D$, for $f_{\textrm{c}}= 0.2$ (orange triangles),
0.3 (green diamonds), 0.5 (red squares) and 1.2 (black circles).
In the strong force limit, i.e. for $f_{\textrm{c}} = 1.2$, $\tau$
does not change appreciably (black circles), while for the weakest force
$f_{\textrm{c}}=0.2$ the translocation time increases as the channel width
decreases (orange triangles). This is confirmed in the inset of panel (b),
where the normalized translocation time $\tau/\tau(D=2)$ is shown as function
of $D$ on log-log scale. The symbols in the inset are the same as those of
the main panel (b).

\begin{figure*}[t]\begin{center}
    \begin{minipage}{0.263\textwidth}
    \begin{center}
    	\includegraphics[width=0.8\textwidth]{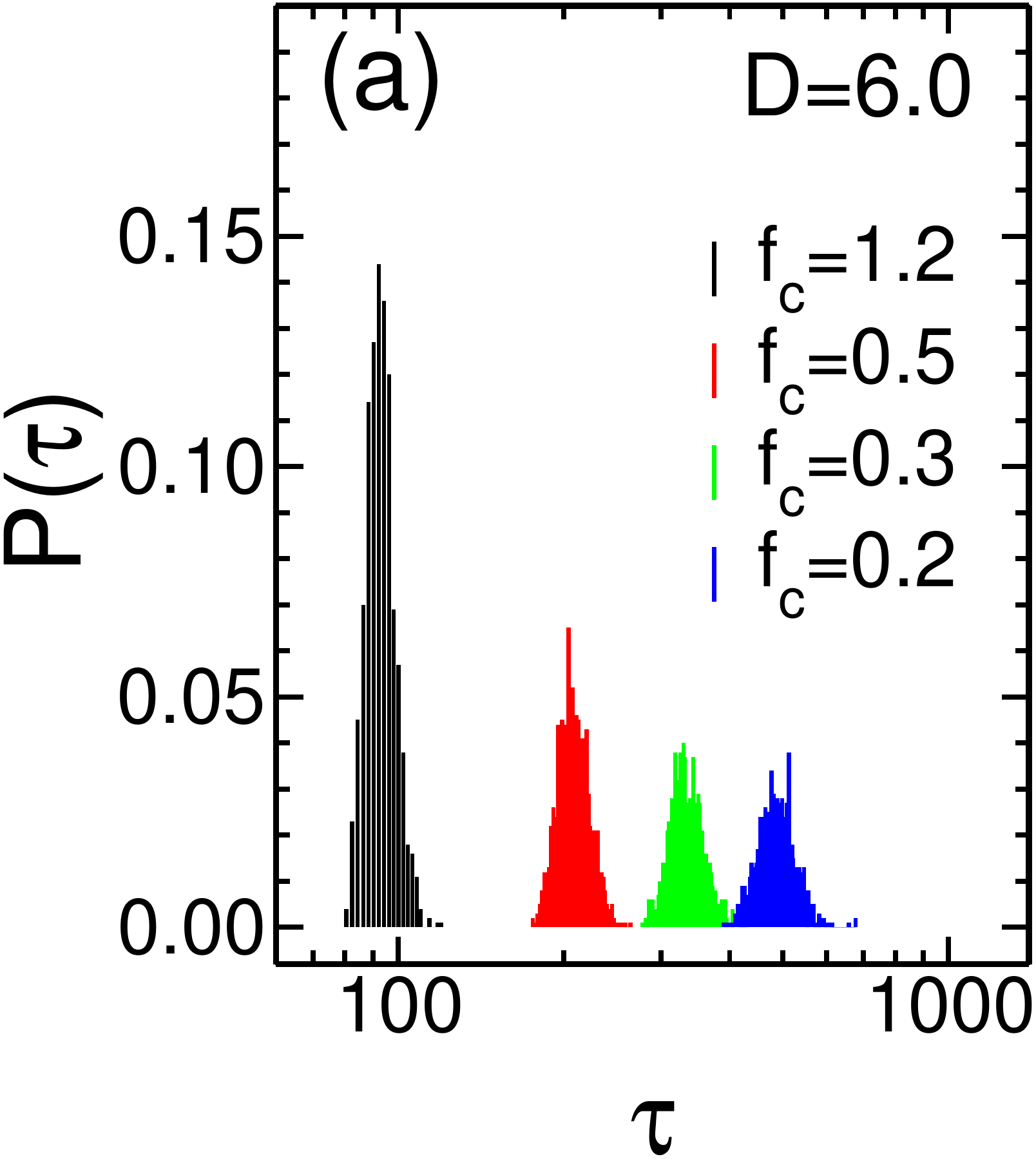}
    \end{center}\end{minipage} \hskip-0.82cm
        \begin{minipage}{0.1945\textwidth}
    \begin{center}
        \includegraphics[width=0.8\textwidth]{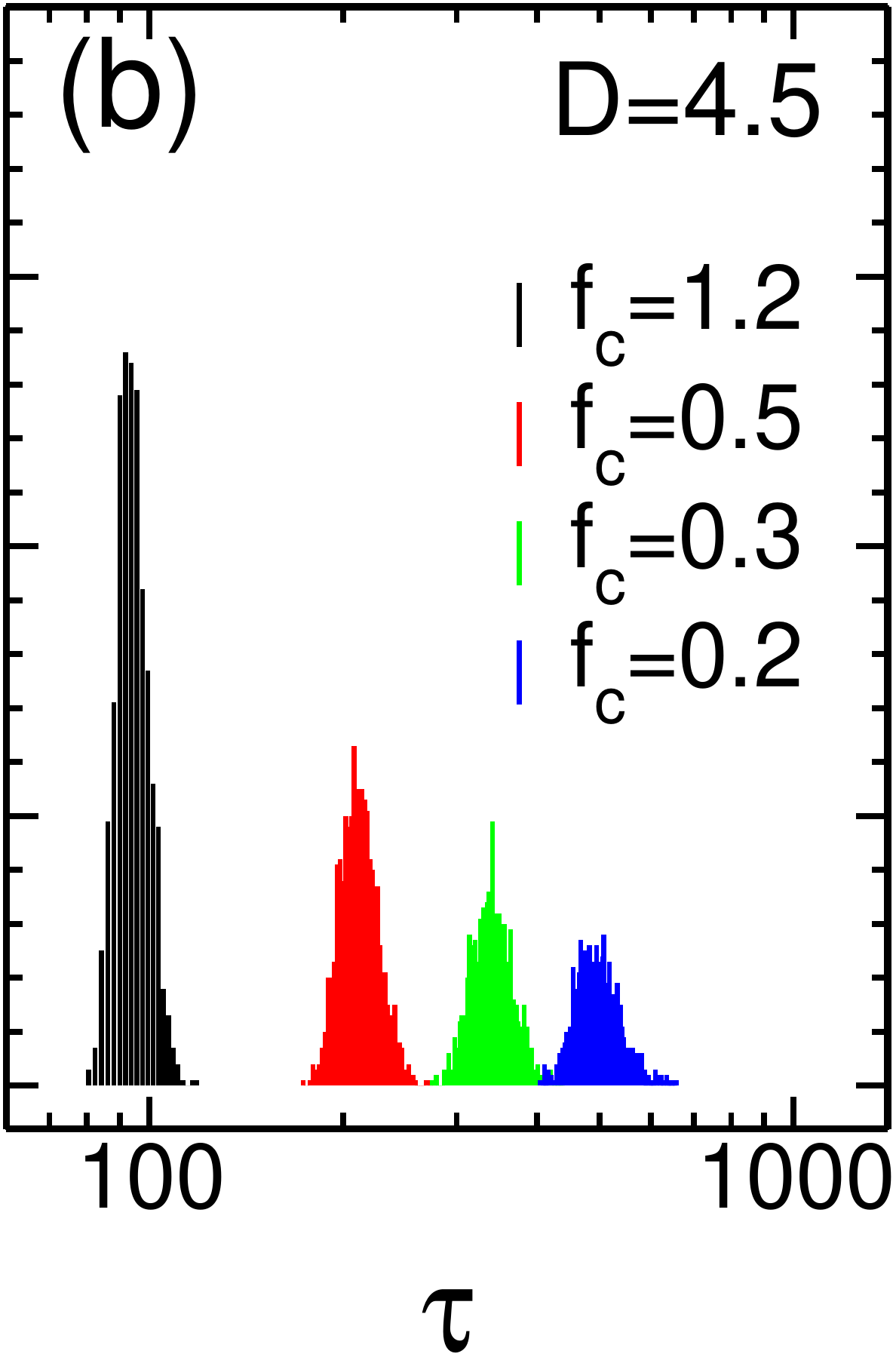}
    \end{center}\end{minipage} \hskip-0.7cm
        \begin{minipage}{0.1945\textwidth}
    \begin{center}
        \includegraphics[width=0.8\textwidth]{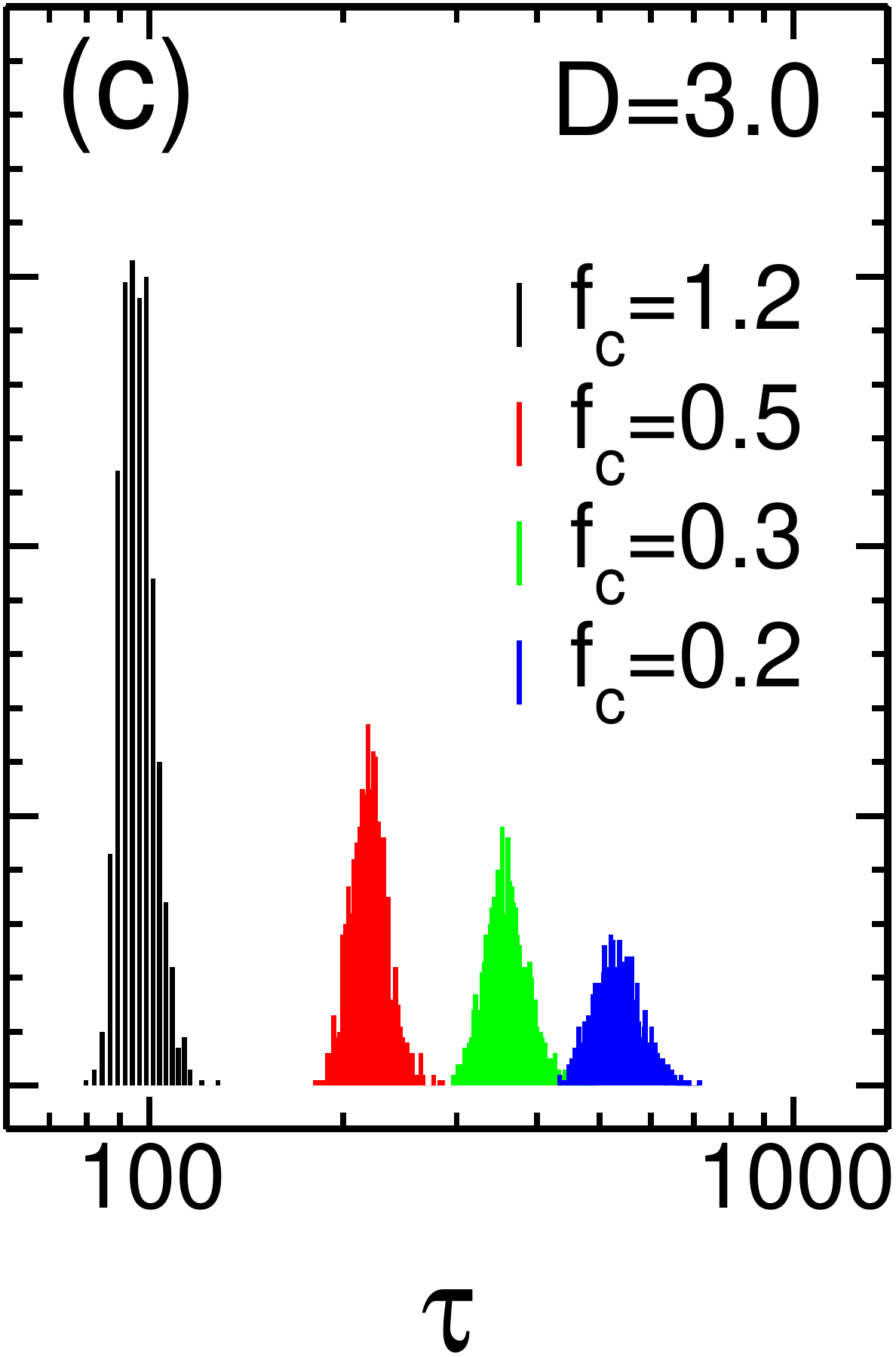}
    \end{center}\end{minipage} \hskip-0.7cm
    	\begin{minipage}{0.1945\textwidth}
    \begin{center}
        \includegraphics[width=0.8\textwidth]{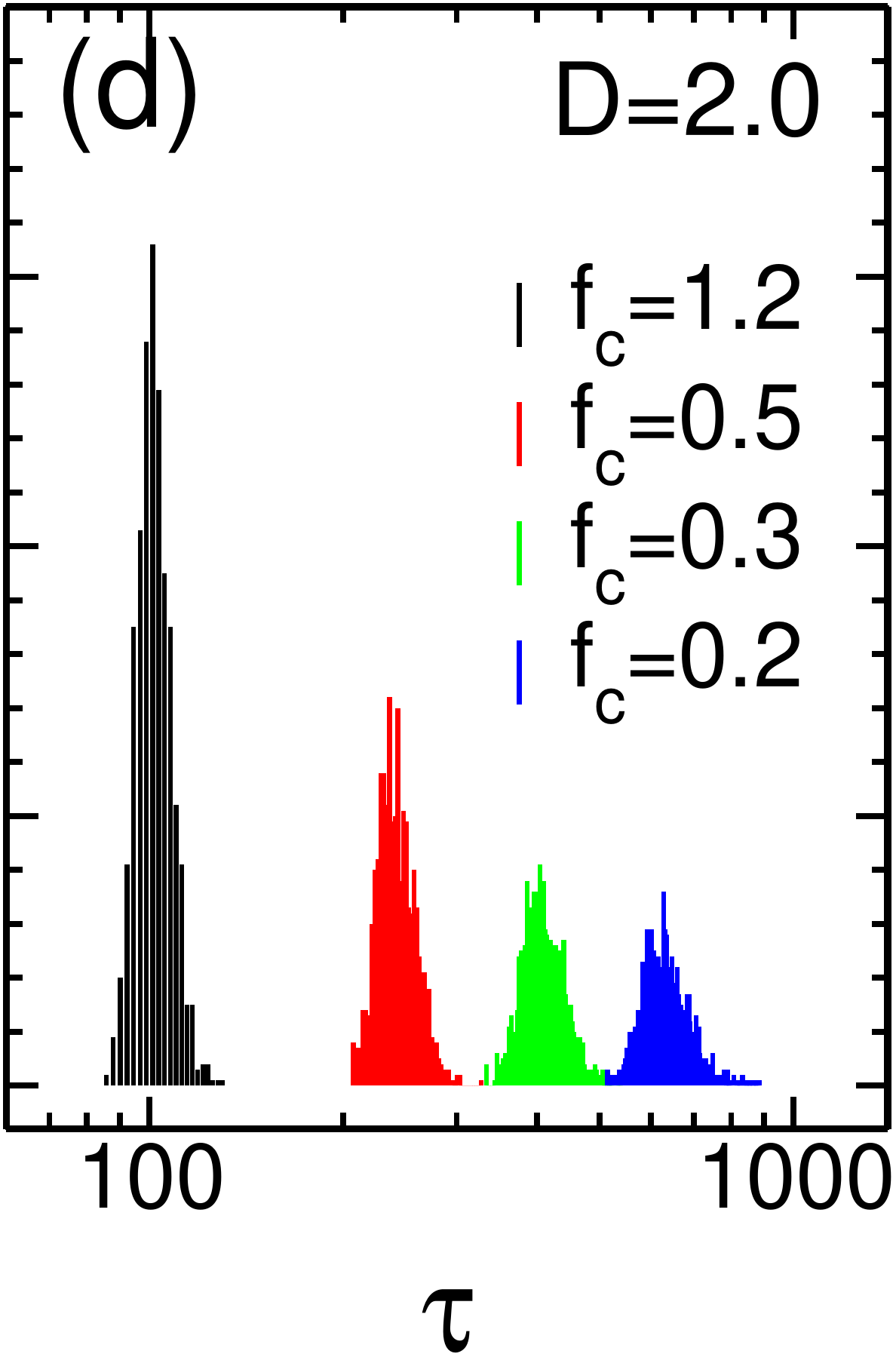}
    \end{center}\end{minipage}
    \hskip-0.7cm
    \begin{minipage}{0.263\textwidth}
    \begin{center}
    	\includegraphics[width=0.785\textwidth]{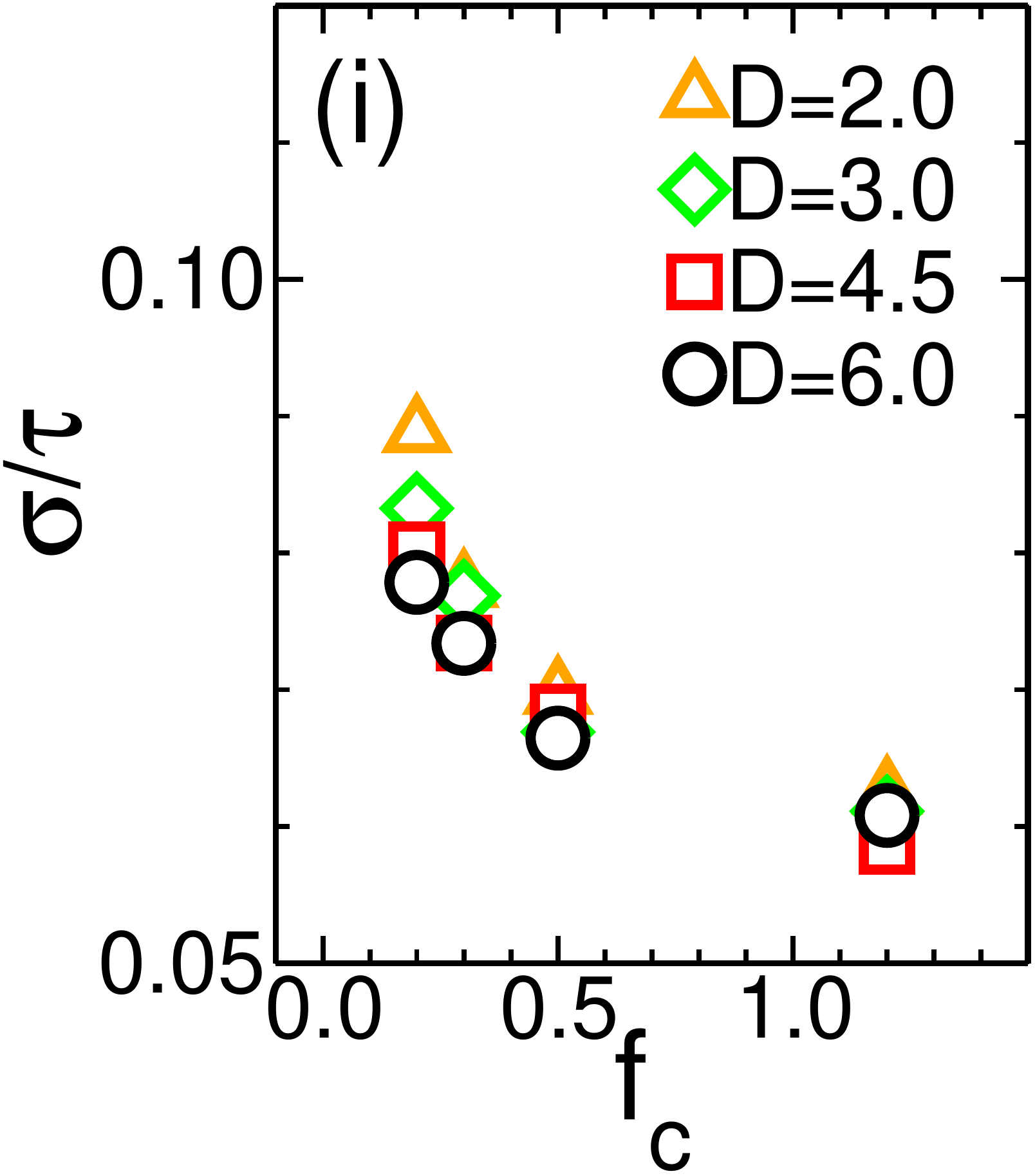}
    \end{center}\end{minipage}
    \begin{minipage}{0.263\textwidth}
    \vspace{+0.1cm}
    \begin{center}
    	\includegraphics[width=0.8\textwidth]{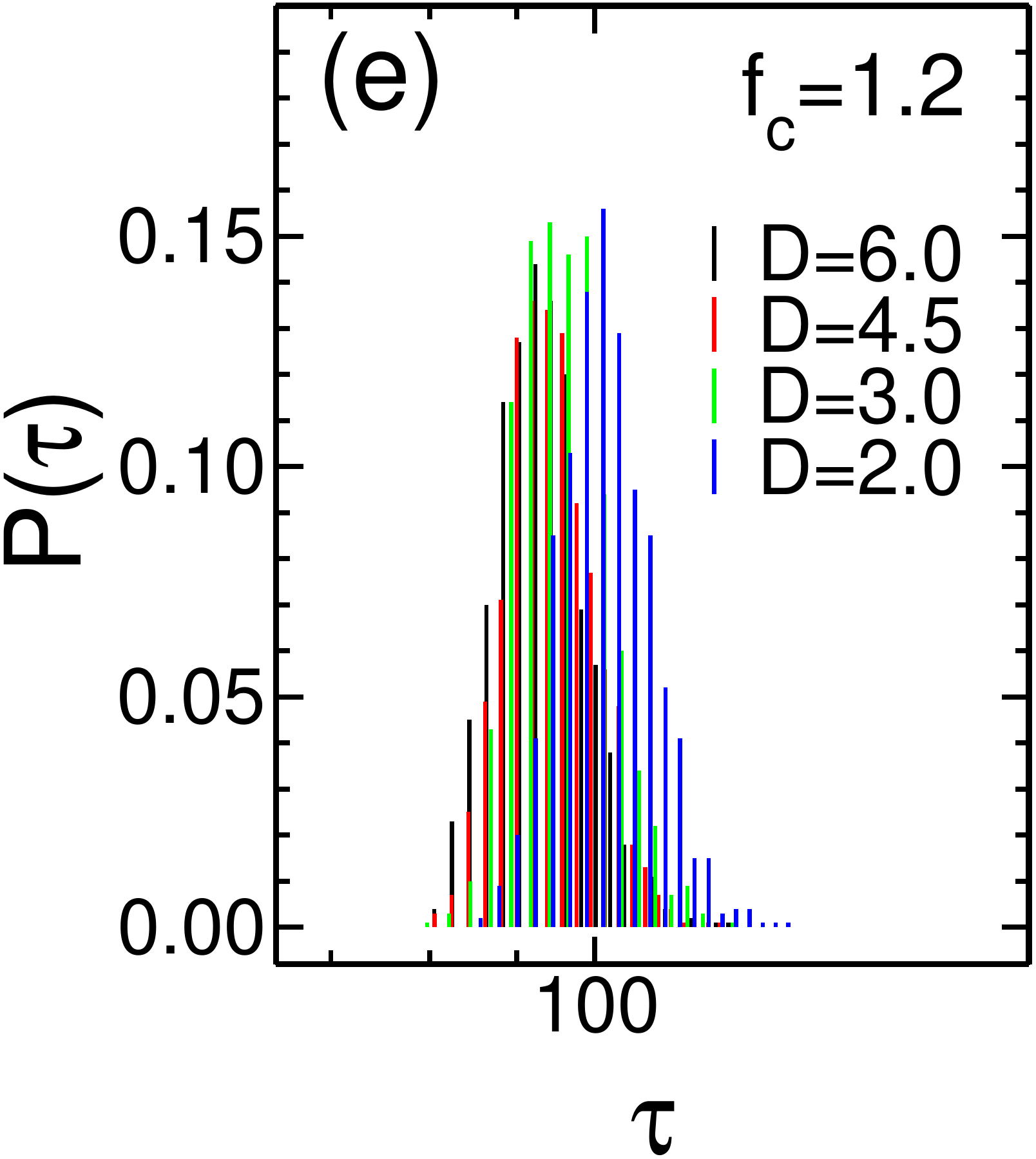}
    \end{center}\end{minipage} \hskip-0.82cm
    \begin{minipage}{0.1945\textwidth}
    \vspace{+0.1cm}
    \begin{center}
        \includegraphics[width=0.8\textwidth]{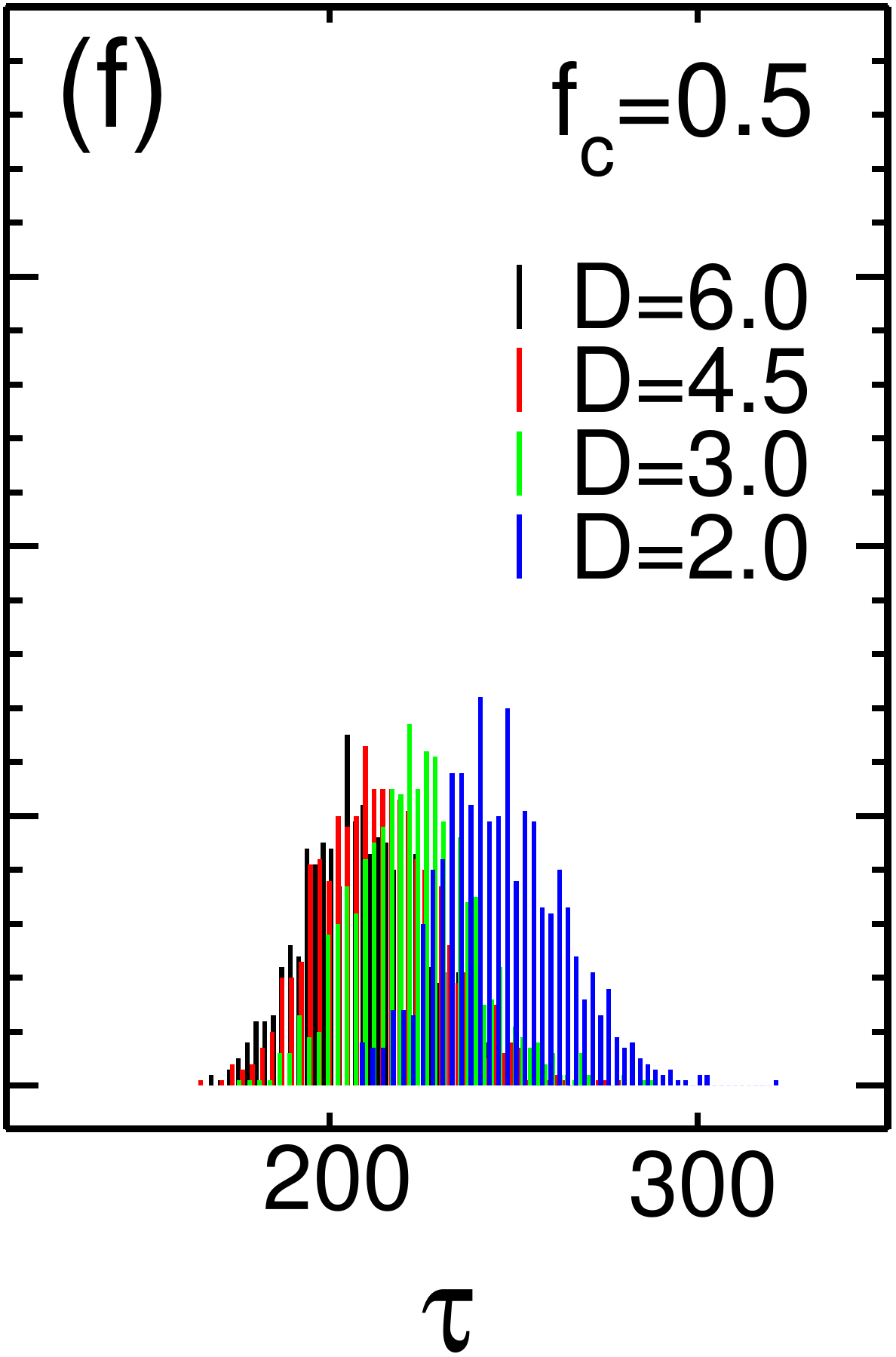}
    \end{center}\end{minipage} \hskip-0.7cm
        \begin{minipage}{0.1945\textwidth}
        \vspace{+0.1cm}
    \begin{center}
        \includegraphics[width=0.8\textwidth]{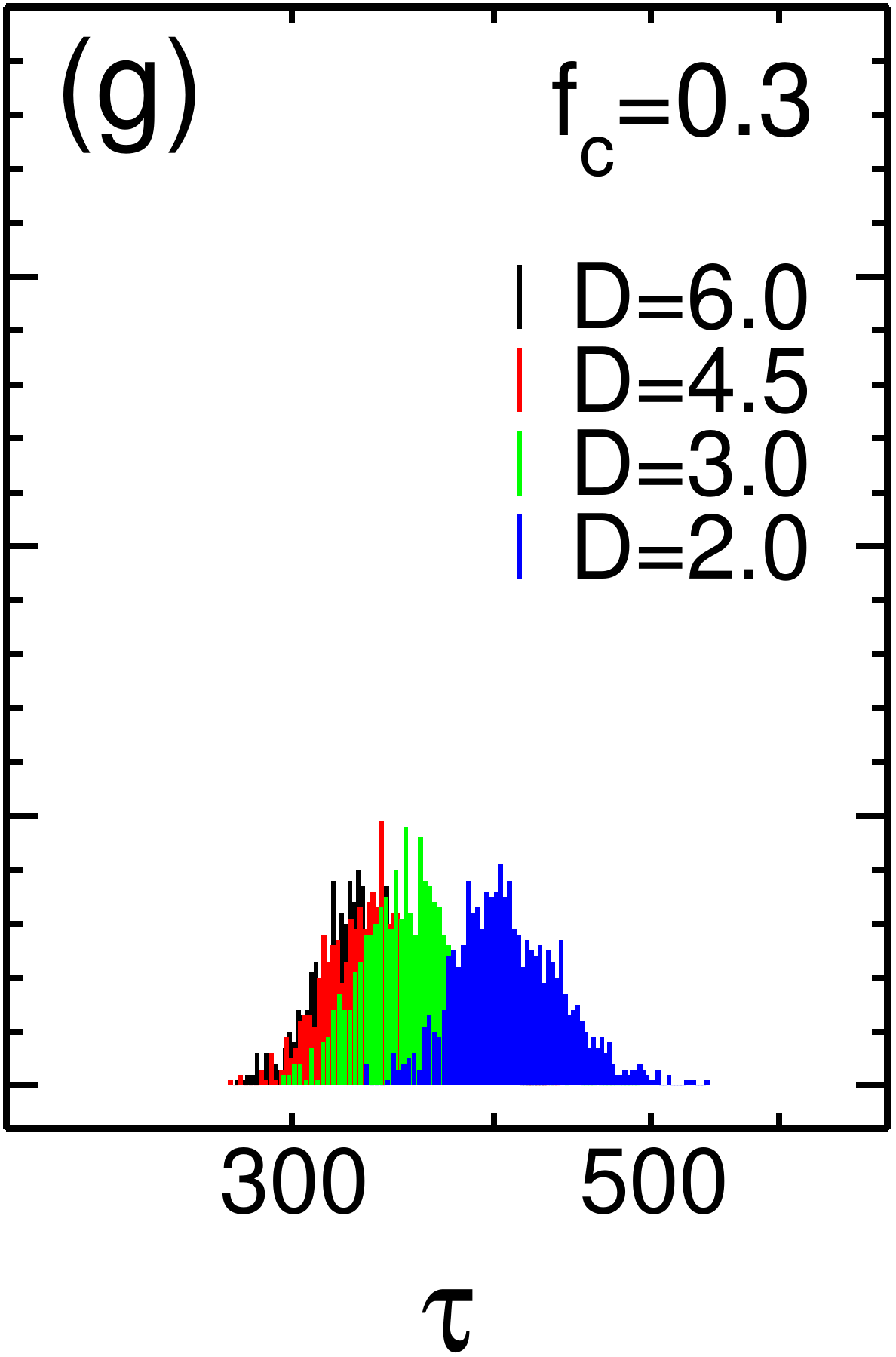}
    \end{center}\end{minipage} \hskip-0.7cm
    	\begin{minipage}{0.1945\textwidth}
    	\vspace{+0.1cm}
    \begin{center}
        \includegraphics[width=0.8\textwidth]{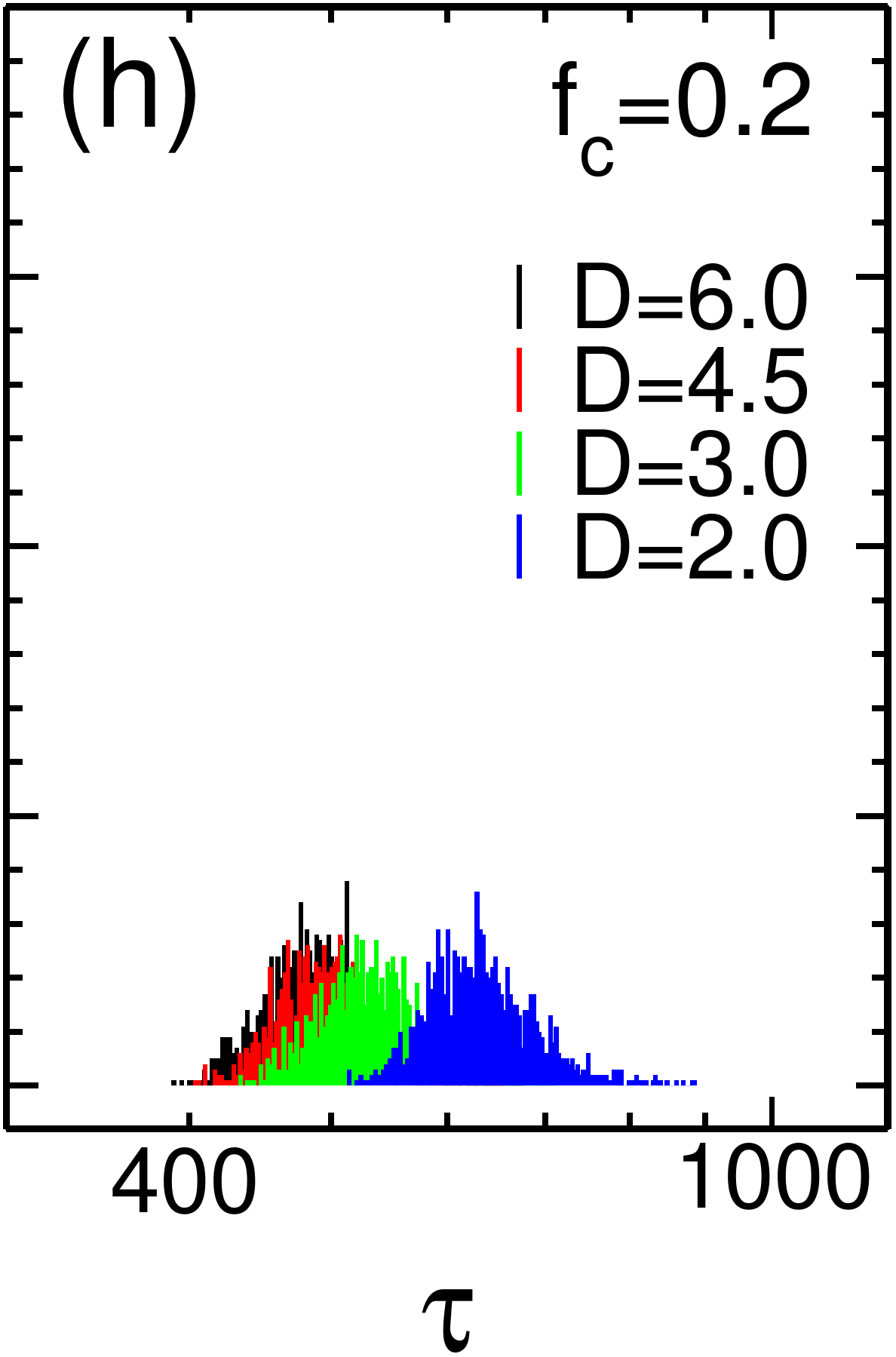}
    \end{center}\end{minipage}
    \hskip-0.7cm
    \begin{minipage}{0.263\textwidth}
    \vspace{+0.1cm}
    \begin{center}
    	\includegraphics[width=0.795\textwidth]{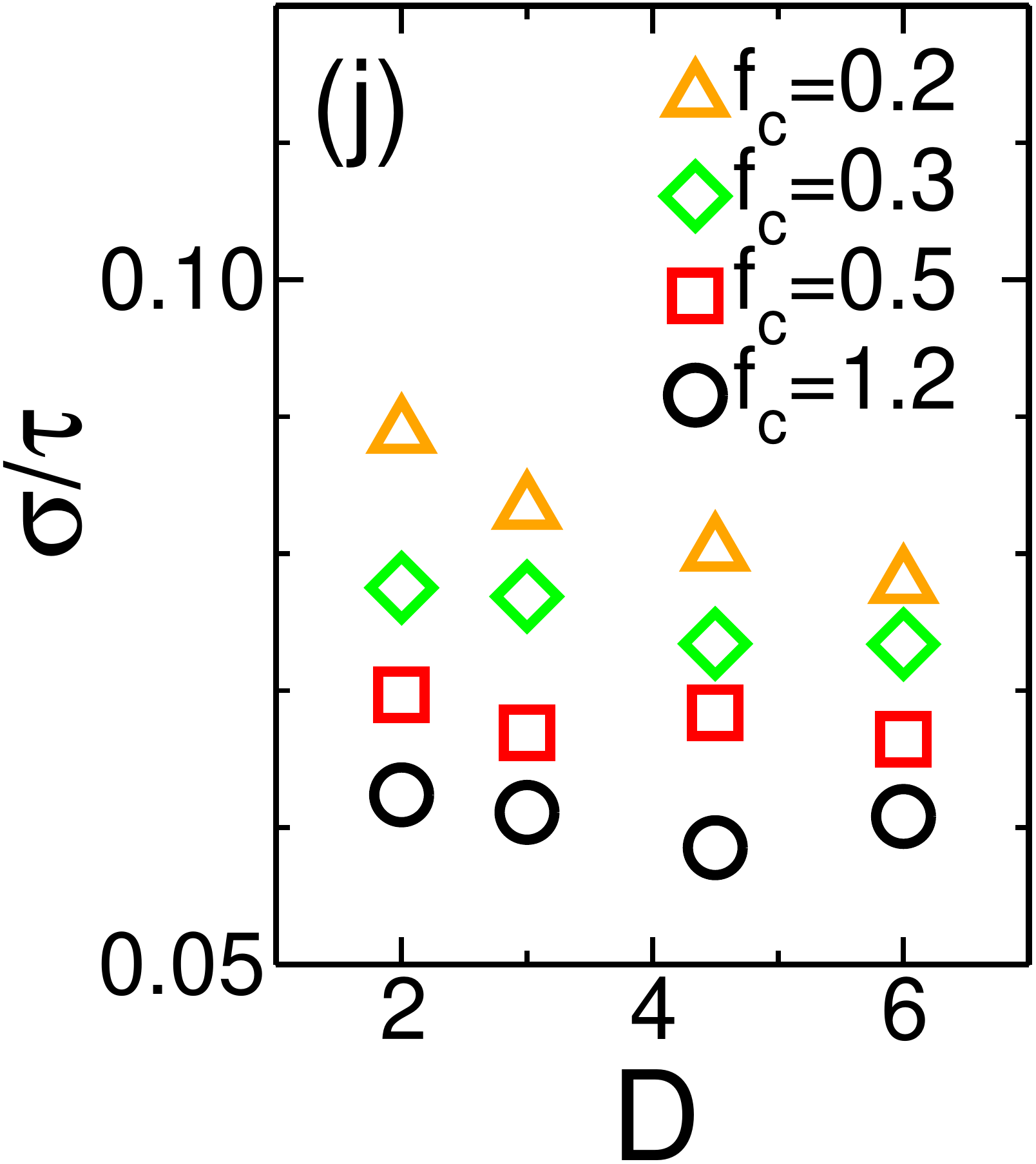}
    \end{center}\end{minipage}
\caption{(a) Probability distribution $P(\tau)$ of translocation times as
a function of translocation time $\tau$, for fixed channel width $D=6.0$ and
channel driving force $f_{\textrm{c}}= 1.2$ (black bars), 0.5 (red bars),
0.3 (green bars) and 0.2 (blue bars). Panels (b), (c), and (d) are the same
as panel (a) but for $D=4.5$, 3.0, and 2.0, respectively. (e) $P(\tau)$ as
function of $\tau$, for fixed $f_{\textrm{c}}=1.2$ and $D = 6.0$ (black bars),
4.5 (red bars), 3.0 (green bars), and 2.0 (blue bars). Panels (f), (g), and
(h) are the same as panel (e) but for $f_{\textrm{c}}=0.5$, 0.3 and 0.2,
respectively. All panels are in log-linear scale. (i) Normalized
standard deviation $\sigma/\tau$ as function of $f_{\textrm{c}}$ for
$D=2.0$ (orange triangles), 3.0 (green diamonds), 4.5 (red squares), and
6.0 (black circles). (j) $\sigma/\tau$ as function of $D$ for $f_{\textrm{c}}
=2.0$ (orange triangles), 0.3 (green diamonds), 0.5 (red squares), and
1.2 (black circles).}
\label{fig:propability}
\end{center}
\end{figure*}

The scaling form for the translocation time can be analytically obtained from IFTP
theory in the SS limit. To this end one needs to integrate $N$ in the TP stage from 0 to
$N_0$, and in the PP stage $\tilde{R}$ should be integrated from $R(N_0)$ to
0. These integrals are performed by combining the mass conservation in the
TP stage, $N=\tilde{s}+\tilde{l}$, and in the PP stage, $N=\tilde{s}+
\tilde{l}=N_0$, with Eq.~(\ref{BD_force}) \cite{jalalJCP2014}. Then,
summing up the TP and PP times the relation for $\tilde{\tau}$ yields
in the form \cite{jalalPRR2021}
\begin{equation}
\int_0^{\tilde{t}_{\textrm{TP}}}\tilde{f}_{\textrm{TP}}(\tilde{t})\textrm{d}
\tilde{t}+\int_{\tilde{t}_{\textrm{TP}}}^{\tilde{\tau}}\tilde{f}_{\textrm{PP}}
(\tilde{t})\textrm{d}\tilde{t}=\int_0^{N_0}\textrm{d}N[\tilde{R}(N)+
\tilde{\eta}_{\textrm{c}}],
\label{sum_f_TP_PP}
\end{equation}
where $\tilde{f}_{\textrm{TP}}(\tilde{t})$ and $\tilde{f}_{\textrm{ {P}P}}
(\tilde{t})$ are determined by Eq.~(\ref{f_TP_PP}). Combining
Eqs.~(\ref{sum_f_TP_PP}) and (\ref{f_TP_PP}) and using the closure
relation $\tilde{R}=A_{\nu} N^{\nu}$, where $A_{\nu}=1.1$ (from LD
data) for the end-to-end distance in the TP stage and $\nu=3/4$
is the Flory exponent in 2D, the translocation time is given by
\cite{jalalPRR2021}
\begin{equation}
\tilde{\tau}=\left[\frac{A_{\nu}N_0^{1+\nu}}{1+\nu}+N_0\tilde{\eta}_{
\textrm{c}}\right]/(G\tilde{f}_{\textrm{c}}),
\label{tau}
\end{equation}
where $G=-0.5(c-a)^2/(b+d)+c-d/2\approx13.7$ for panel (a) in
Fig.~\ref{fig:force}. Equation~(\ref{tau}) shows that the translocation time
scales with the channel driving force as $\tilde{\tau}\propto\tilde{f}
_{\textrm{c}}^{-1}$, i.e., for the SS regime the force exponent is $\beta
=-1$. Moreover, the  {effective} translocation exponent $\alpha$ { {that is defined via $\tilde{\tau}\propto N_0^{\alpha}$} is bounded between $1$
and $1+\nu$, i.e., $1<\alpha<1+\nu$, due to the competition between the
two terms in the brackets of Eq.~(\ref{tau}), in which the first and the
second terms originate from the {\it cis}-side and channel frictions,
respectively. The lower and the upper bounds are assumed in the very short
and very long chain limits, respectively \cite{jalalJCP2014}.

To compare the scaling result of Eq.~(\ref{tau}) with the LD simulation
data, the inset of Fig.~\ref{fig:tau}(a) shows the force exponent $\beta$
from LD as a function of the channel width $D$. As the
channel width decreases $\beta$ moves closer to $-1$. The reason for this
is that as an increasing channel diameter allows more pronounced spatial
fluctuations of the {\it trans}-side subchain, leading to an increased
entropic force. Consequently, the force exponent $\beta$ changes as a
function of the channel diameter.  For the minimum channel width $D=2$
the force exponent assumes its ideal value $\beta=-1$, due to negligible
configurational fluctuations of the subchain on the {\it trans} side (see
Fig.~\ref{fig:tau}(a)). This is the value in Eq.~(\ref{tau}) from IFTP theory.
Moreover, as shown in the inset of Fig.~\ref{fig:tau}(a), increasing the
channel width changes the value of the force exponent toward its
asymptotic value $\beta\sim-0.9$ for a very wide channel, corresponding
to a half-space with infinite $D$ \cite{ikonen2012b}. It should mentioned
that at high channel driving forces the force exponent approaches $\beta=-1$
regardless of the channel width, as in this limit the value of the driving
force by far exceeds the entropic one.

To investigate the effect of the entropic force on the translocation time the inset of Fig.~\ref{fig:tau}(b)
shows the normalized translocation time $\tau/\tau (D=2)$ as function $D$
for various values of $f_{\textrm{c}}$. As can be seen, the deviation of
$\tau$ from the value for $D=2$ (when the entropic force is negligible due
to the small configurational fluctuations of the {\it trans} side subchain)
grows with increasing $D$. This is more pronounced at weaker driving forces,
again due to the fact that by increasing $D$ the entropic force plays a more
important role.

\subsection{Distribution of translocation times}
\label{DistributionTime}

We next study the effect of the channel driving force $f_{\textrm{c}}$ as
well as the channel diameter $D$ on the distribution $P(\tau)$ of
translocation times $\tau$. Figure~\ref{fig:propability}(a) shows $P(\tau)$
as function of $\tau$ for fixed contour length $N_0=100$ and channel width
$D=6$, for the channel driving force values $f_{\textrm{c}}=1.2$ (black
bars), 0.5 (red bars), 0.3 (green bars), and 0.2 (blue bars). Panels (b),
(c), and (d) are the same as panel (a), but for channel widths $D=4.5$, $3$,
and 2, respectively. As can be seen in each panel, decreasing channel
driving force increases the mean translocation time, and $P(\tau)$
becomes wider (note the log scale on the horizontal axis). This is again due to
the fact that decreasing $f_{\textrm{c}}$ allows more substantial spatial
fluctuations of the mobile subchain in the {\it cis} and on the {\it trans}
sides.
Consequently, the fluctuations of translocation times increase as well. In panel (e) $P(\tau)$ is shown
as function of $\tau$, for fixed value of $f_{\textrm{c}}=1.2$ and for
channel widths $D=6.0$ (black bars), 4.5 (red bars), 3.0 (green bars), and
2.0 (blue bars). Panels (f), (g) and (h) are the same as panel (e) but for
channel driving forces $f_{\textrm{c}}=0.5$, 0.3, and 0.2, respectively. As
can seen from panel (e) to panel (h) decreasing $f_{\textrm{c}}$ leads to
increasing separation of the distributions in each panel. 
 
To quantify the distinctions, in panel (i) the normalized
standard deviation $\sigma/\tau$ is shown as function of $f_{\textrm{c}}$
for the channel widths $D=2.0$ (orange triangles), 3.0 (green diamonds),
4.5 (red squares), and 6.0 (black circles). Moreover, panel (j) shows
$\sigma/\tau$ as function of $D$ for the driving forces $f_{\textrm{c}}=0.2$
(orange triangles), 0.3 (green diamonds), 0.5 (red squares), and 1.2 (black
circles). As can be clearly seen, in panel (i) at constant $D$ the change in
$\sigma/\tau$ is more pronounced by increasing $f_{\textrm{c}}$ as compared to
its change at constant $f_{\textrm{c}}$ when increasing $D$ in panel (j).

\subsection{Mean-square displacement}
\label{MSD}

\begin{figure*}\begin{center}
    \begin{minipage}{0.24\textwidth}
    \begin{center}
    		\includegraphics[width=1.0\textwidth]{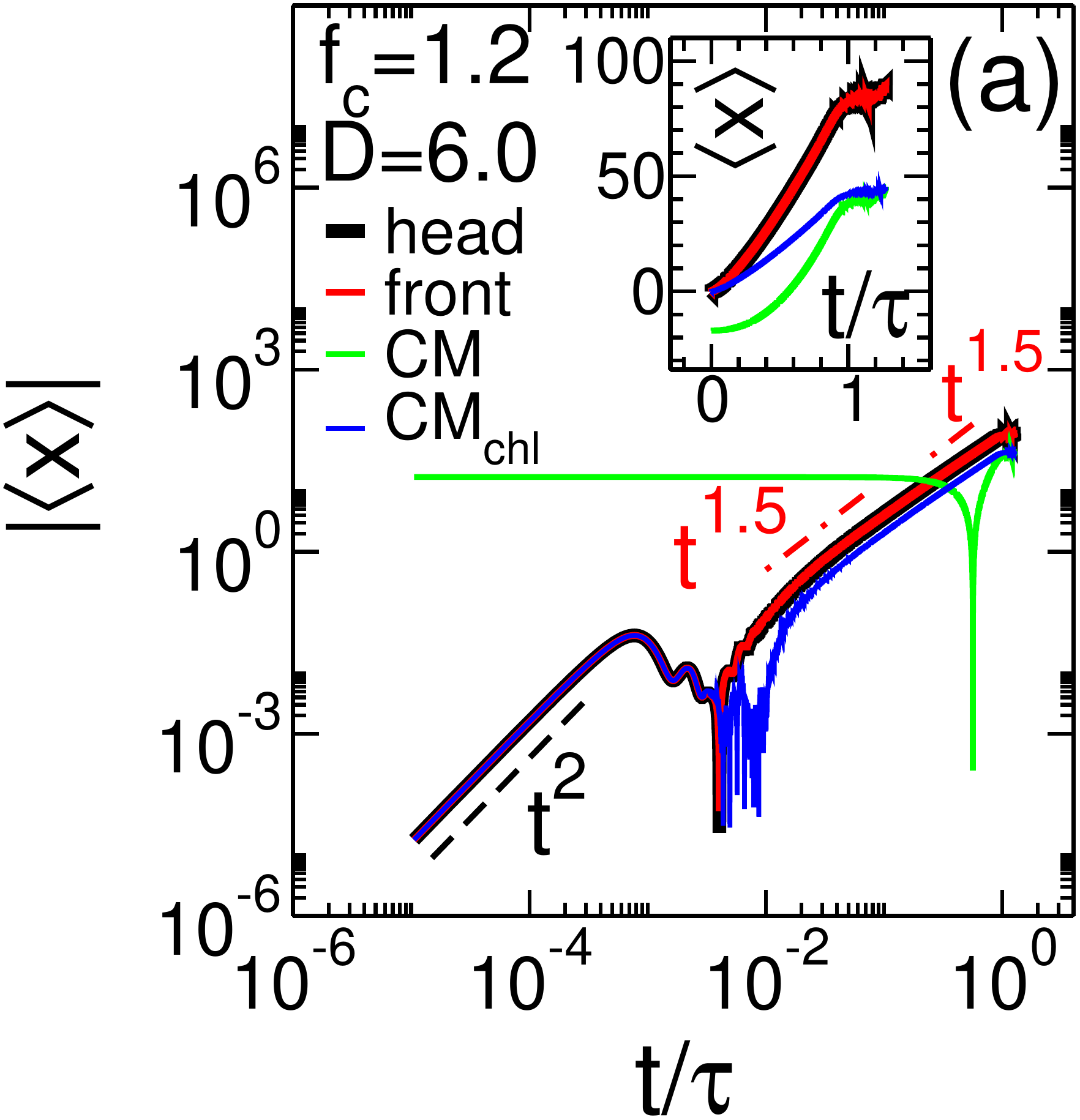}
    \end{center}\end{minipage} \hskip0.0cm
        \begin{minipage}{0.176\textwidth}
    \begin{center}
        \includegraphics[width=1.0\textwidth]{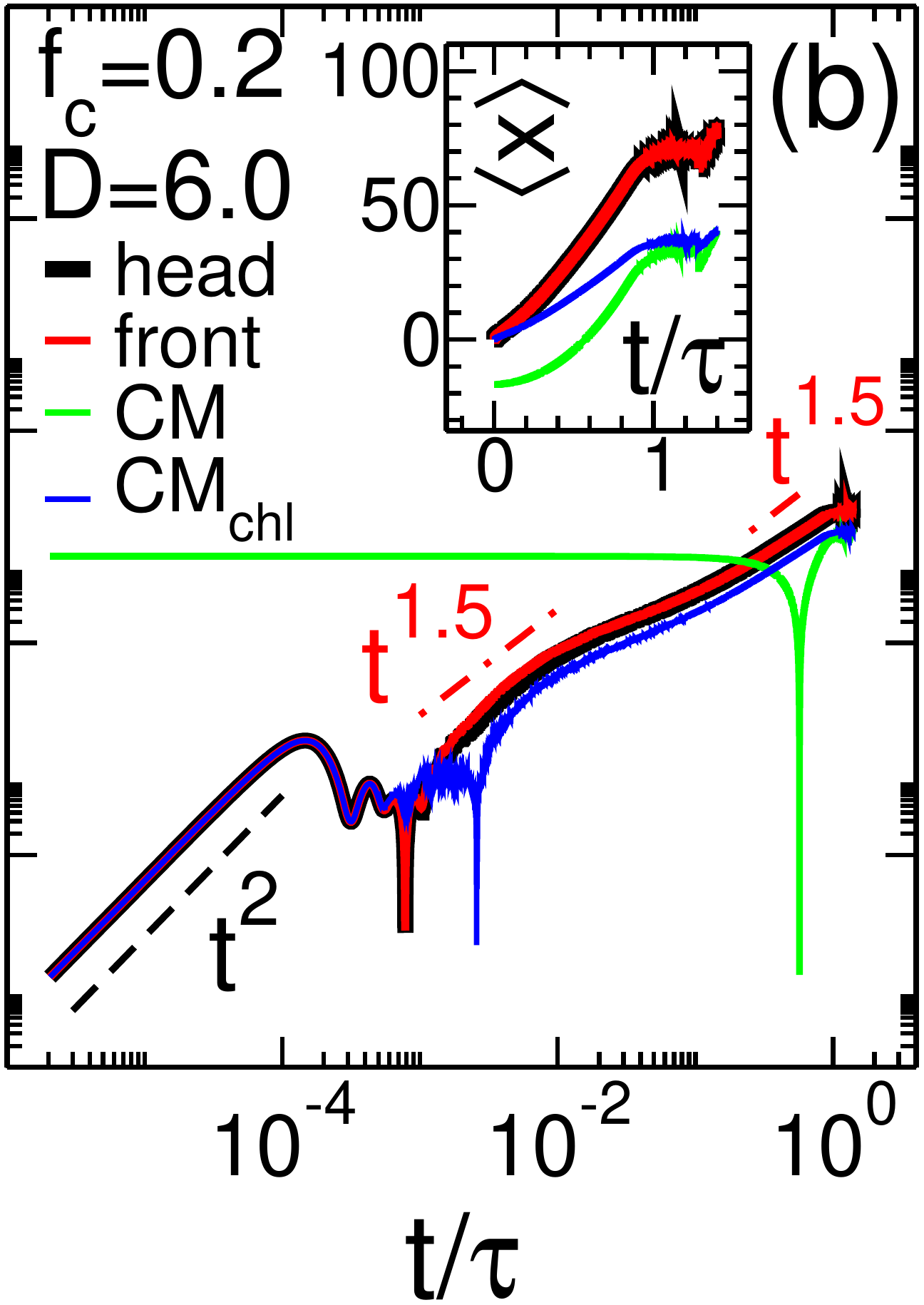}
    \end{center}\end{minipage} \hskip0.0cm
        \begin{minipage}{0.176\textwidth}
    \begin{center}
        \includegraphics[width=1.0\textwidth]{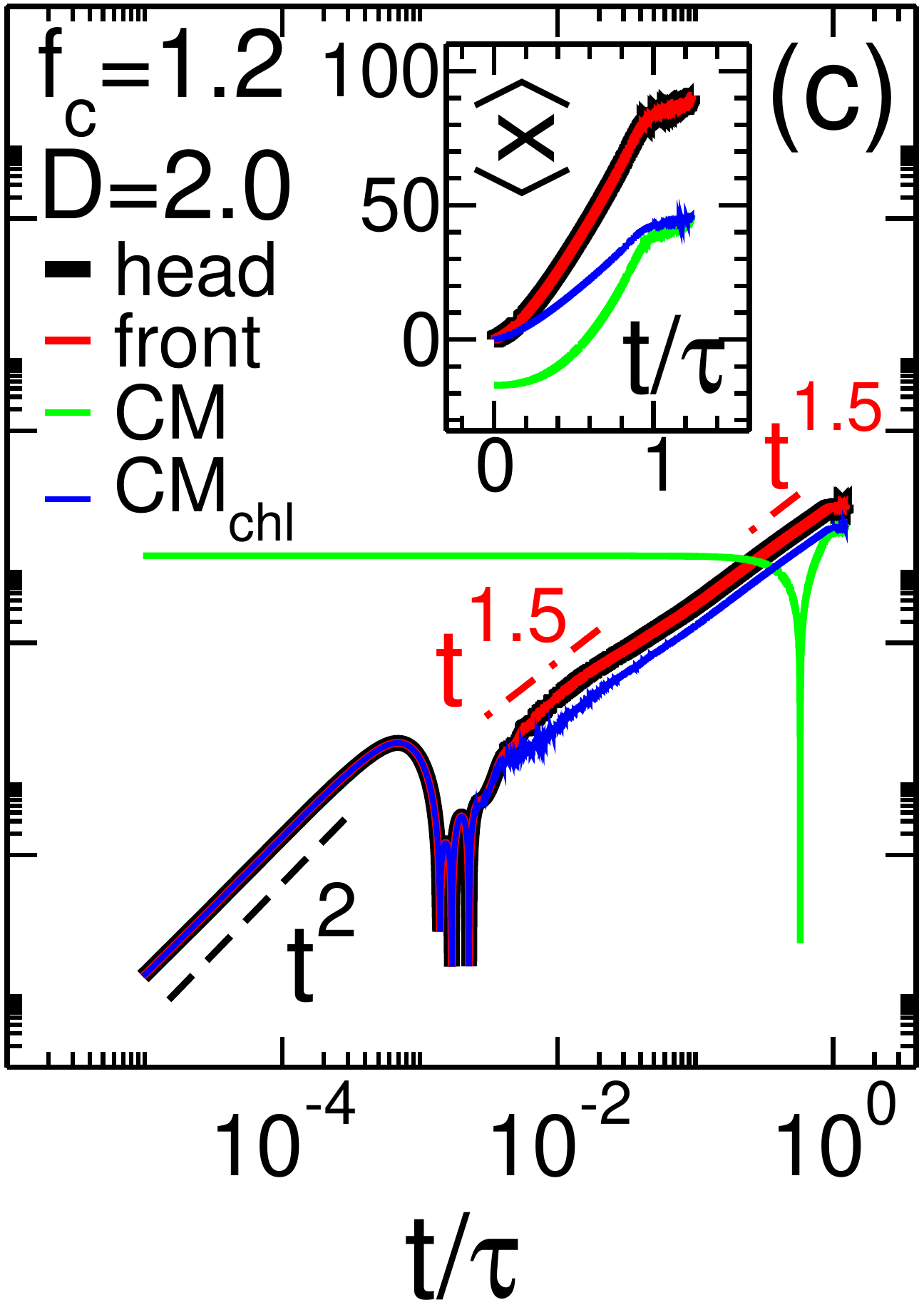}
    \end{center}\end{minipage} \hskip0.0cm
    	\begin{minipage}{0.176\textwidth}
    \begin{center}
        \includegraphics[width=1.0\textwidth]{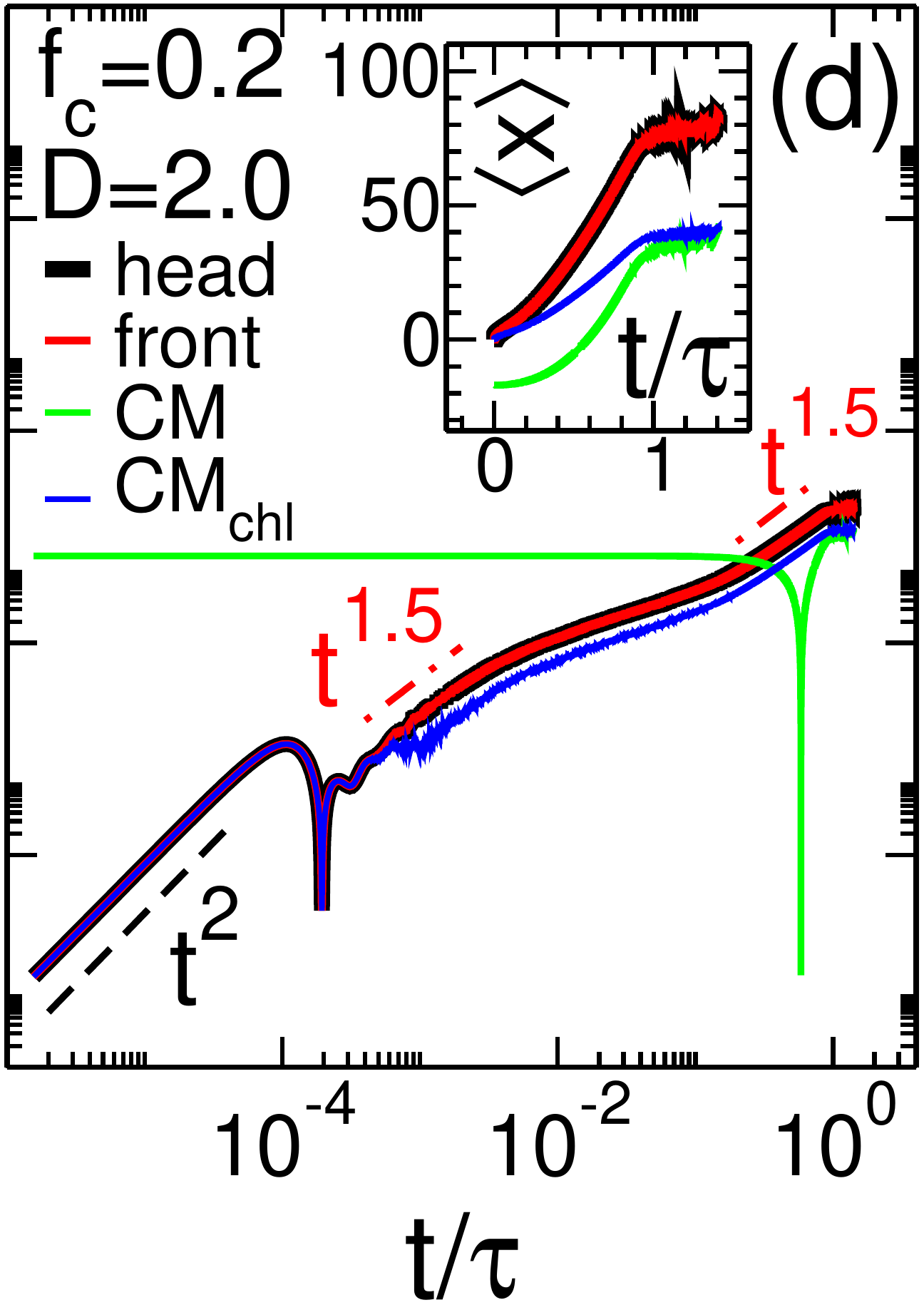}
    \end{center}\end{minipage}
    \vspace{+0.2cm}
    \begin{minipage}{0.24\textwidth}    
    \begin{center}
        \includegraphics[width=1.0\textwidth]{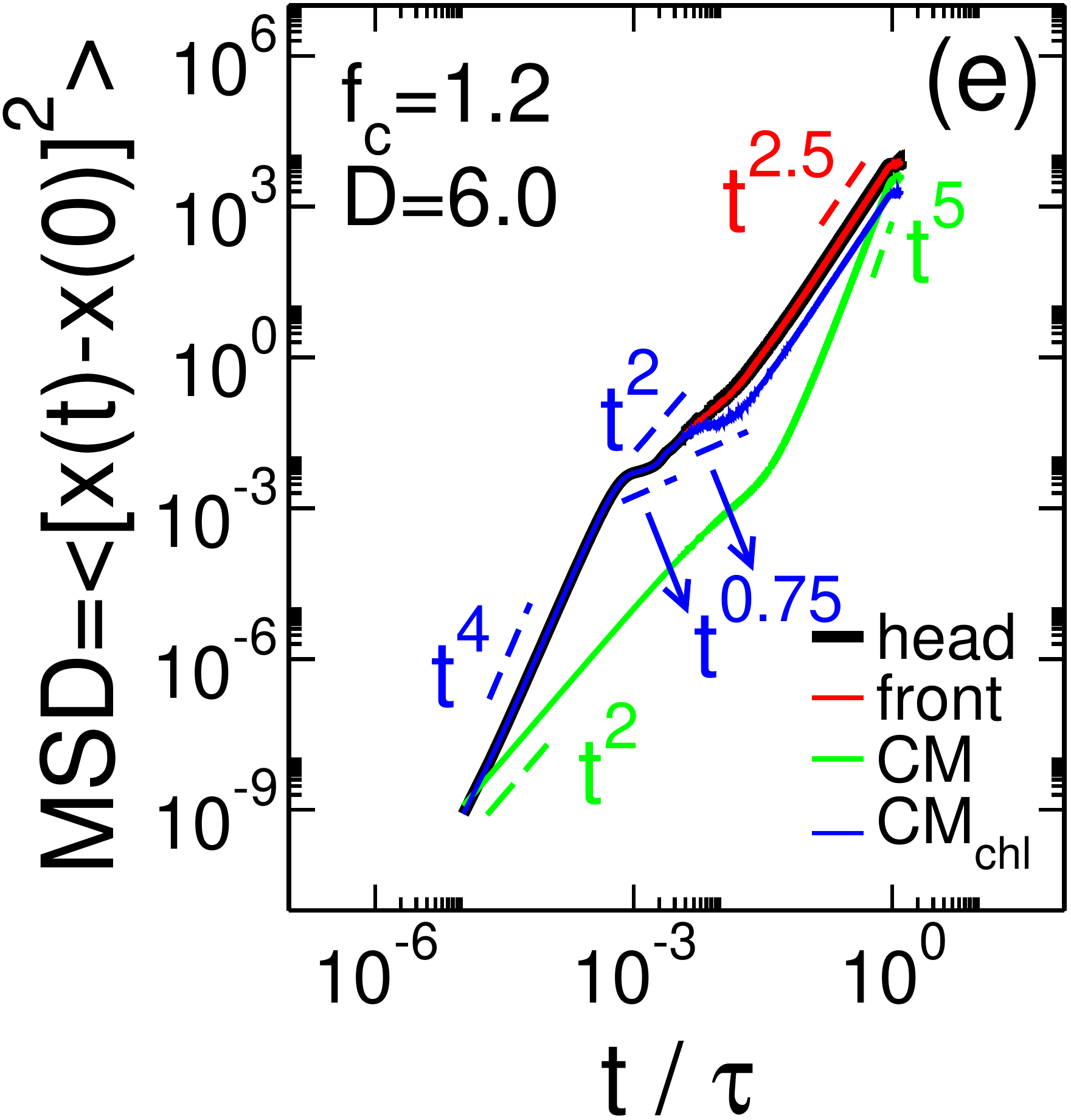}
    \end{center}\end{minipage} \hskip0.0cm
        \begin{minipage}{0.176\textwidth}
    \begin{center}
        \includegraphics[width=1.0\textwidth]{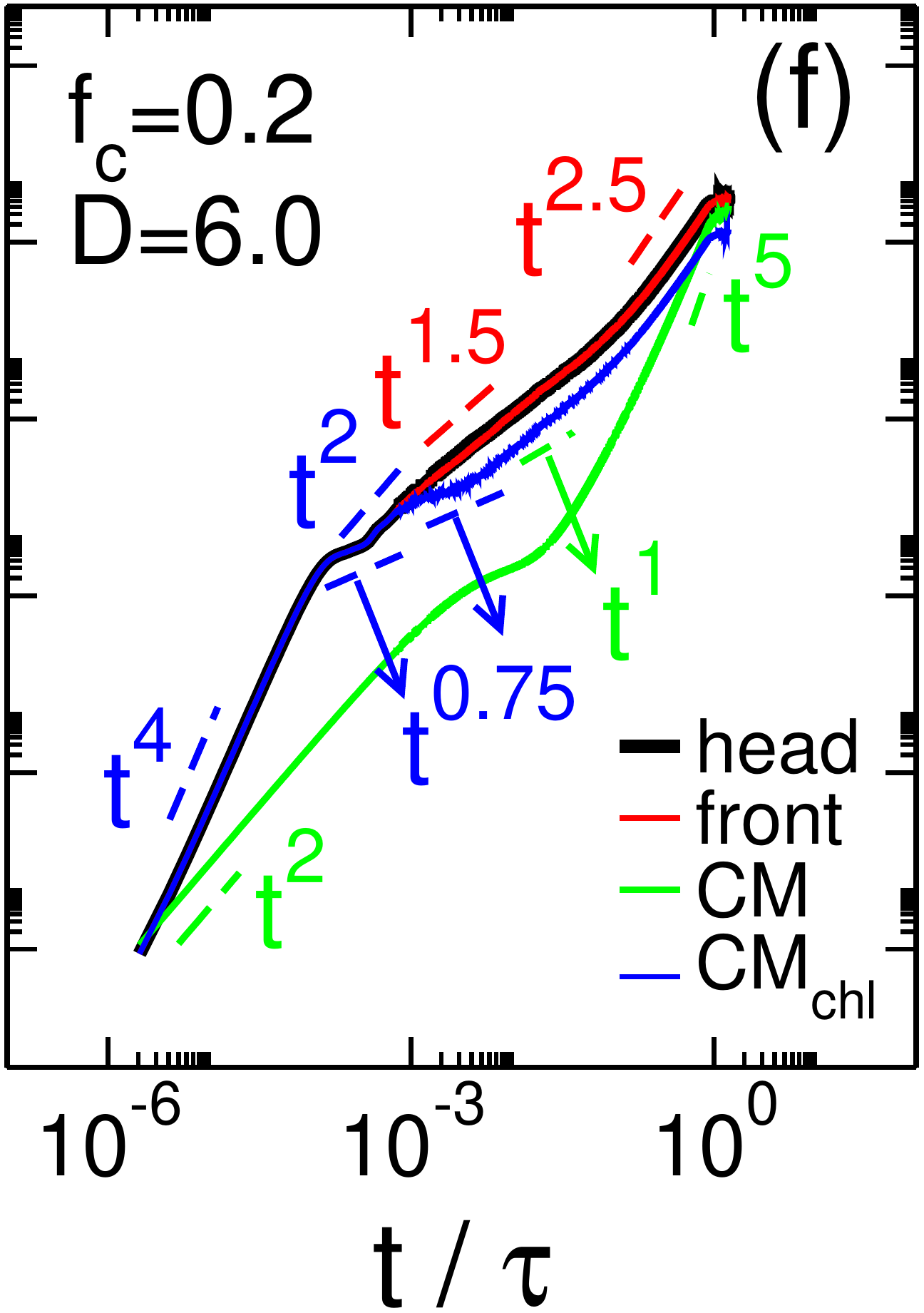}
    \end{center}\end{minipage} \hskip0.0cm
        \begin{minipage}{0.176\textwidth}
    \begin{center}
        \includegraphics[width=1.0\textwidth]{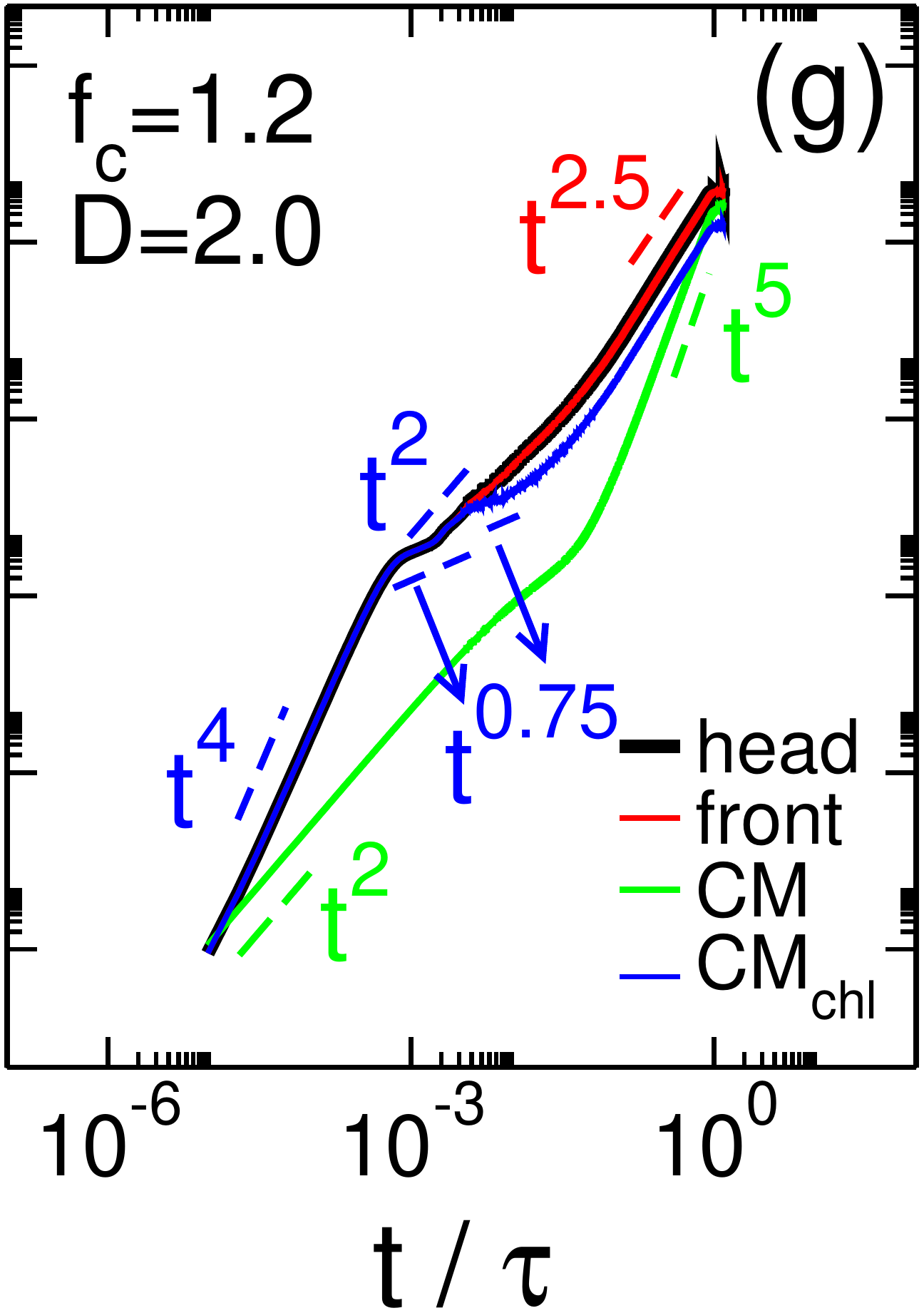}
    \end{center}\end{minipage} \hskip0.0cm
    	\begin{minipage}{0.176\textwidth}
    \begin{center}
        \includegraphics[width=1.0\textwidth]{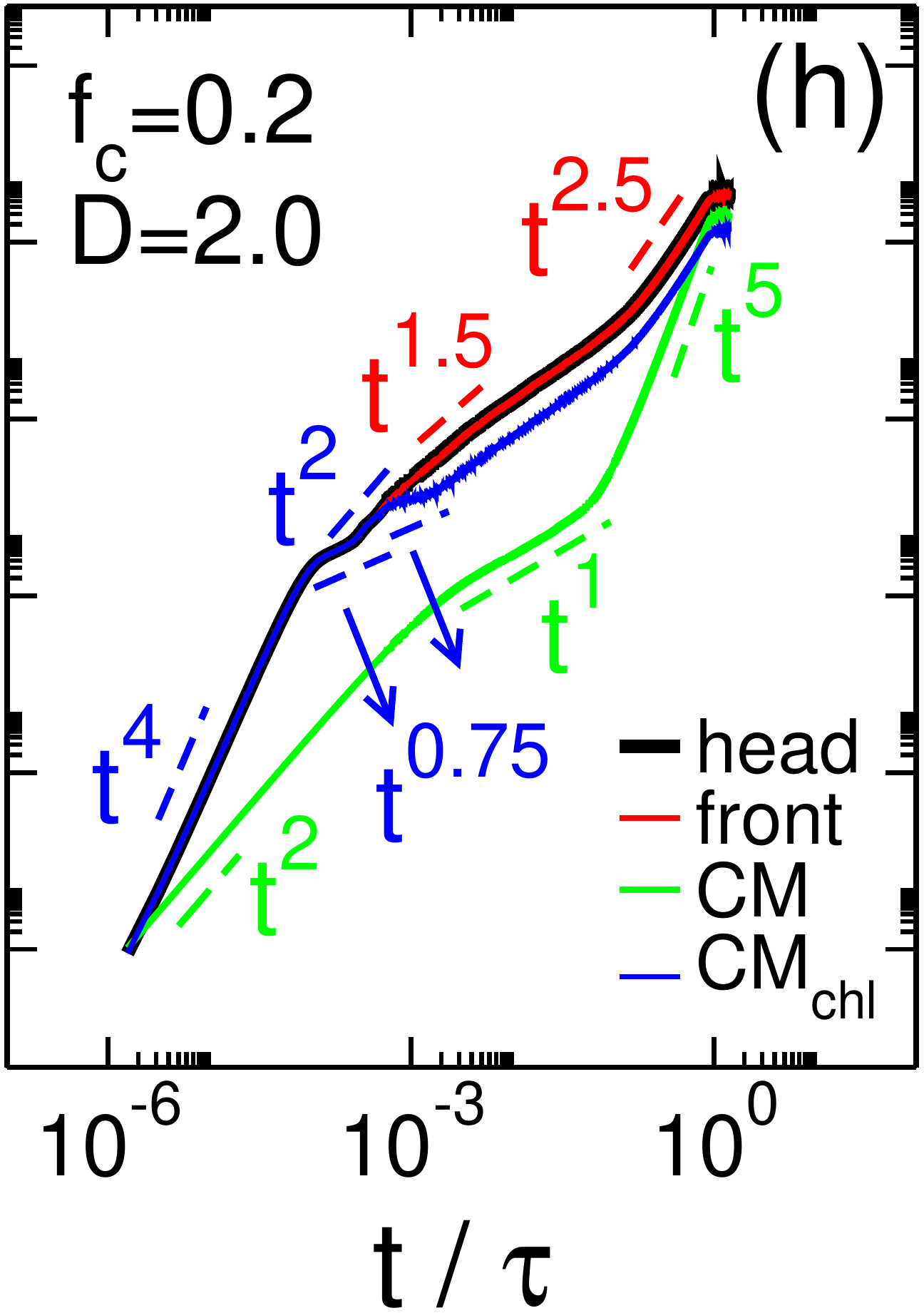}
    \end{center}\end{minipage}
    \vspace{+0.2cm}
    \begin{minipage}{0.24\textwidth}    
    \begin{center}
        \includegraphics[width=1.0\textwidth]{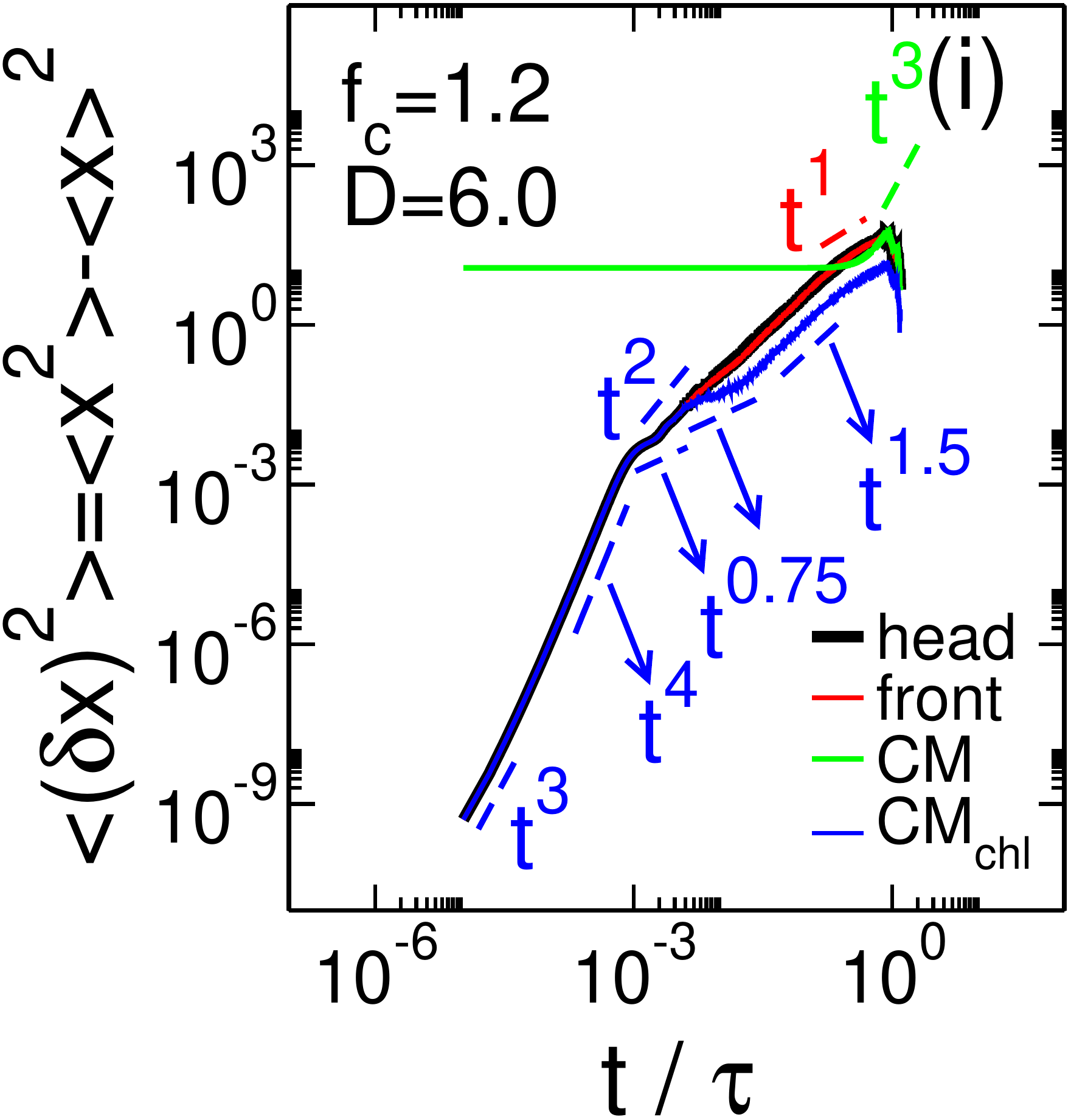}
    \end{center}\end{minipage} \hskip0.0cm
        \begin{minipage}{0.176\textwidth}
    \begin{center}
        \includegraphics[width=1.0\textwidth]{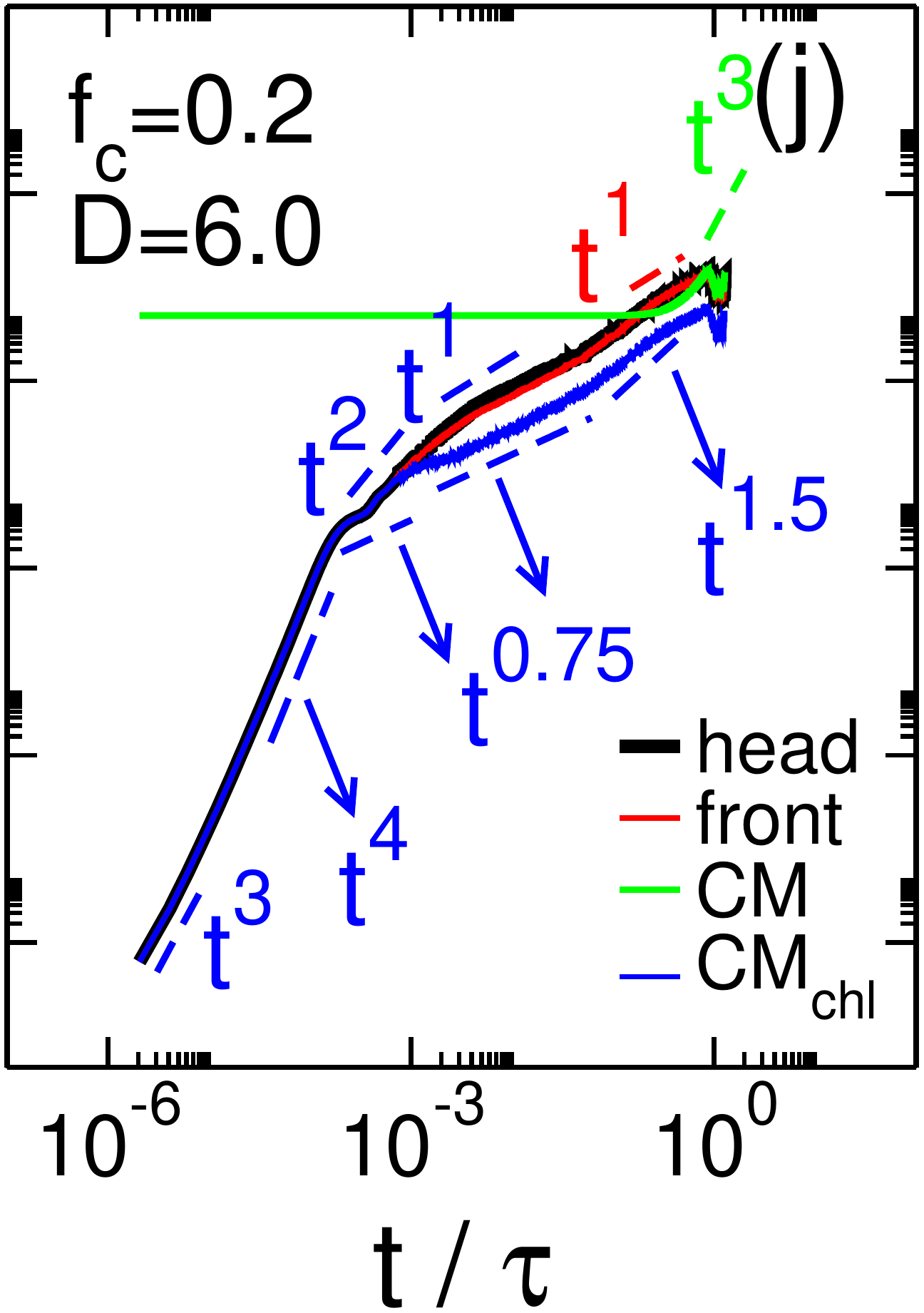}
    \end{center}\end{minipage} \hskip0.0cm
        \begin{minipage}{0.176\textwidth}
    \begin{center}
        \includegraphics[width=1.0\textwidth]{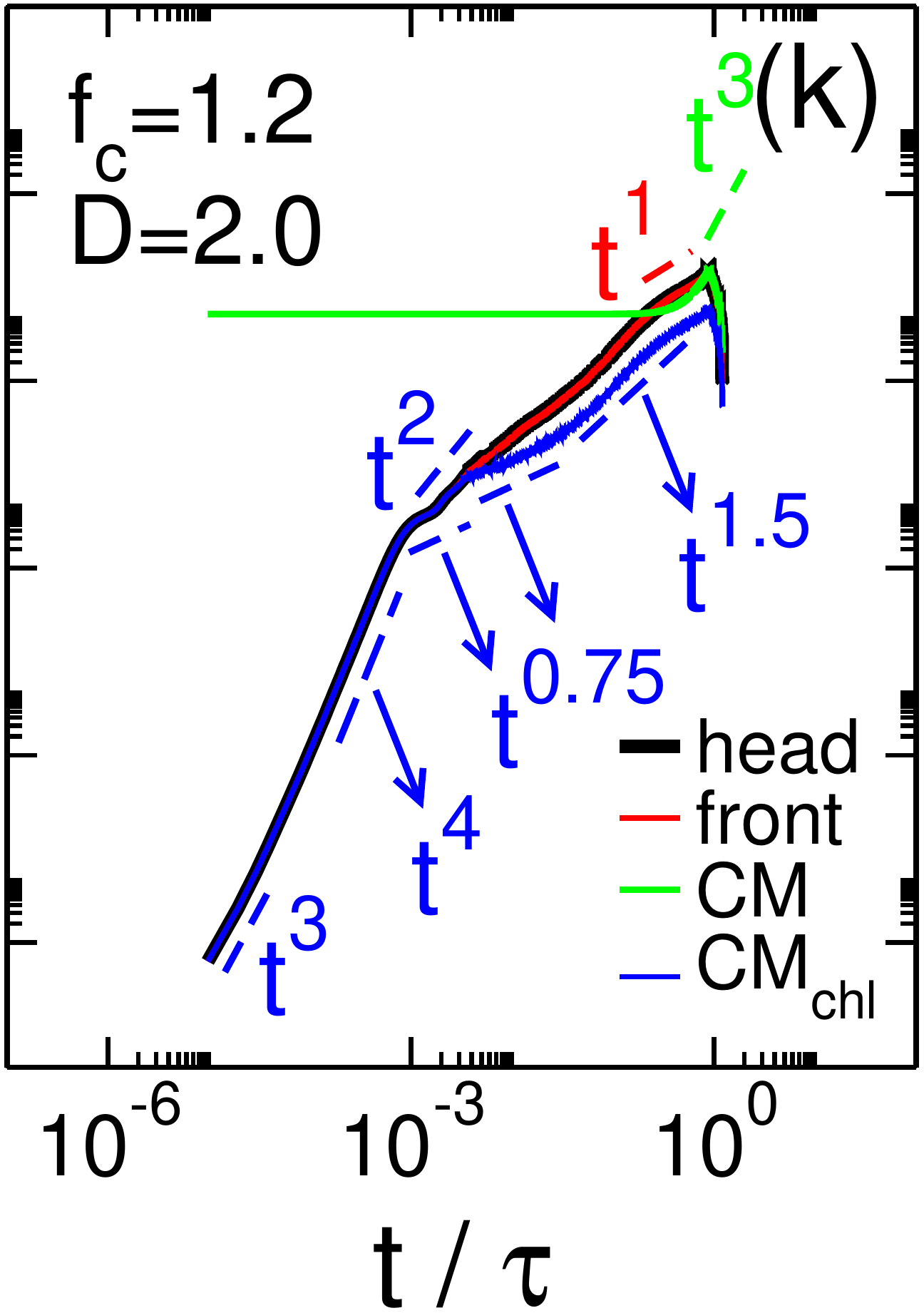}
    \end{center}\end{minipage} \hskip0.0cm
    	\begin{minipage}{0.176\textwidth}
    \begin{center}
        \includegraphics[width=1.0\textwidth]{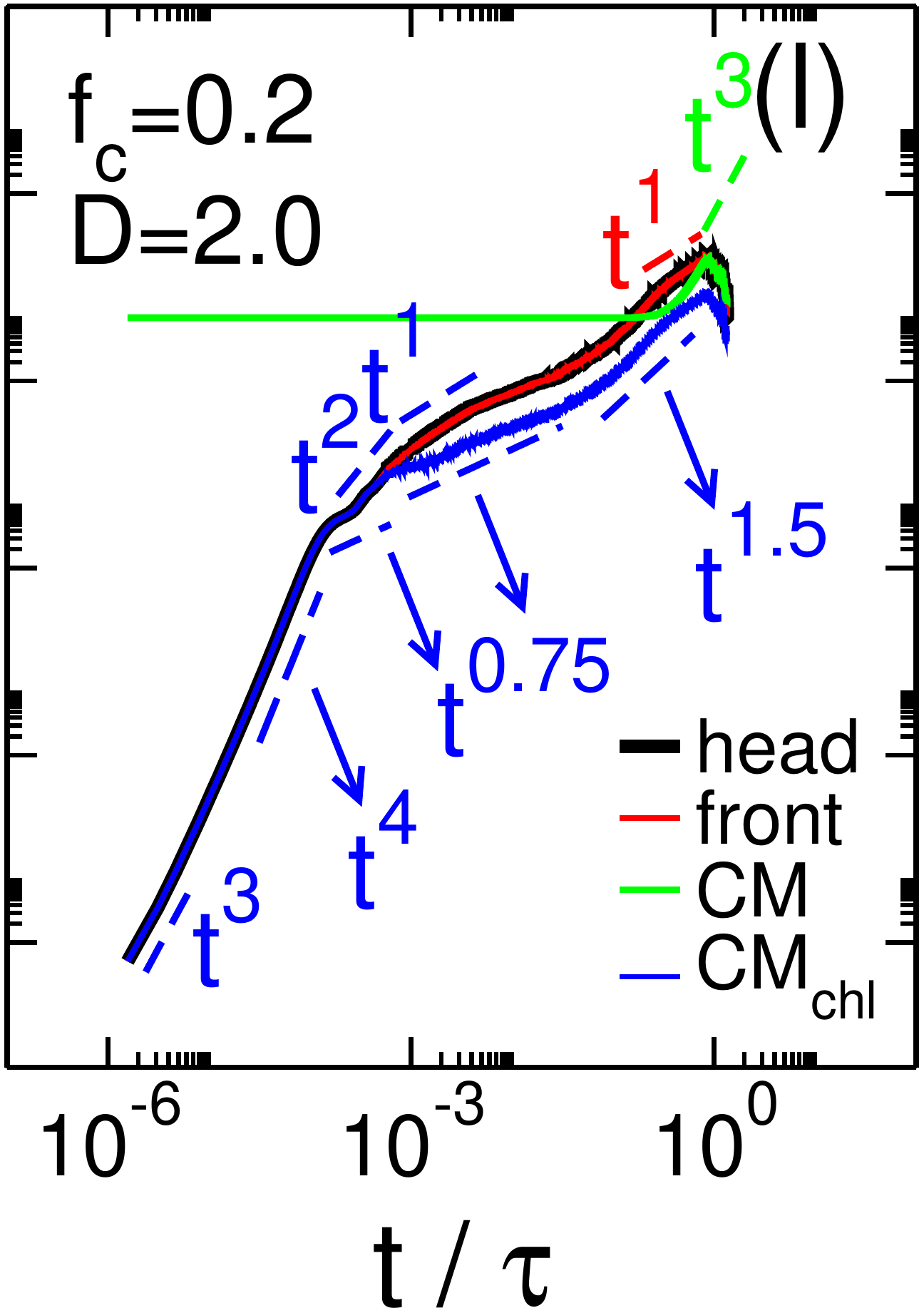}
    \end{center}\end{minipage}
    \begin{minipage}{0.24\textwidth}    
    \begin{center}
        \includegraphics[width=1.0\textwidth]{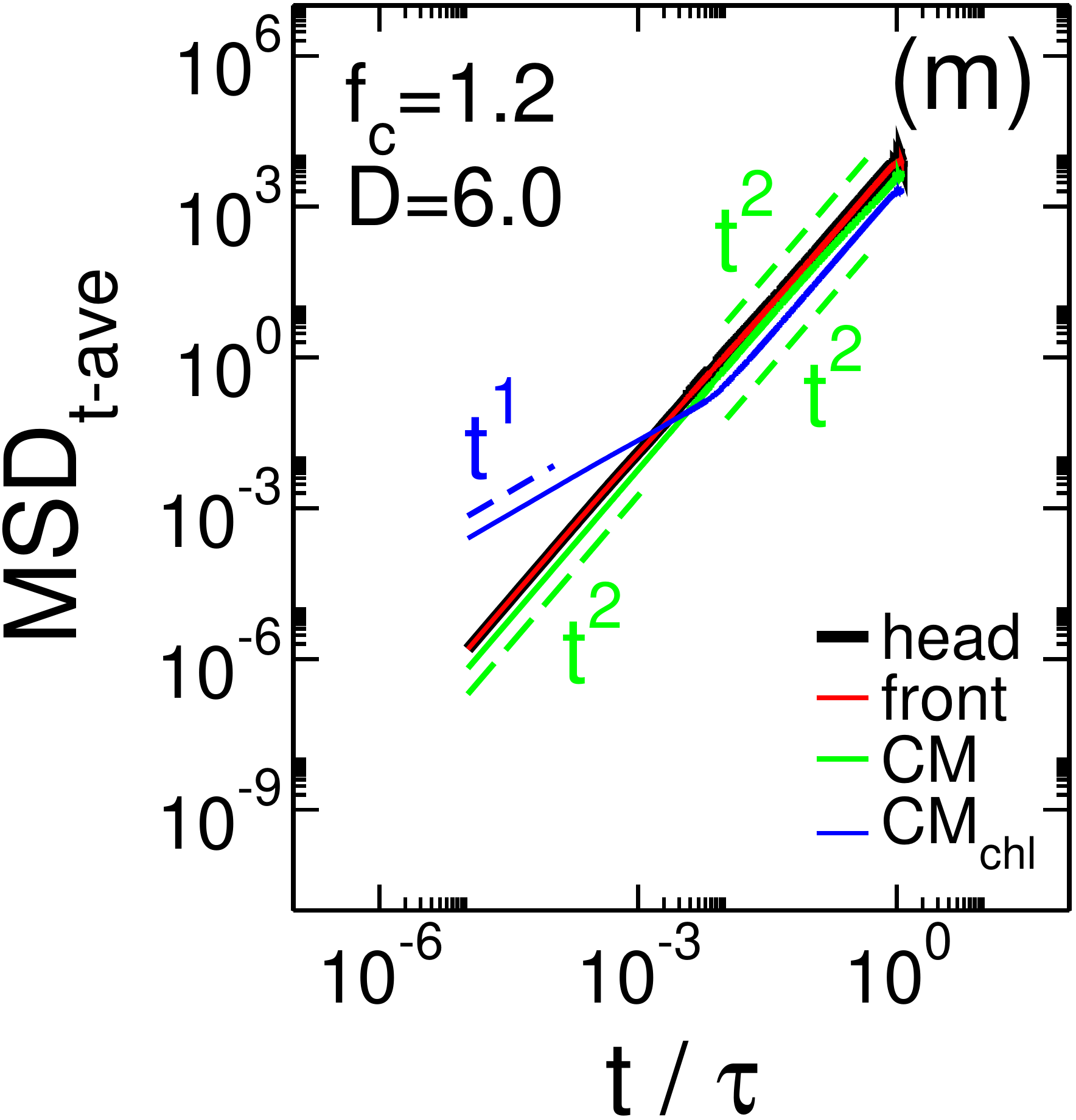}
    \end{center}\end{minipage} \hskip0.0cm
        \begin{minipage}{0.176\textwidth}
    \begin{center}
        \includegraphics[width=1.0\textwidth]{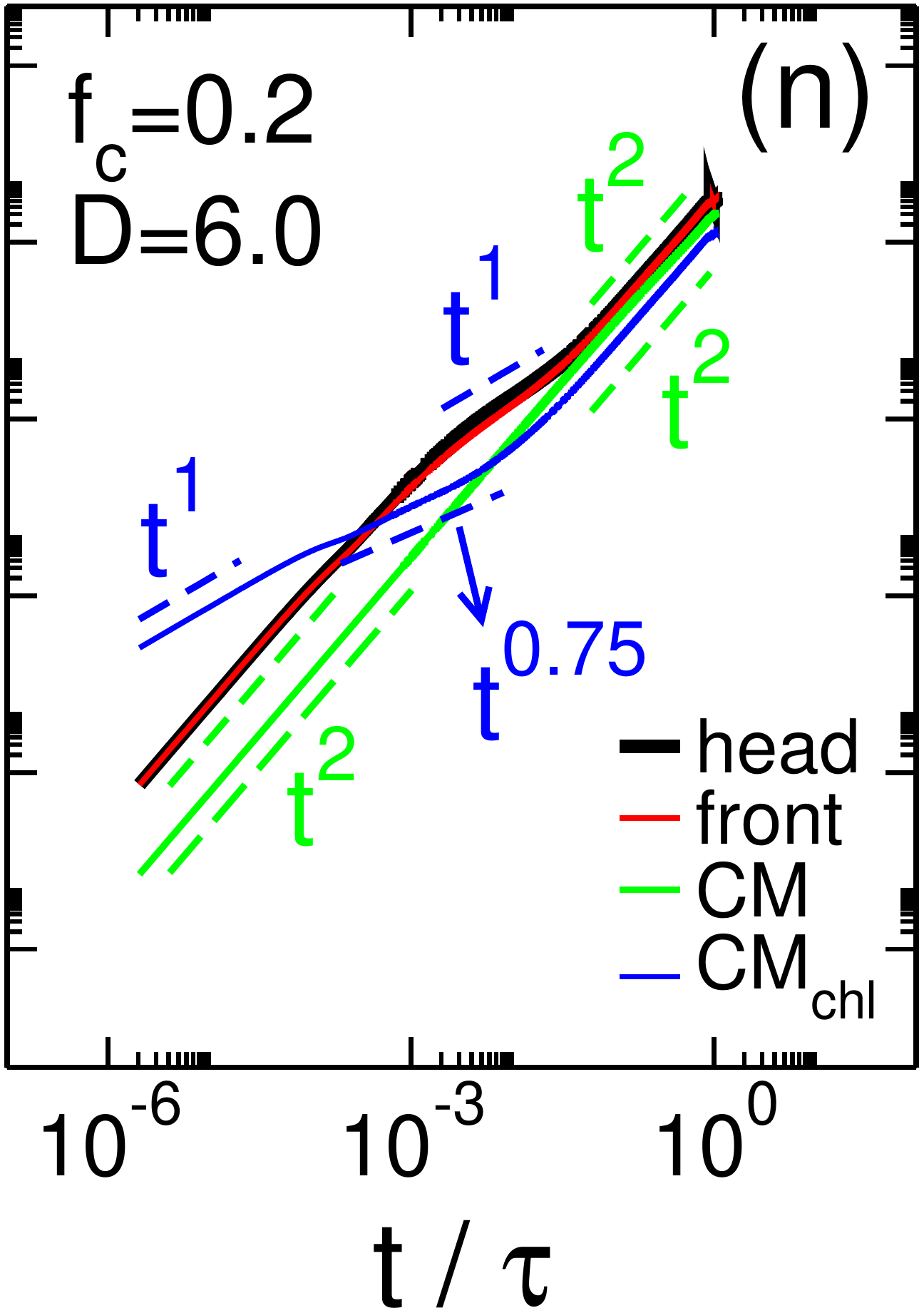}
    \end{center}\end{minipage} \hskip0.0cm
        \begin{minipage}{0.176\textwidth}
    \begin{center}
        \includegraphics[width=1.0\textwidth]{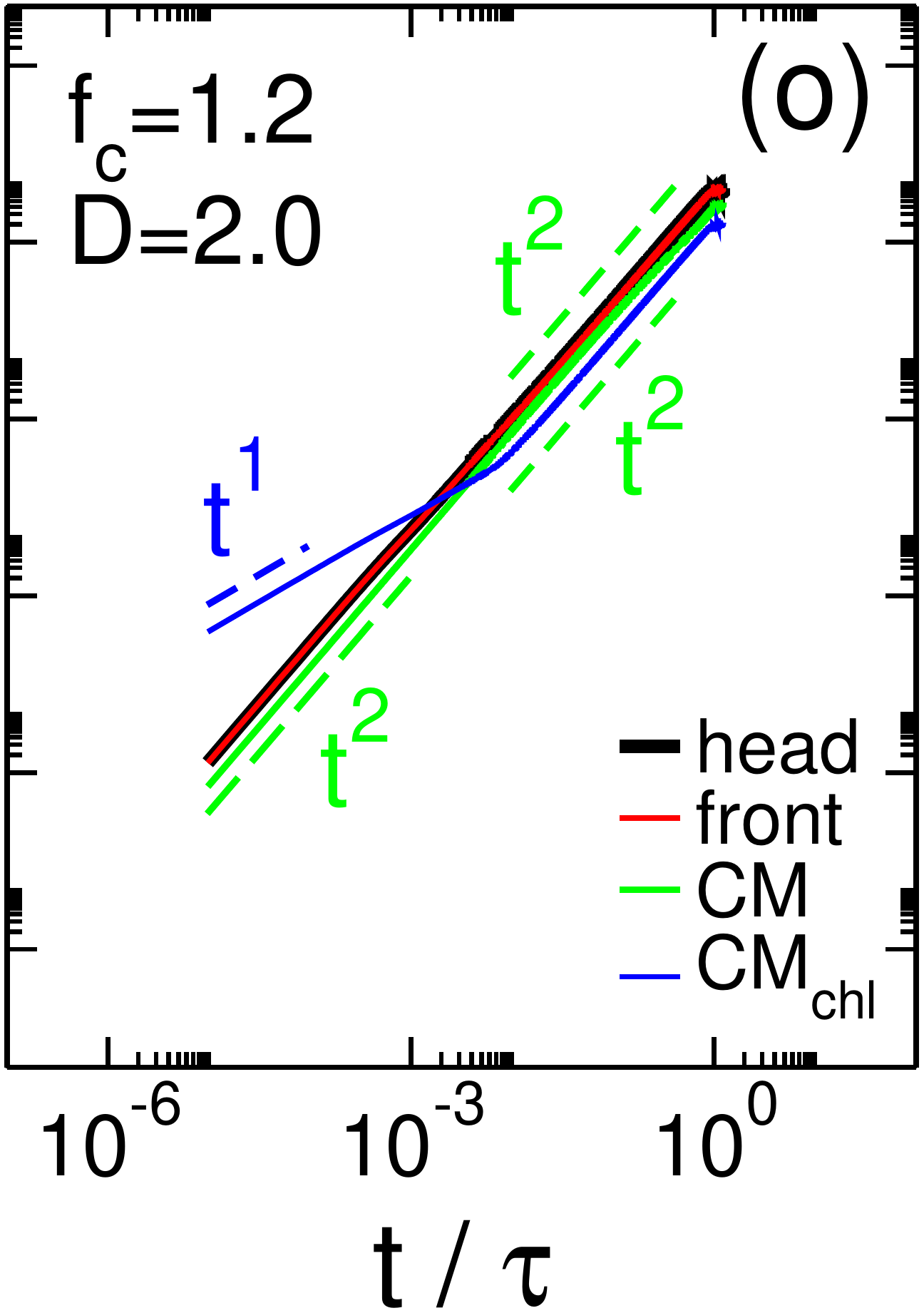}
    \end{center}\end{minipage} \hskip0.0cm
    	\begin{minipage}{0.176\textwidth}
    \begin{center}
        \includegraphics[width=1.0\textwidth]{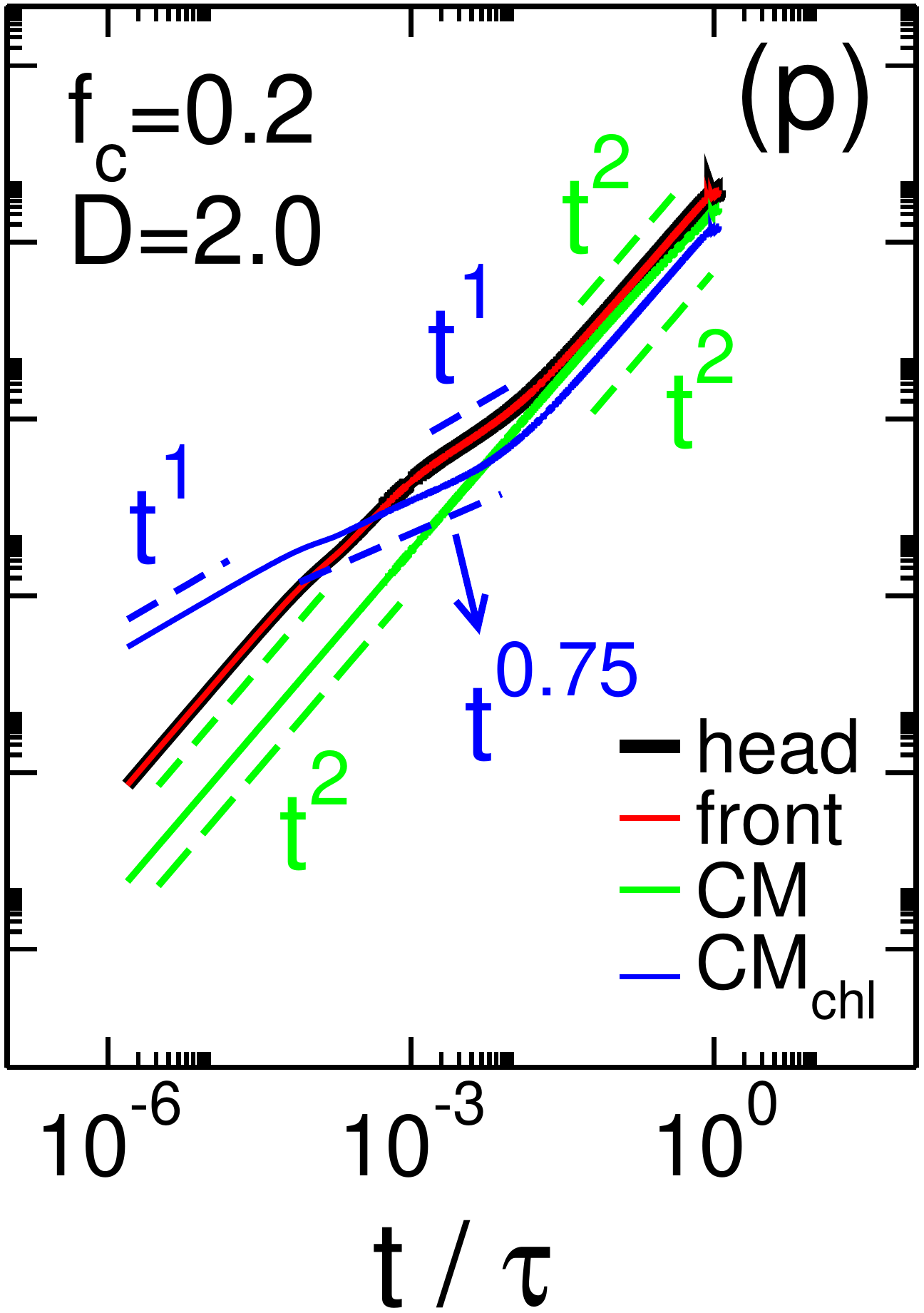}
    \end{center}\end{minipage}
\caption{(a)  {Absolute value of the} horizontal {{ensemble average of the}} position $|\langle x \rangle|$ of the head (black squares), front
(red circles), center of mass of the whole chain (CM, green diamonds) and
center of mass for the {\it trans}-side subchain ($\mathrm{CM}_{\mathrm{
chl}}$ blue triangles), as function of the normalized time $t/\tau$  {in log--log scale} for
fixed values of the channel force $f_{\textrm{c}} =1.2$ and channel diameter
$D=6.0$.  {The inset shows $\langle x \rangle $ as a function of $t/\tau$ 
in linear-linear scale for the same values of the parameters as in the main panel.} 
Panel (b) is the same as panel (a) but for $f_{\textrm{c}}=0.2$.
Panels (c) and (d) are the same as panels (a) and (b), respectively, but
for $D=2.0$. (e) ensemble-averaged MSD as a function of normalized time $t/
\tau$ for head (black line), front (red line), CM (green line), and
$\mathrm{CM}_{\mathrm{chl}}$ (blue line) as function of $t/\tau$ for
$f_{\textrm{c}}=1.2$ and $D=6.0$. Panel (f) is the same as panel (e) but
for $f_{\textrm{c}}=0.2$. Panels (g) and (h) are the same as panels (e)
and (f), respectively, but for $D=2.0$. The dashed lines and the slopes
serve as guide to the eye. Panels (i), (j), (k), and (l) present the
variance $\langle(\delta x)^2\rangle=\langle x^2\rangle-\langle x
\rangle^2$ as function of $t/\tau$. Panels (m), (n), (o), and (p) show
the time-averaged MSD, $\textrm{MSD}_{\textrm{t-ave}}$, as function of
$t/\tau$.  {All slopes in panels (a)--(p) are a guide to the eye.}}
\label{fig:msd}
\end{center}
\end{figure*}

Next we consider the dynamics of the translocation process in more
detail by monitoring different parts of the chain motion as function of
time. To this end the $x$ component of the positions of the head monomer
(the first monomer of the chain in terms of the chemical distance along
the chain), the front monomer (leading edge of the chain in physical
space, i.e., the monomer with the largest $x$ coordinate), the center of
mass of the whole chain (CM), and {the center of mass 
for the {\it trans}-side subchain} ($\textrm{CM}_{\textrm{chl}}$), as well
as the corresponding mean-square displacements (MSDs) are studied.  {The main p}anel
(a) in Fig.~\ref{fig:msd} shows 
{ {the absolute value of the ensemble average of} the $x$ component 
 {of the position ($| \langle x \rangle |$)} of the head monomer
(black squares), front monomer (red circles), CM (green diamonds), and
$\textrm{CM}_{\textrm{chl}}$ (blue triangles) as function of normalized
time $t/\tau$  {in log--log scale} and for fixed
$f_{\textrm{c}}=1.2$ and $D=6.0$. Panel (b) is the same as
panel (a) but for fixed value of $f_{\textrm{c}}=0.2$. Panels (c) and
(d) are the same as panels (a) and (b), respectively, but for fixed
$D=2.0$.
 {As $\langle x \rangle$ has negative values due to the local retracting movement of the monomers to the {\it cis} side at short times, this feature appears in the main panels (a)--(d) as sharp valleys. These fluctuations clearly show that during the translocation process the monomers have back-and-forth movements at the entrance of the channel. Insets of panels (a)--(d) present $\langle x \rangle $ as a function of $t/\tau$ on linear-linear scale for the same values of parameters as in the corresponding main panels,}
%
for the head
monomer (black squares) and front monomer (red circles) coincide with
each other. Moreover, at short times the values of $\textrm{CM}_{
\textrm{chl}}$ (blue triangles) are the same as of the data for the
head and front monomers, however, at long times its behavior is similar
to that of CM.

In Fig.~\ref{fig:msd}(e) the MSD $\langle [x(t)-x(0)]^2 \rangle$ is
shown as function of the normalized time $t/\tau$ with fixed values
$f_{\textrm{c}}=1.2$ and $D=6.0$, for the head monomer (black solid
line), front monomer (red solid line), CM (green solid line), and
$\textrm{CM}_{\textrm{chl}}$ (blue solid line). Panel (f) is the same as
panel (e) but for fixed $f_{\textrm{c}}=0.2$. Panels (g) and
(h) are the same as panels (e) and (f), respectively, but for fixed
$D=2.0$. As seen in panels (e)--(h) the behavior of the MSD for the head,
front and $\textrm{CM}_{\textrm{chl}}$ are similar to each other in almost
all time regimes. At the very short times their local slope is 4, followed
by 0.75 in the short time regime, then 2 and 1.5 at intermediate times,
and finally the slope is 2.5 at long times.
In contrast, the MSD for CM is different from the other MSDs in different
time regimes. Indeed, the MSD slopes for CM are 2, 1, and 4 at very short,
intermediate, and long time regimes, respectively.

Panel (i) shows the variance of the $x$ component $\langle(\delta x)^2
\rangle=\langle x^2\rangle-\langle x\rangle^2$ as function of
normalized time $t/\tau$, with fixed $f_{\textrm{c}}=1.2$ and $D=6.0$,
for the head monomer (black solid line), front monomer (red solid line),
CM (green solid line) and $\textrm{CM}_{\textrm{chl}}$ (blue solid line).
Panel (j) is the same as panel (i) but for fixed $f_{\textrm{c}}=0.2$.
Panels (k) and (l) are the same as panels (i) and (j), respectively, but
for fixed $D=2.0$. The behavior of the variance for the head, front, and
$\textrm{CM}_{\textrm{chl}}$ in panels (e)--(h) are again similar to each
other. The slopes of the variance are 3, 4, 0.75, 2, and 1.5 at very
short, short, intermediate, and long times, respectively. Conversely,
for CM (green line) the slope is zero and becomes 3 at long times. 
 {All slopes in panels (a)-(p) in Fig.~\ref{fig:msd} 
are a guide to the eye.}

\begin{figure*}
\begin{center}
        \begin{minipage}{1.0\textwidth}
    \begin{center}
        \includegraphics[width=0.45\textwidth]{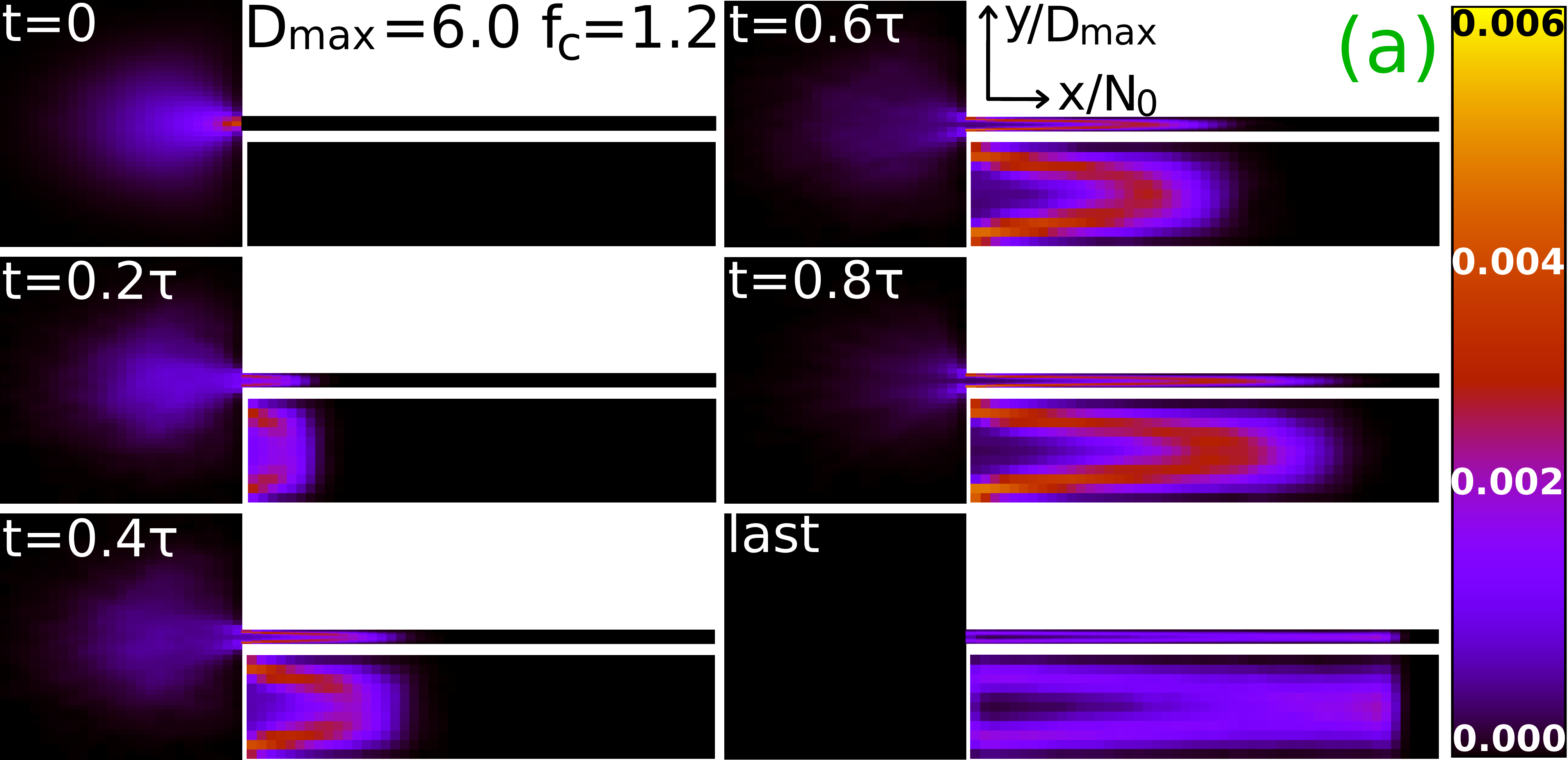}
        \hspace{+0.3cm}
        \includegraphics[width=0.45\textwidth]{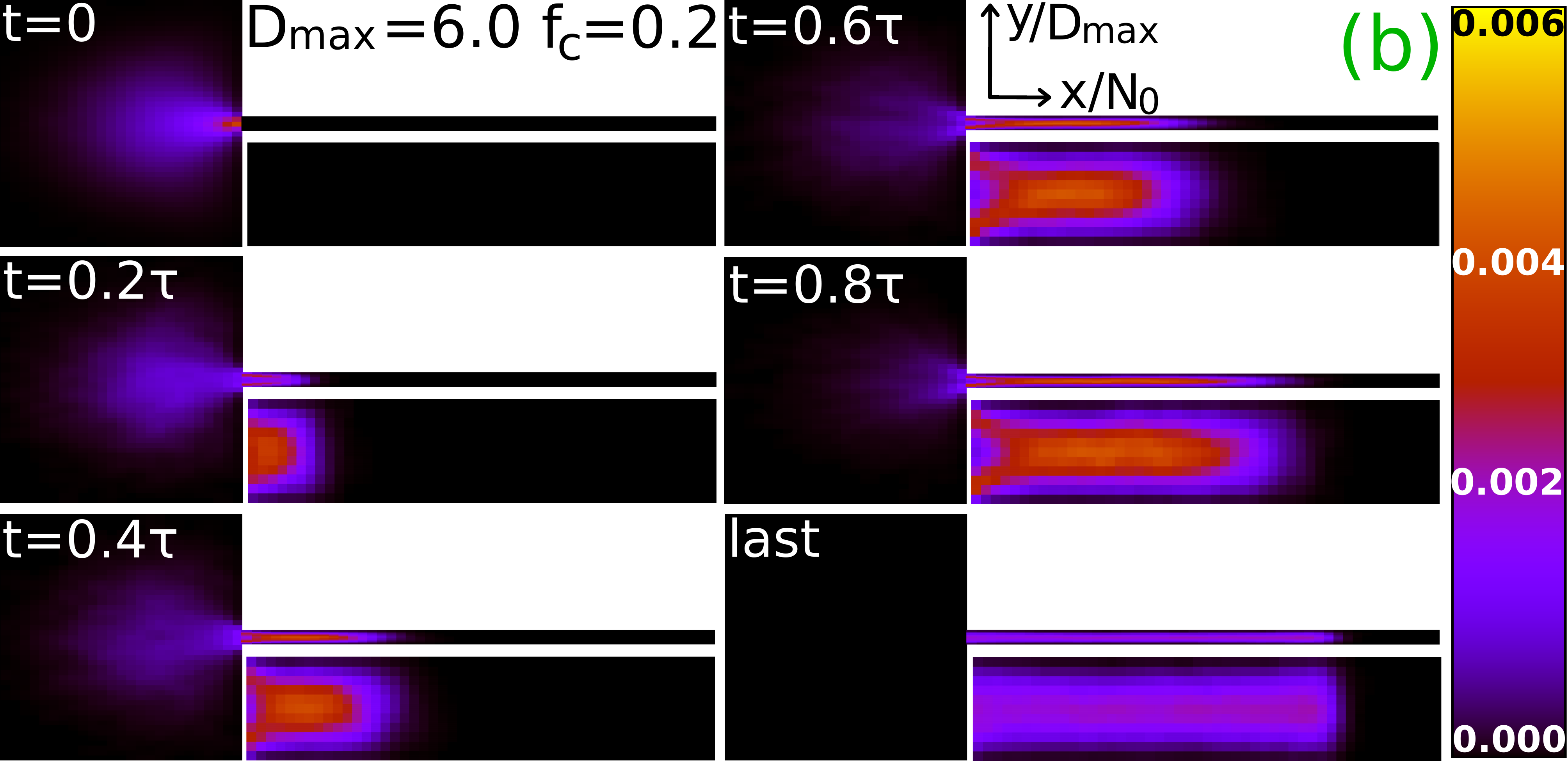}
    \end{center}\end{minipage}  
    \begin{minipage}{1.0\textwidth}
    \begin{center}
    	\vspace{+0.5cm}
        \includegraphics[width=0.45\textwidth]{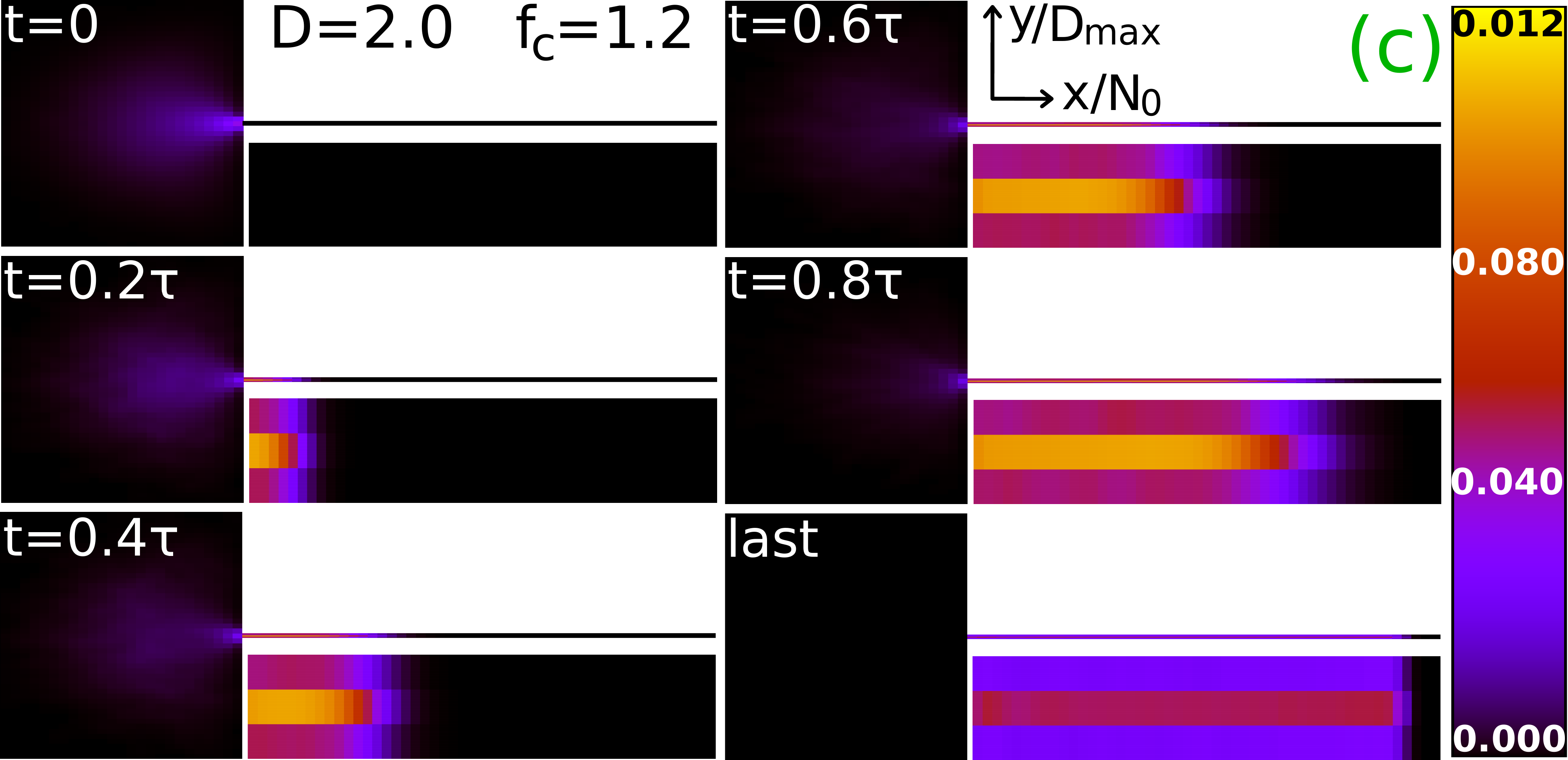}
        \hspace{+0.3cm}
        \includegraphics[width=0.45\textwidth]{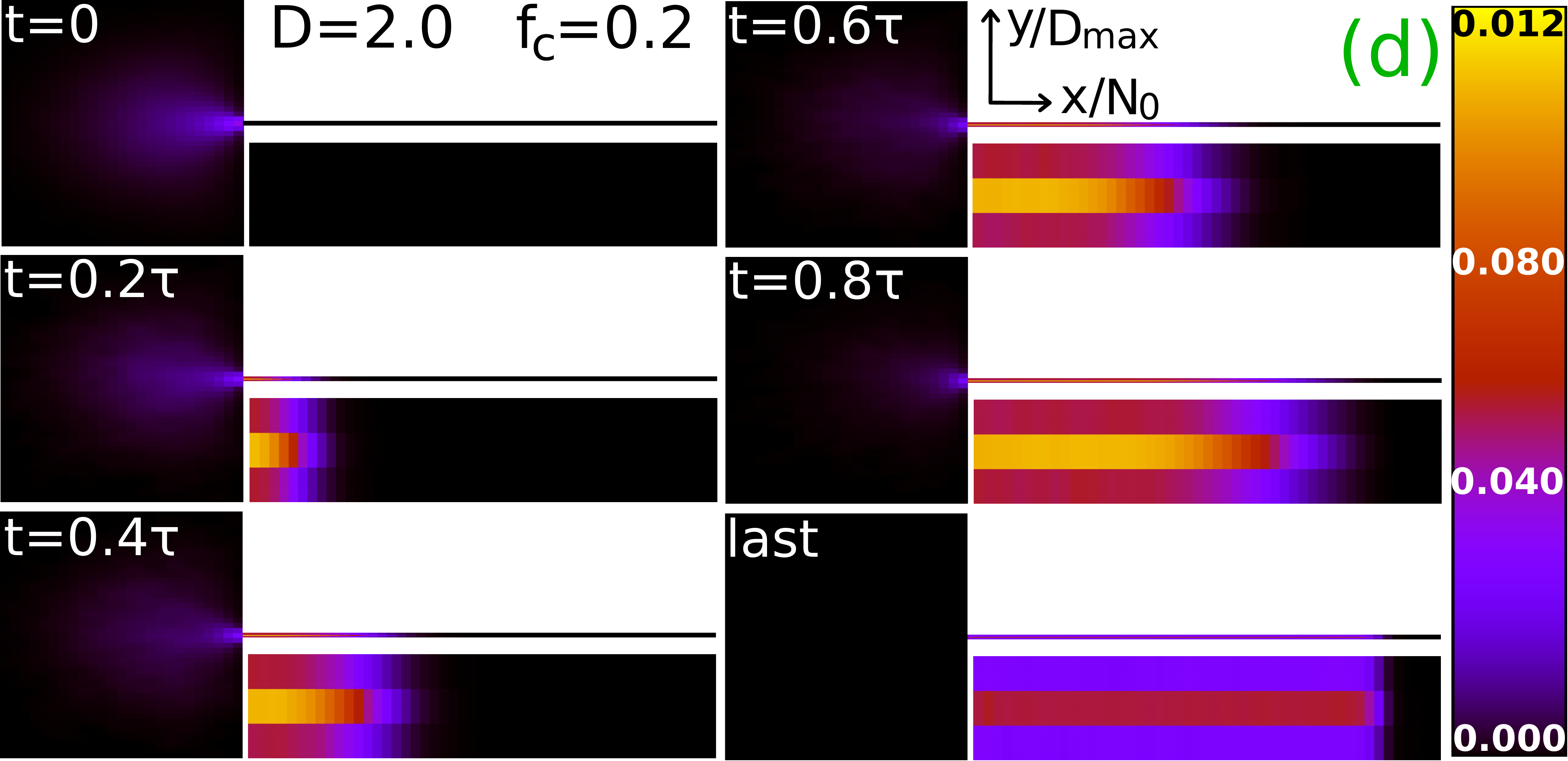}
    \end{center}\end{minipage} 
\caption{(a) 2D monomer density of the chain on the {\it cis} as well as
{\it trans} side at different times $t/\tau=0$, 0.2, 0.4, 0.6, and 0.8,
as well as the final configurations of the chain, for fixed values  {of the channel width} $D=6.0$ 
and $f_{\textrm{c}}=1.2$. Inside the channel the horizontal (parallel to
the channel axis) and vertical distances are normalized by $N_0=100$ and
$D_{\textrm{max}}=6.0$  {(where $D_{\textrm{max}}$ is the maximum channel width)}, respectively. The monomer density of the {\it
trans} side (channel) is also shown in a separate, magnified panel below
the actual, much narrower channel for a better visibility. Panel (b) is
the same as panel (a) but for the weakest channel driving force in our
LD simulations ($f_{\textrm{c}}=0.2$). Panels (c) and (d) are the same
as panels (a) and (b), respectively, but for the narrowest channel in
the LD simulations ($D=2.0$).}
\label{fig:density}
\end{center}
\end{figure*}

Finally, we consider the time-averaged MSD $\textrm{MSD}_{\textrm{
t-ave}}$ computed as a running average over single trajectories
\cite{pt,pccp}. Sampling positions as the time series $x(0),x(\Delta
t),x(2\Delta t),x(3\Delta t),\ldots x((n-1)\Delta t)$ with time
step $\Delta t$, $\textrm{MSD}_{\textrm{t-ave}}$ at ("lag") time $j
\Delta t$ is defined as
\begin{equation}
\textrm{MSD}_{\textrm{t-ave}}(j \Delta t)=\frac{1}{(n-j)}\sum_{i=0}^{
n-1-j}\big[x\big((i+j)\Delta t\big)-x\big(i\Delta t\big)\big]^2,
\label{MSD_t_ave}
\end{equation}
where $j=1,2,3,\ldots,n-1$. In equilibrium systems\footnote{Up to a
factor of 2 occurring naturally due to the definition of $\textrm{MSD}_{
\textrm{t-ave}}$ \cite{pre12}.} and systems with stationary increments
$\textrm{MSD}_{\textrm{t-ave}}$ converges to the ensemble-averaged MSD,
$\textrm{MSD}=\left<\big[x(t)-x(0)\big]^2\right>$. Given small numbers
of independent trajectories, the statistics of $\textrm{MSD}_{\textrm{
t-ave}}$ for sufficiently long time series is often better than the
ensemble-averaged MSD. For finite trajectories the time-averaged MSD
fluctuates from one trajectory to the next, as often measured in terms
of the ergodicity breaking parameter \cite{yhe}. In "weakly non-ergodic"
systems \cite{bouchaud,yhe,pccp} the time-averaged MSD may differ from
the ensemble-averaged MSD even in the limit of long measurement times.

To check the above argument, in panel (m) $\textrm{MSD}_{\textrm{t-ave}}$
is shown as function of normalized time $t/\tau$ with fixed $f_{\textrm{c}}
=1.2$ and $D=6.0$, for the head monomer (black solid line), front monomer
(red solid line), CM (green solid line), and $\textrm{CM}_{\textrm{chl}}$
(blue solid line). Panel (n) is the same as panel (m) but for fixed value
$f_{\textrm{c}}=0.2$. Panels (o) and (p) are the same as panels (m) and (n),
respectively, but for fixed value of $D=2.0$.

As panels (m)--(p) clearly show in the long time limit the data on all curves
scale like $t^2$. At very short times for $\textrm{CM}_{\textrm{chl}}$
the time-averaged MSD scales as $\textrm{MSD}_{\textrm{t-ave}}\sim t^1$,
while it scales as $t^2$ for the head, front, and CM. Moreover, in panels
(n) and (p) in the weak force limit ($f_{\textrm{c}}=0.2$) another
intermediate scaling regime appears in which the time-averaged MSD for
$\textrm{CM}_{\textrm{chl}}$ scales as $t^{0.75}$. The time-averaged
MSD analysis may thus provide additional information on the details
of translocation processes, however, a more detailed study
of the relation between ensemble and time-averaged MSDs will be the
topic of future work.

\subsection{Monomer density}
\label{density}

Finally, we study the time evolution of the monomer density. It reveals the average spatial configuration
of the polymer during the translocation process. In Fig.~\ref{fig:density}(a)
the monomer density is shown for the maximum channel width $D_{\mathrm{max}}=6$
and channel driving force $f_{\textrm{c}}=1.2$ at the times $t=0$, $0.2\tau$,
$0.4\tau$, $0.6\tau$, $0.8\tau$ during the translocation process, and for
the final configuration. Averaging has been performed over $1000$ independent
trajectories. For the sake of better visibility a blow-up of the channel is
added underneath the actual narrow channel. Inside the channel the size
of the cells in the $y$ and $x$ directions is normalized by $D_{\textrm{max}}$
and $N_0$, respectively. At time zero the whole chain is on the {\it cis}
side, and as the first bead is fixed at the entrance of the channel the
polymer possesses an average pear-like shape. Over time more monomers traverse
the channel entrance. At all moments of the translocation process
the density of monomers around the corners of the channel entrance is higher
than that along the channel axis due to the high driving force. At constant
channel with $D_{\textrm{max}}=6$ when the channel driving force decreases
to $f_{\textrm{c}}=0.2$ (panel (b)), the monomer density is more pronounced
along the channel axis.

Panels (c) and (d) are the same as panels (a) and (b), respectively, but for
channel width $D=2$. Comparing panels (c) and (d) reveals that by decreasing
the channel width to $D=2$, the narrowest one in the current study, as the
monomers inside the channel do not have enough space in the $y$ direction
to perform significant spatial fluctuations, one finds them sharply around
the channel axis, regardless of the value of the channel driving force.
Moreover, it should be noted that the value of the monomer density in panels
(c) and (d) is twice that in panels (a) and (b) (see the color bar).

\section{Summary and conclusions}
\label{conclusion}

In this work we have combined IFTP theory with extensive LD simulations to unravel
the dynamics of polymer
translocation into a long channel. The axial driving
force $f_{\textrm{c}}$ in this "chain sucker" scenario \cite{RalfJCP2011}
acts on all monomers inside the channel, away from the pore (entry point
of the channel), and the monomers inside the channel thus pull the other
monomers on the {\it cis} side along. When the channel width $D$ is wide
enough to allow the polymer to fluctuate in the channel, an entropic force due to the
spatial fluctuations of the {\it trans}-side subchain starts to play a
role in the translocation dynamics when its magnitude becomes
comparable to that of the driving force. Our analysis demonstrates
how the {\it trans}-side entropic force depends on the values of
both $D$ and $f_{\textrm{c}}$, as the spatial configuration of the {\it
trans}-side subchain is determined by both. 

To further verify this, in
Fig.~\ref{fig:Rex}(a) the horizontal size of the last configuration of the
chain inside the channel $R_{\parallel}$ is shown as function of $D$ for
the channel driving force $f_{\textrm{c}}=1.2$ (turquoise open circles),
0.5 (orange open squares), 0.3 (green open diamonds), and 0.2 (pink open
triangles). The black solid curve is shown as a guide to the eye and depicts
the scaling exponent for the equilibrium configuration of the chain inside a
channel with width $D$. From blob theory the equilibrium axial chain size
exponent is $-1/3$ in the absence of any channel driving
force $R_{\parallel}\propto D^{-1/3}$ \cite{DeGennesBook}.  As the driving
force increases the axial chain-size exponent deviates from its equilibrium
value of $-1/3$. Moreover panel (b) shows that at constant $D$, the value of
$R_{\parallel}$ also depends on $f_{\textrm{c}}$. Therefore, panels (a) and
(b) clearly confirm that $R_{\parallel}$ depends on both $f_{\textrm{c}}$ and
$D$, and it is hard to separate the effect of their individual contributions
to $R_{\parallel}$. In particular, the contributions of $f_{\textrm{c}}$
and $D$ to the entropic force are coupled to each other. 

The only case in
the current study, wherein the {\it trans}-side entropy does not play any
significant role in the translocation dynamics is for the narrow channel width
$D=2$, and the corresponding LD result for the force exponent is $\beta=-1$
(orange dashed line in Fig.~\ref{fig:tau}(a)), as predicted by IFTP theory
wherein the contribution of the entropic force is not taken into account,
see Eq.~(\ref{tau}). As the value of $D$ increases, the {\it trans}-side
subchain possesses more spatial configurations, and entropic force effects on
the dynamics of translocation process are more pronounced, and consequently
$\beta$ deviates from $-1$ (see inset of Fig.~\ref{fig:tau}(a)).

Our study has unveiled a rich behavior of the observed translocation dynamics
as function of the system parameters of channel driven translocation.
Such scenarios are important for experimental systems such as long micro- 
or nano-channels that confine the DNA and apply flows along the channel 
\cite{Tegenfeldt_ABC2004,Tegenfeldt_PNAS2004,Tegenfeldt_PNAS2005,%
Tegenfeldt_book,Tegenfeldt_CSR2010}.

\section{Acknowledgments}
Computational resources from the Center for Scientific Computing Ltd. and 
from the Aalto University School of Science “Science-IT” project are
gratefully acknowledged. T.A-N. was supported in part by the Academy of
Finland through its PolyDyna (Grant No. 307806) and QFT Center of
Excellence Program grants (Grant No. 312298). R.M. acknowledges the
German Research Foundation (DFG, grant ME 1535/12-1) and the Foundation
for Polish Science (FNP, Humboldt Polish Honorary Research Scholarship)
for support.

\begin{figure}[b]\begin{center}
    \begin{minipage}{0.49\columnwidth}
    \begin{center}
        \includegraphics[width=1.067\columnwidth]{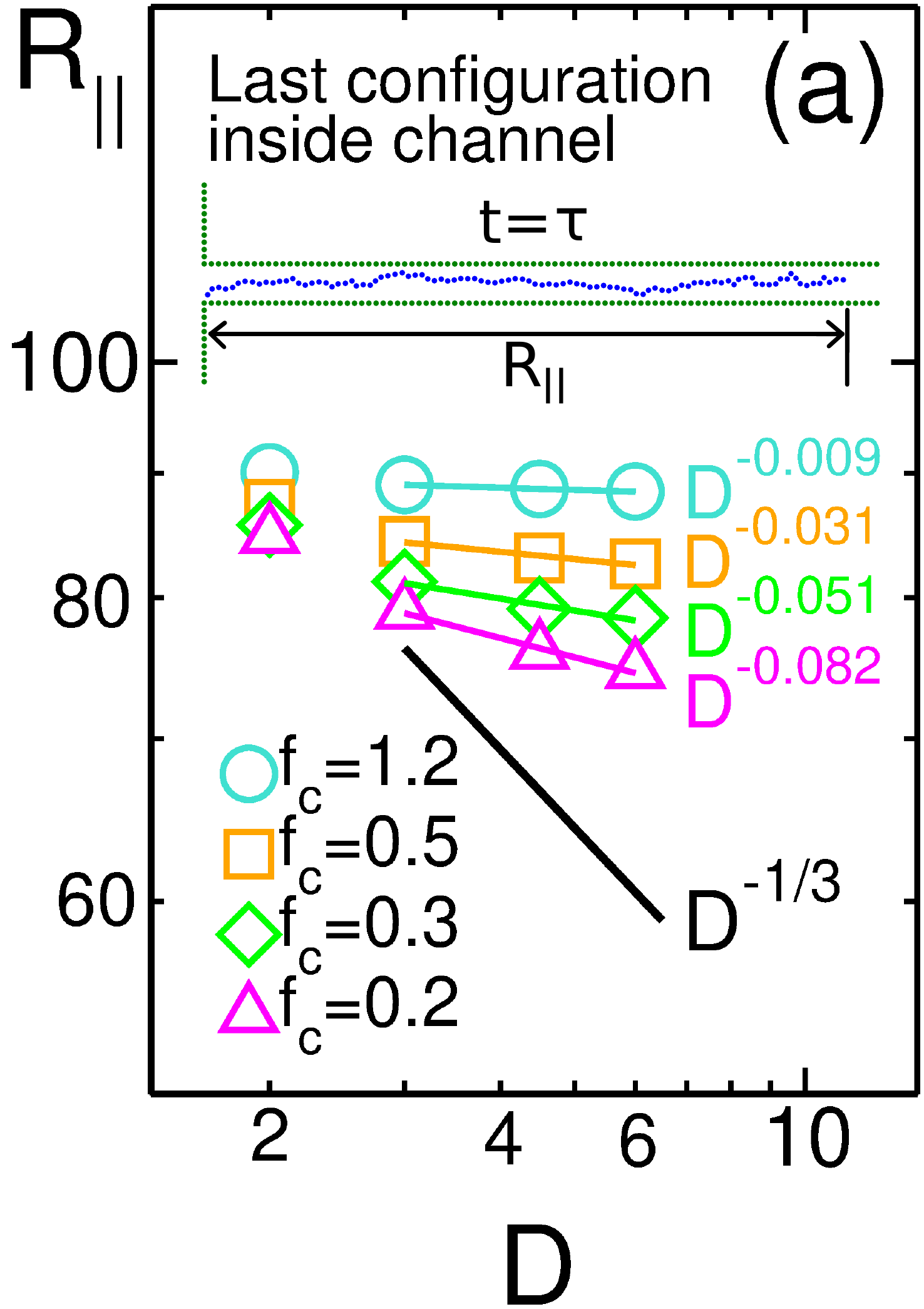}
    \end{center}\end{minipage}  \hskip-0.0cm
    \begin{minipage}{0.49\columnwidth}
    \begin{center}
        \includegraphics[width=0.9\columnwidth]{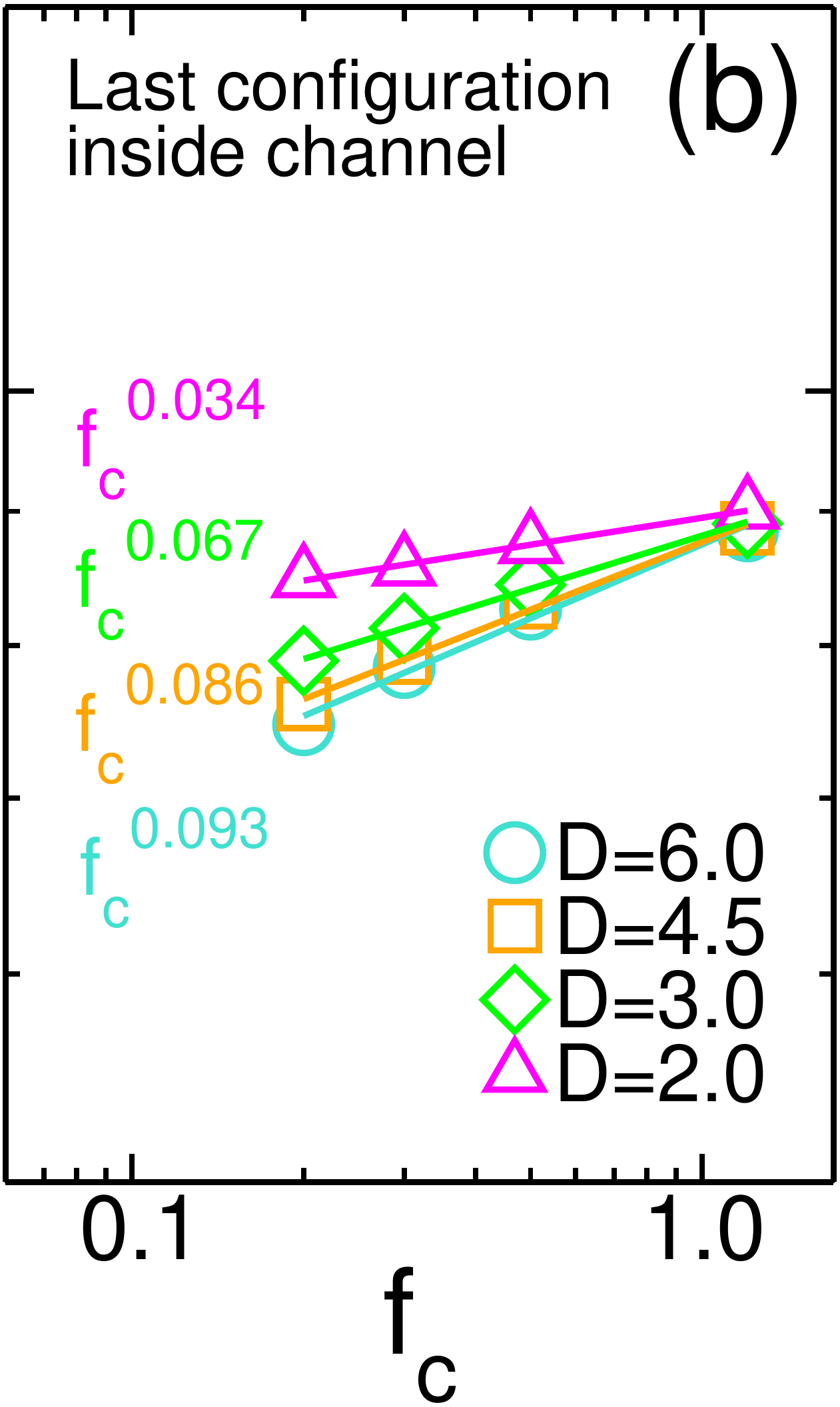}
    \end{center}\end{minipage}
\caption{(a) Horizontal size $R_{\parallel}$ of the final chain configuration
at the end of the translocation process, as function of the channel width
$D$, for the channel driving forces $f_{\textrm{c}}=1.2$ (turquoise open
circles), 0.5 ( {o}range open squares), 0.3 (green open diamonds), and 0.2
(pink open triangles). As can be seen all fitted slopes are far from the
equilibrium behavior (for $f_{\textrm{c}}=0$) with size exponent $-1/3$
(black solid curve as guide to the eye). (b) $R_{\parallel}$ as function
of $f_{\textrm{c}}$ for channel widths $D=6$ (turquoise open circles), 4.5
(Orange open squares), 3 (green open diamonds) and 2 (pink open triangles),
with the corresponding values of the exponents 0.034, 0.067, 0.86 and
0.093, respectively.}
\label{fig:Rex}
\end{center}
\end{figure}

\appendix

\section{Bond length distribution}
\label{blength}

One of the quantities that shows the dynamics of the translocation process at
the monomer level is the bond length distribution. In Fig.~\ref{fig:bond}
the normalized bond length $l_{\textrm{b}}/l_{\textrm{b,equil}}$ is shown
as function of the bond index $\textrm{b}$, for constant
channel driving force $f_{\textrm{c}} = 1.2$ and different channel
widths $D=6.0$ (solid filled symbols), 4.5 (vertical hashed filled), 3.0
(horizontal hashed filled), and 2.0 (cross hashed filled), at the
times $t/\tau=0.2$ (black circles), 0.4 (red squares), 0.6 (green
diamonds), 0.7 (blue triangles), and 0.8 (orange upside triangles). Here
$l_{\textrm{b,equil}} \approx 0.96$ is the average value of the bond
length at equilibrium, and $\tau$ is the average translocation time for the
corresponding set of parameters. Panels (b), (c) and (d) are the same as panel
(a) but for channel driving forces $f_{\textrm{c}}=0.5$, $0.3$, and $0.2$,
respectively. As can be seen, at the beginning
of the translocation process fewer bonds are affected by the tension force
as compared to later times. At the TP
time of about $t_{\textrm{TP}}=0.7\tau$, all bonds in the polymer
on average are stretched by the tension force. This is clearly shown by the
blue curves in all panels of Fig.~\ref{fig:bond}. As the channel driving
force gets weaker (from panel (a) to (d)) the bonds are less stretched by
the tension force. This is another confirmation for the existence of TP
along the backbone of the chain. Each empty circle approximately shows the
bond index inside the entrance of the channel at the corresponding moment.

\begin{figure*}[t]\begin{center}
    \begin{minipage}{0.284\textwidth}
    \begin{center}
        \includegraphics[width=1.0\textwidth]{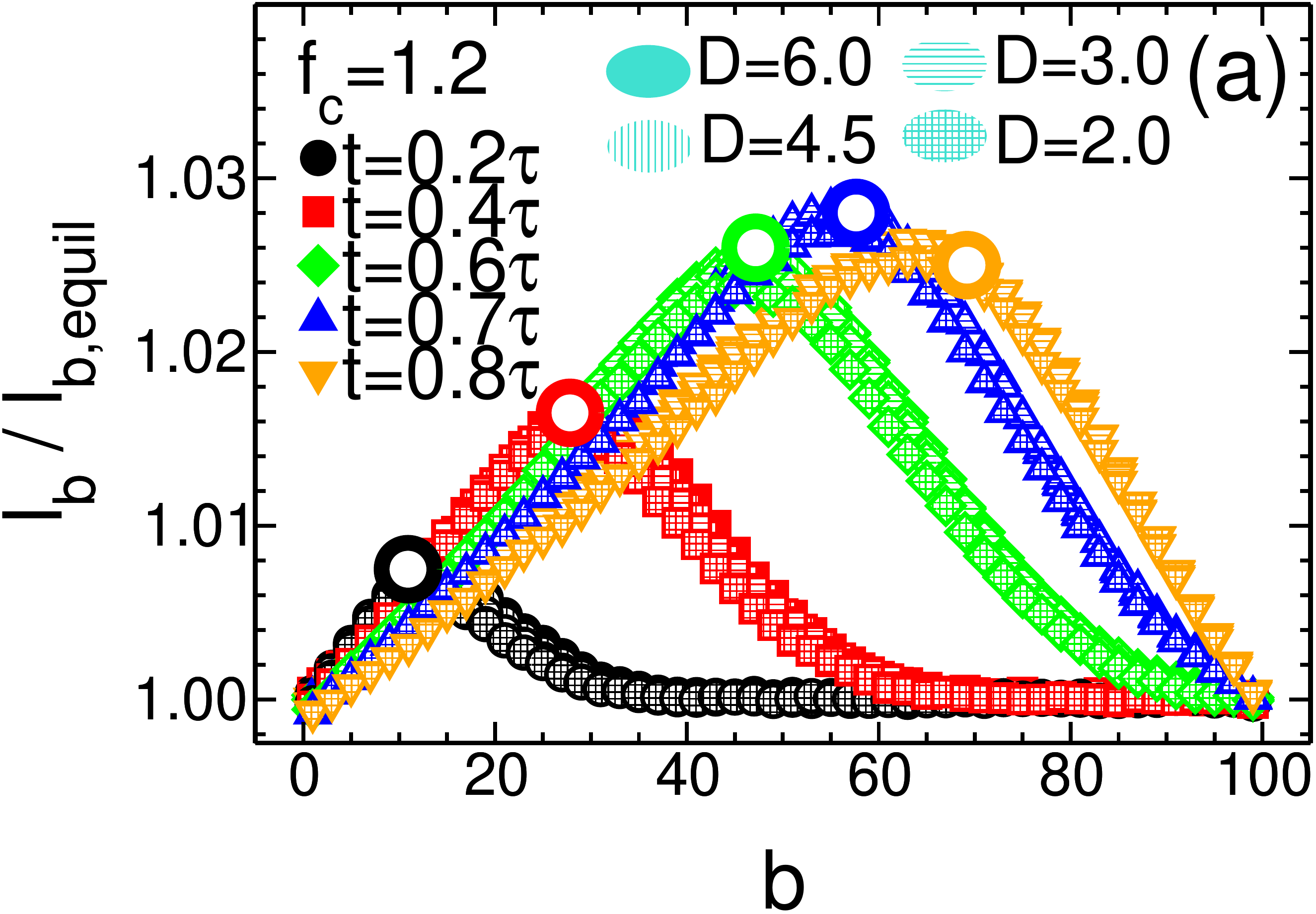}
    \end{center}\end{minipage} \hskip-0.1cm
        \begin{minipage}{0.255\textwidth}
    \begin{center}
        \includegraphics[width=1.0\textwidth]{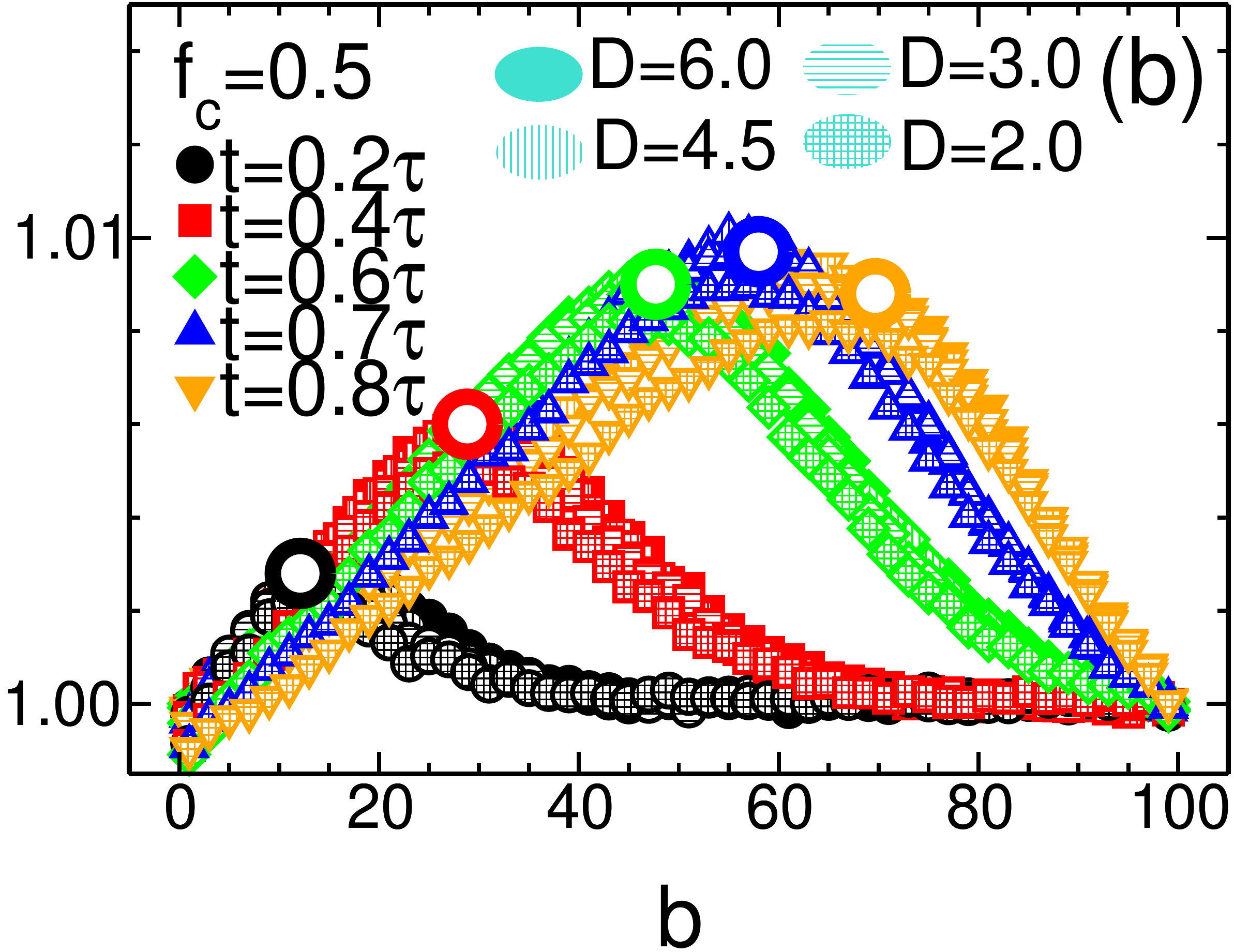}
    \end{center}\end{minipage} \hskip-0.1cm
        \begin{minipage}{0.229\textwidth}
    \begin{center}
        \includegraphics[width=1.0\textwidth]{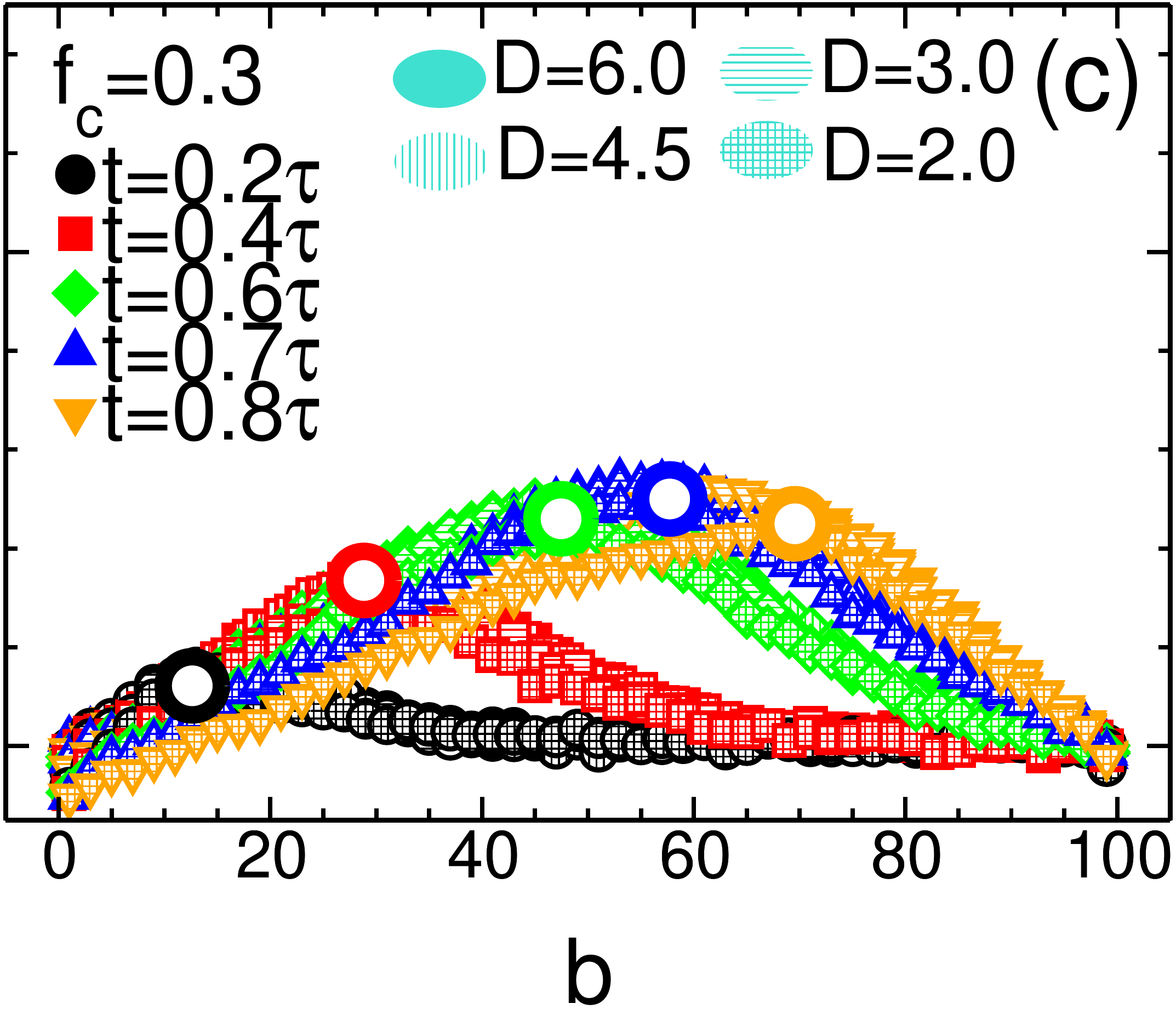}
    \end{center}\end{minipage} \hskip-0.1cm
    	\begin{minipage}{0.229\textwidth}
    \begin{center}
        \includegraphics[width=1.0\textwidth]{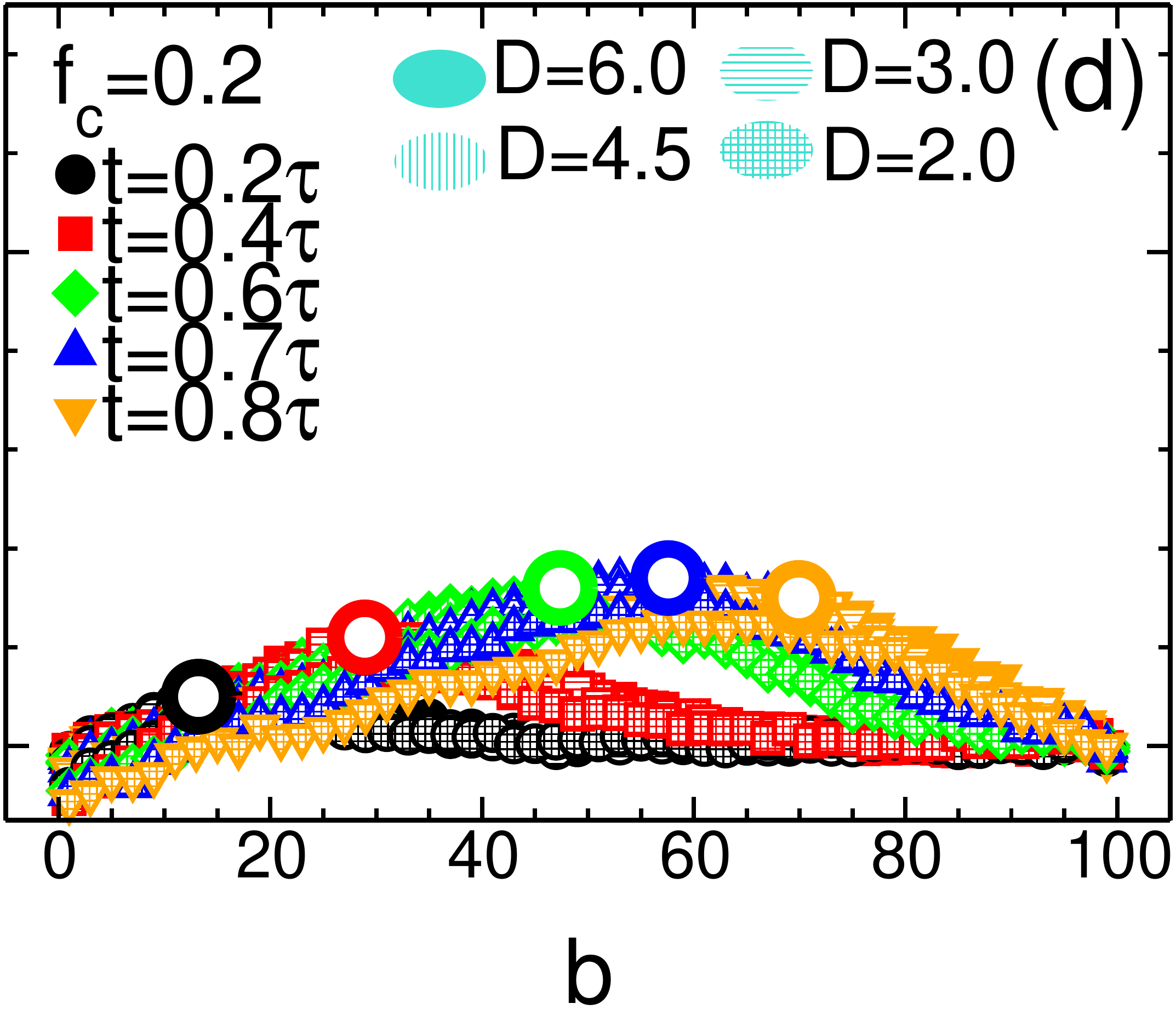}
    \end{center}\end{minipage}
\caption{(a) Normalized bond length $l_{\textrm{b}}/l_{\textrm{b,equil}}$
as function of the bond index $b$ for fixed channel driving force $f_{
\textrm{c}}=1.2$ and channel widths $D=6.0$ (solid filled symbols), 4.5
(vertical hashed filled), 3.0 (horizontal hashed filled), and 2.0 (cross
hashed filled), at times $t/\tau=0.2$ (black circles), 0.4 (red squares),
0.6 (green diamonds), 0.7 (blue triangles) and 0.8 (orange upside triangles).
Panels (b), (c) and (d) are the same as panel (a), but for $f_{\textrm{c}}
=0.5$, $0.3$, and $0.2$, respectively. Empty circles approximately show
the bond index inside the entrance of the channel at the corresponding
times. $l_{\textrm{b,equil}}\approx 0.96$ is the equilibrium average
value of the bond length.}
\label{fig:bond}
\end{center}
\end{figure*}

\section{Velocity  {distributions for monomers}}
\label{velocity}

We consider the monomer velocity distribution here. Its $x$ component
is obtained from averaging over $1000$ uncorrelated
trajectories for each set of parameters. In Fig.~\ref{fig:velocity}(a)
the normalized $x$ component of the monomer velocity $v_x/f_{\textrm{c}}$
is shown as function of the monomer index $m$, for fixed value of
the channel driving force $f_{\textrm{c}}=1.2$ and channel widths
$D=6.0$ (solid filled symbols), 4.5 (vertical hashed filled),
3.0 (horizontal hashed filled), and 2.0 (cross hashed filled),
at the different times $t/\tau = 0.2$ (black circles),
0.4 (red squares), 0.6 (green diamonds), 0.7 (blue triangles), and 0.8
(orange upside triangles). Panels (b), (c) and (d) are the same as panel
(a) but for channel driving forces $f_{\textrm{c}}=0.5$, 0.3, and 0.2,
respectively. To assess how the driving force affects the monomer
velocity distribution, the normalized velocity has been plotted in
all panels. As can be seen the behavior of the curves at any given moment
is similar for different values of the driving force. As
the value of $f_{\textrm{c}}$ decreases (from panel (a) to panel (d)) the
fluctuations in the curves increase originating from two facts. The
first one is that by decreasing the force the value of the normalized velocity
increases (note that $f_{\textrm{c}}$ is in the denominator of $v_x /
f_{\textrm{c}}$) and this increases the fluctuations of the
normalized velocity. The second reason is that by decreasing the driving
force, the tension force along the backbone of the chain becomes weaker,
giving rise to more pronounced spatial fluctuations of the chain and
consequently higher fluctuations of the monomer velocity. To have the same
amount of monomer velocity fluctuations the average must be taken over more
trajectories as the channel driving force decreases.

\begin{figure*}[t]\begin{center}
    \begin{minipage}{0.2893\textwidth}
    \begin{center}
        \includegraphics[width=1.0\textwidth]{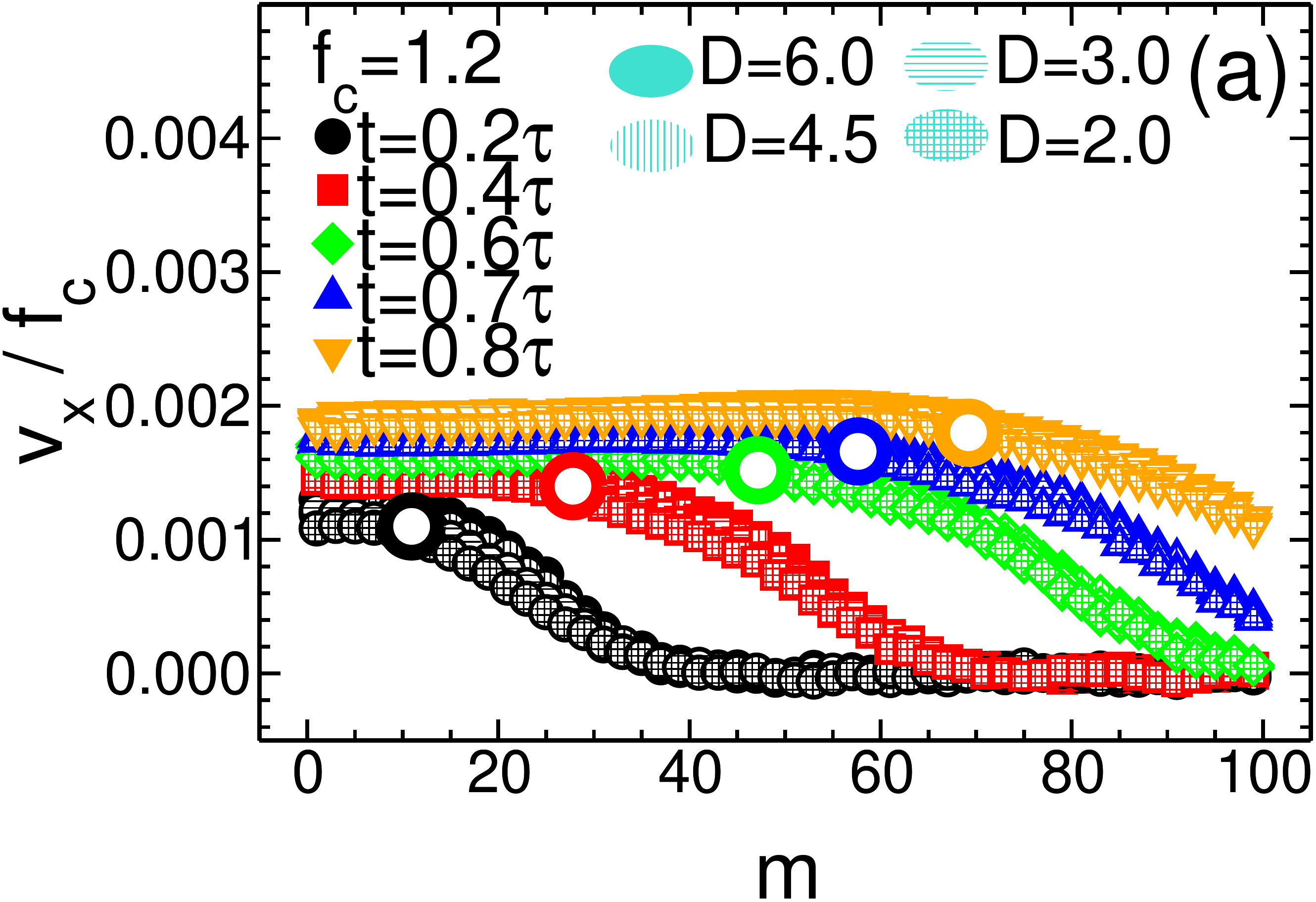}
    \end{center}\end{minipage} \hskip-0.1cm
        \begin{minipage}{0.235\textwidth}
    \begin{center}
        \includegraphics[width=1.0\textwidth]{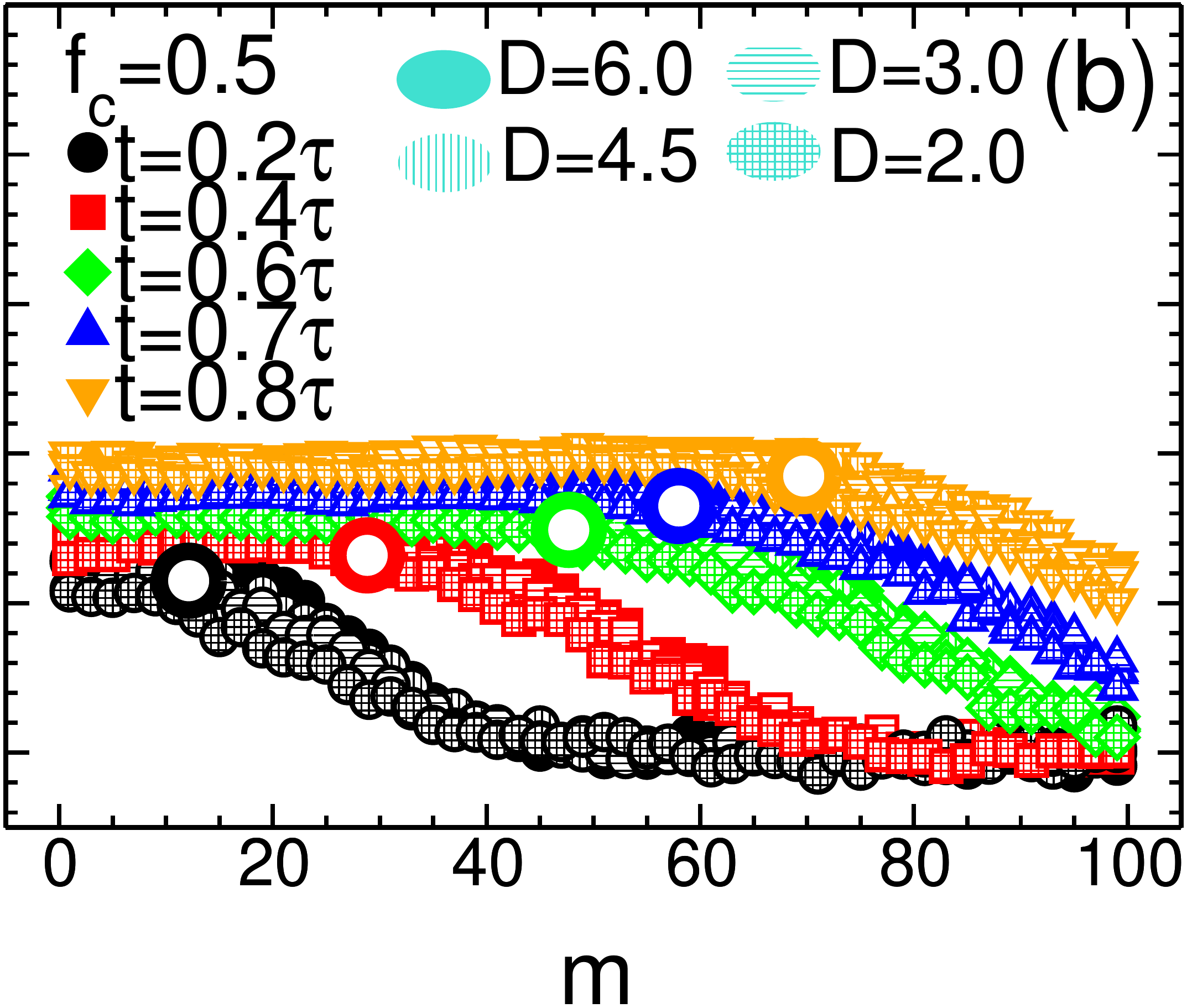}
    \end{center}\end{minipage} \hskip-0.1cm
        \begin{minipage}{0.235\textwidth}
    \begin{center}
        \includegraphics[width=1.0\textwidth]{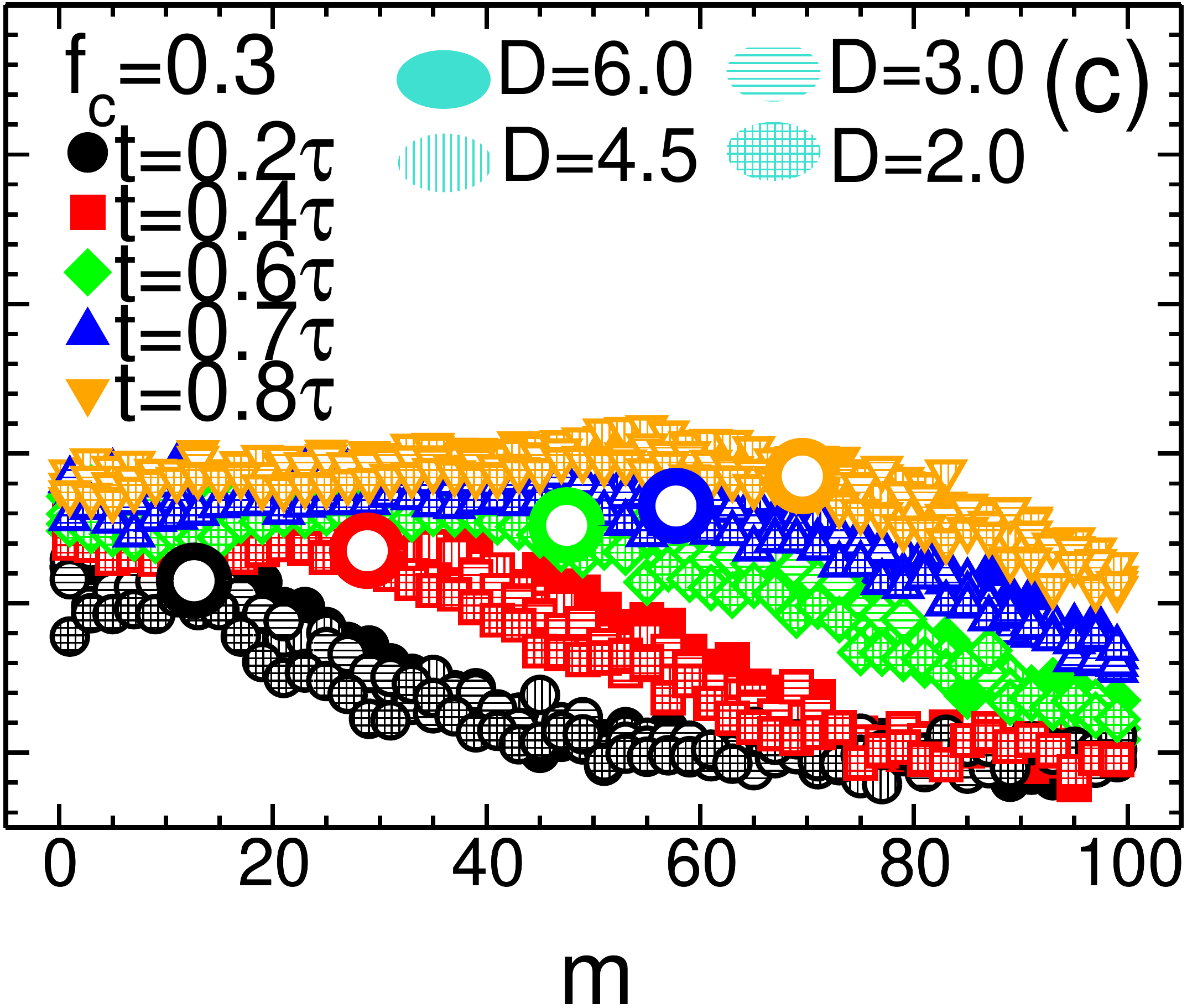}
    \end{center}\end{minipage} \hskip-0.1cm
    	\begin{minipage}{0.235\textwidth}
    \begin{center}
        \includegraphics[width=1.0\textwidth]{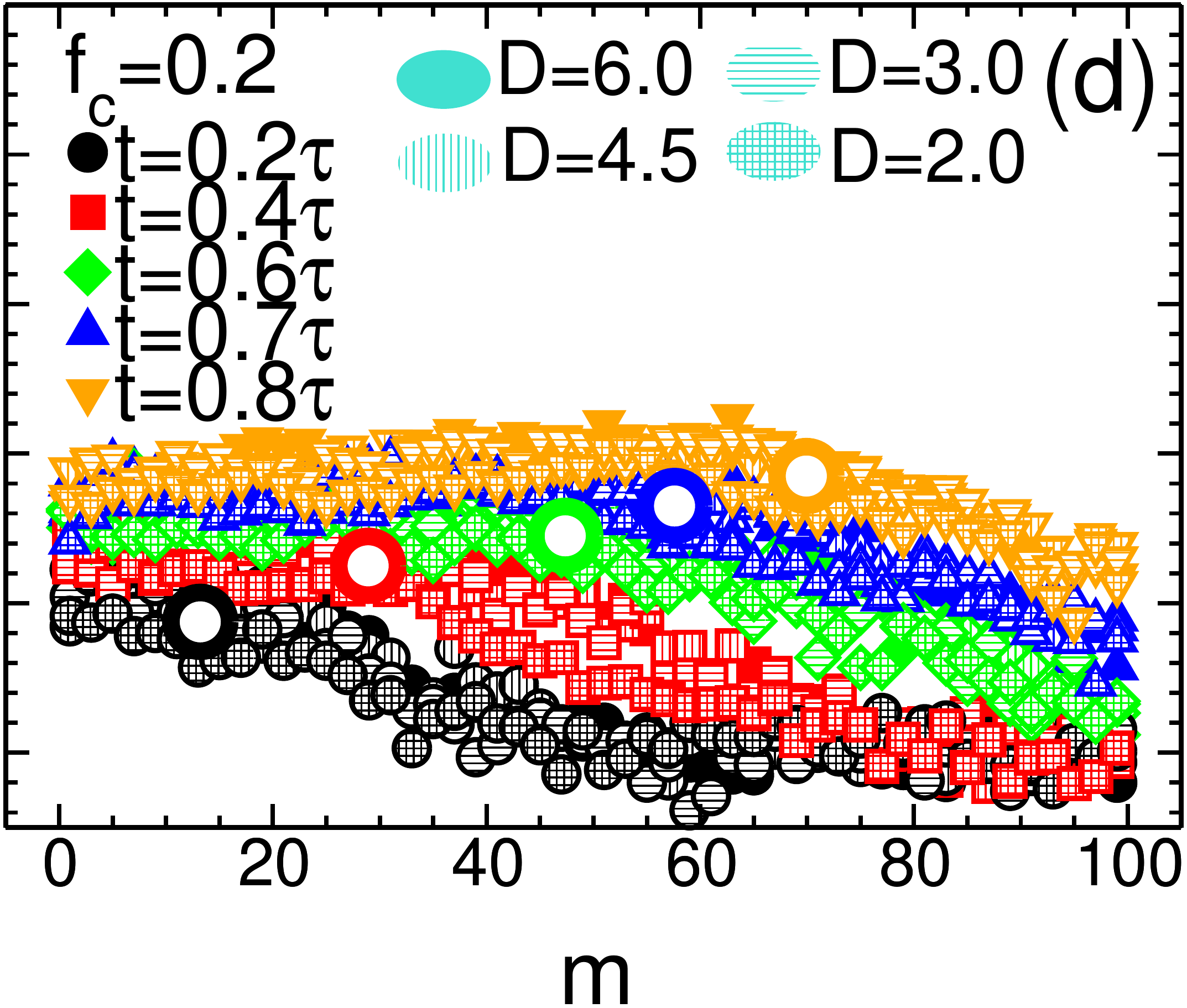}
    \end{center}\end{minipage}
\caption{(a) Normalized $x$ component of the monomer velocity $v_x/
f_{\textrm{c}}$ as function of the monomer index $m$, for fixed
channel driving force $f_{\textrm{c}}=1.2$ and channel widths $D=6.0$
(solid filled symbols), 4.5 (vertical hashed filled), 3.0 (horizontal
hashed filled), and 2.0 (cross hashed filled), during the translocation
process at times $t/\tau=0.2$ (black circles), 0.4 (red squares), 0.6
(green diamonds), 0.7 (blue triangles) and 0.8 (orange upside triangles).
The empty circles approximately specify the monomer index
inside the entrance of the channel at the corresponding moments. Panels (b),
(c) and (d) are the same as panel (a) but for channel driving forces
$f_{\textrm{c}}= 0.5$, 0.3 and 0.2, respectively.}
\label{fig:velocity}
\end{center}
\end{figure*}

\end{document}